\documentclass[twocolumn]{aastex631}

\usepackage{graphicx}	
\usepackage{amsmath}	
\usepackage{amssymb}	
\usepackage{xcolor}
\usepackage{soul}

\def\omA{\omega_{\rm A}}
\def\tA{t_{\rm A}}
\def\kA{k_{\rm A}}
\def\kd{k_{\rm d}}
\def\tb{t_{\rm b}}
\def\tiw{t_{\Omega}}

\newcommand{\om}{2\Omega}

\def\bu{\boldsymbol{u}}
\def\bB{\boldsymbol{B}}
\def\bb{\boldsymbol{b}}
\def\bOm{\boldsymbol{\Omega}}
\def\bk{\boldsymbol{k}}
\def\bfL{\boldsymbol{f}_B}
\def\bj{\boldsymbol{j}}
\def\be{\boldsymbol{e}}

\def\kTI{k_{\rm TI}}

\def\omm{\hat{\omega}}
\def\omnu{\hat{\omega}_\nu}
\def\ometa{\hat{\omega}_\eta}
\def\omkap{\hat{\omega}_\kappa}

\def\Eq{Equation}
\def\Eqs{Equations}
\def\beq{\begin{equation}}
\def\eeq{\end{equation}}

\begin{document}

\title{Tayler Instability Revisited}

\author{Valentin A. Skoutnev}
\affiliation{Physics Department and Columbia Astrophysics Laboratory, Columbia University, 538 West 120th Street New York, NY 10027,USA}

\author{Andrei M. Beloborodov}
\affiliation{Physics Department and Columbia Astrophysics Laboratory, Columbia University, 538 West 120th Street New York, NY 10027,USA}
\affiliation{Max Planck Institute for Astrophysics, Karl-Schwarzschild-Str. 1, D-85741, Garching, Germany}



\begin{abstract}
Tayler instability of toroidal magnetic fields $B_\phi$ is broadly invoked as a trigger for turbulence and angular momentum transport in stars. This paper presents a systematic revision of the linear stability analysis for a rotating, magnetized, and stably stratified star. For plausible configurations of $B_\phi$, instability requires diffusive processes: viscosity, magnetic diffusivity, or thermal/compositional diffusion. Our results reveal a new physical picture, demonstrating how different diffusive effects independently trigger instability of two types of waves in the rotating star: magnetostrophic waves and inertial waves. It develops via overstability of the waves, whose growth rate sharply peaks at some characteristic wavenumbers. We determine instability conditions for each wave branch and find the characteristic wavenumbers. The results are qualitatively different for stars with magnetic Prandtl number $Pm\ll 1$ (e.g. the Sun) and $Pm\gg 1$ (e.g. protoneutron stars). The parameter dependence of unstable modes suggests a non-universal scaling of the possible Tayler-Spruit dynamo.
\end{abstract}



\section{Introduction}

Turbulence triggered by fluid instabilities in stars can amplify magnetic fields and facilitate the transport of angular momentum across stellar interiors. Angular momentum transport in turn affects chemical mixing and stellar evolution \citep{maeder2000evolution}. Hydrodynamic processes alone are insufficient to explain the observed slow rotation of stars, which implies a key role of magnetohydrodynamic (MHD) processes \citep{eggenberger2012angular,ceillier2013understanding,marques2013seismic,fuller2014angular,cantiello2014angular,den2019constraining,ouazzani2019gamma,schurmann2022spins}.

One popular candidate mechanism for both generation of magnetic fields and angular momentum transport via Maxwell stresses is the Tayler-Spruit dynamo \citep{spruit2002dynamo}. The mechanism can operate in a stably stratified fluid and invokes turbulence triggered by the Tayler instability (TI) of a toroidal magnetic field \citep{tayler1973adiabatic}. Although rotation tends to suppress the instability and can completely choke it in ideal MHD \citep{pitts1985adiabatic}, the presence of non-ideal processes turns out to enable the TI even when rotation is fast \citep{spruit1999differential}. The Tayler-Spruit dynamo has been applied to stellar evolution models using two versions of nonlinear turbulence saturation \citep{spruit2002dynamo,fuller2019slowing}, improving the agreement of models with observations. It has also been discussed as a possible dynamo mechanism in nascent neutron stars \citep{barrere2022,margalit2022angular}.

The first basic step in building a dynamo model is a linear stability analysis that shows the most unstable modes at which turbulence is driven. In this paper we present a systematic revision of the stability analysis. Our results reveal a new physical picture of TI, showing how different diffusive effects independently trigger instabilities in different wave branches. Our analysis includes viscosity $\nu$ as well as magnetic diffusivity $\eta$ and thermal or compositional diffusion $\kappa$. This allows us to examine TI for any magnetic Prandtl number $Pm=\nu/\eta$. We find qualitatively different behavior in stars with $Pm\ll 1$ and $Pm\gg 1$.

We will investigate the stability of rotating configurations with initially axisymmetric toroidal magnetic fields. Linear perturbations are characterized by an azimuthal number $m$ (it turns out that $m=1$ is the most relevant mode) and a wavevector $\bk_{\rm pol}$ in the poloidal plane. We examine perturbations varying on a global scale in azimuthal angle $\phi$ and on small scales in the poloidal plane: $k_{\rm pol}\gg r^{-1}$, where $r$ is the distance from the magnetic (and rotation) axis. The general dispersion relation for such perturbations was first established by \cite{acheson1978instability}, and a simpler version is obtained in the Boussinesq approximation.

We will derive the approximate analytical solutions of the dispersion relation and verify them numerically. This analytical approach leads to a clear picture of the instability growth rate $\gamma(k)$, with well defined maxima at characteristic wavenumbers in each wave branch. It also reveals how the conditions for instability and the most unstable modes depend on the parameters of the problem: the star's rotation rate $\Omega$, Brunt-V\"ais\"al\"a frequency $N$, viscosity $\nu$, magnetic diffusivity $\eta$, thermal/compositional diffusivity $\kappa$, and the toroidal magnetic field $B_\phi$. 

A convenient parameter equivalent to $B_\phi$
is the Alfv\'en frequency $\omA\equiv B_\phi/(4\pi\rho r^2)^{1/2}$, where $\rho$ is the local fluid density. The primary regime of interest is that of fast rotation, $\omA\ll\Omega$. For completeness, we also discuss the non-rotating limit $\Omega\ll\omA$. It is useful to see how the instability growth rate found in non-rotating stars, $\gamma_{\Omega=0}(k)$, is reduced by rotation and how this reduction shapes the peaks of $\gamma(k)$.

In the limit of negligible rotation ($\Omega\ll\omA$), the TI proceeds via direct onset with a pure imaginary frequency. It has been thoroughly studied using both the ideal MHD energy principle \citep{tayler1973adiabatic,goossens1978hydromagnetic,goossens1980sigma,goossens1980unstably,goossens1981additional,goldstein2019tayler} and local WKB analysis \citep{spruit1999differential}, as well as numerical simulations \citep{gellert2008helicity,bonanno2012breakdown,weber2015tayler,guerrero2019global}. 

The rotating regime is qualitatively different because the Coriolis force gives perturbations a real part to their frequency. The perturbations become oscillating modes, which include inertial waves (IW) and magnetostrophic waves (MW) (e.g. see \cite{lehnert1954magnetohydrodynamic,greenspan1968theory}). TI develops as an overstability of the waves. 

Several previous works presented important analytical studies of the rotating TI \citep{pitts1985adiabatic,spruit1999differential,zhang2003nonaxisymmetric,bonanno2013rotational,ma2019angular} and its numerical simulations \citep{braithwaite2006differential,zahn2007magnetic,arlt2011amplification,ibanez2015stability,monteiro2023global,ji2023magnetohydrodynamic,petitdemange2023spin,barrere2023numerical}. However, the analytical works are incomplete because they did not solve the dispersion relation and missed the presence of multiple unstable wave branches, which lead to confusion. In addition, no effects of viscosity $\nu$ were included. Comparison of our results with the previous work will be given in the main text and Appendix~\ref{ap:PreviousProofs}.

The paper is organized as follows. The linear stability problem is set up in Section \ref{sec:setup}, including a simple derivation of the dispersion relation. The non-rotating case is reviewed in Section~\ref{sec:nonrotatingTI}. Section~\ref{sec:RotatingTI} then analyzes the rapidly rotating case and contains our main results with analytical expressions for the instability growth rate. Section~\ref{sec:CondInst} summarizes the necessary and sufficient instability criteria in all regimes. The results and their implications are discussed in Section~\ref{sec:discussion}.

\section{Setup of stability analysis}\label{sec:setup}

\subsection{Background state}\label{sec:backgroundstate}

We are interested in rotating and magnetized stars and consider their zones with stable stratification (thermal or compositional). Due to fast horizontal transport of angular momentum, a good approximation in stably stratified regions is the ``rotating shell'' approximation \citep{zahn1992circulation}: the rotation rate $\Omega$ is a function only of the spherical radius $R$, and differential rotation is quantified by $q=d\ln \Omega/d\ln R$. Following the symmetry of the problem, we will use both spherical coordinates $(R,\theta,\phi)$ and cylindrical coordinates $(r,\phi,z)$. 

Similar to the previous works, we will examine the stability of a magnetized equilibrium state (the ``background state'') to small perturbations. An equilibrium state must satisfy a balance between gravity, the pressure gradient, and the Lorentz forces. Such equilibria in general exist for a non-barotropic fluid (ideal gas with two independent  parameters, pressure and temperature), even with pure toroidal magnetic fields \citep{reisenegger2009stable}.\footnote{For a fluid with hydrostatic pressure $P\gg B^2/8\pi$, the irrotational part of the Lorentz force can be balanced by a slight adjustment of the pressure gradient $\nabla P$, and the freedom in choosing a small temperature variation may be used to adjust density $\rho$ (and hence gravity force) to balance the solenoidal part of the Lorentz force. Note also that the solenoidal force component disappears in cylindrical geometry that describes the fluid near the polar axis; this polar region is also most relevant for TI, since the instability is fastest near the axis.}
We will consider a background toroidal field of the form 
\begin{equation}
\label{eq:Bphi}
    \bB=B_\phi\boldsymbol{e}_\phi, \qquad B_\phi\propto r^p \cos \theta, \qquad  r=R\sin\theta,
\end{equation}
where $\boldsymbol{e}_\phi$ is the unit toroidal vector. Near the axis where $\cos^2\theta\approx 1$, the field configuration approaches the cylindrical setup studied in \cite{spruit1999differential}. Note that any finite electric current density on the axis requires $p=1$  by Stokes theorem. Since the TI is known to be most unstable near the axis, $p=1$ is the main case of interest. However, different $\theta$-profiles of $B_\phi$ could in principle give a larger local $p\equiv d\ln B_\phi/d\ln r$ away from the axis. We incorporate this possibility in our local stability analysis at $\theta\neq 0$ by using the profile~(\ref{eq:Bphi}) with a free parameter $p$.
Magnetic configurations with more general profiles of $B_\phi$ may also be considered. However, it turns out that the latitudinal dependence of $B_\phi$ dominates the TI dispersion relation (see Appendix \ref{ap:GeneralDispDer}); therefore we keep the simple form of $B_\phi$ with the single parameter $p$.

In a differentially rotating star, a simple configuration of $B_\phi(R,\theta)$ may originate from toroidal winding of an axisymmetric poloidal field with radial component $B_R\ll B_\phi$. In particular, winding of a dipole radial component $B_R\propto\cos\theta$ gives $B_\phi\propto \sin\theta \cos \theta$ according to the azimuthal component of the induction equation:
\begin{align}
\label{eq:winding}
    \frac{\partial B_\phi}{\partial t}=q\Omega B_R\sin\theta.
\end{align}
This form of the generated toroidal field corresponds to the aforementioned case $p=1$ \citep{spruit1999differential,zahn2007magnetic,ma2019angular} and represents the lowest spherical harmonic $l=2$ for a general axisymmetric toroidal vector field. 

Apart from sourcing the growth of $B_\phi$, the poloidal magnetic field can affect the fluid stability. It can suppress the TI because it introduces an additional energy cost to perturbations \citep{braithwaite2009axisymmetric}. At the same time, it can enable a different MHD instability: the magneto-rotational instability (MRI) \citep{velikhov1959stability,balbus1991powerful}. MRI occurs only if the fluid rotation rate decreases with cylindrical radius, $(\partial\ln \Omega/\partial\ln r)_z=q\sin^2\theta<0$ where $r^2=R^2-z^2$. It can have a large growth rate $\gamma_{\rm MRI}\sim q \Omega  \sin^2 \theta$ away from the polar regions (if not suppressed by stratification) and become dominant over an underlying TI \citep{wheeler2015role}. Since our interest is the TI, we focus below on pure toroidal field configurations, which are most unstable to the TI. Our analysis applies to cases where $B_R$ is sufficiently weak to not affect the TI and where the MRI is not present (in particular where $q\geq0$ or where stratification suppresses MRI if $q< 0$). The role of the poloidal field is described in more detail in Section \ref{sec:radialfield}.

\subsection{Dynamical equations for small perturbations}

Suppose the background configuration is perturbed with some infinitesimal velocity field $\bu$ (measured in the background fluid rest frame). In general, the velocity $\bu$ is accompanied by perturbations of fluid density, pressure, and magnetic field; these perturbations will be denoted by $\rho'$,  $P'$, and $\bb$, respectively. The change $\bB\rightarrow\bB+\bb$ implies a perturbation of the Lorentz force $\bfL=[(\nabla\times \bb)\times\bB + (\nabla\times \bB)\times\bb]/4\pi$.

The perturbation evolution obeys the MHD equations. For small-scale subsonic perturbations in a stratified fluid, the equations can be simplified using the Boussinesq approximation (see e.g. \cite{garaud2018double} for a review). It employs expansion in the limit of a large sound speed (compared to the perturbation velocities and the Alfv\'en speed) and a large hydrostatic scale-height (compared to the radial  wavelength of perturbations). Then the dynamical equations become \citep{spiegel1960boussinesq}
\begin{align}
\partial_t \boldsymbol{u}
=&-\frac{\nabla P'}{\rho_m}
 +\vartheta\, \boldsymbol{e}_R
 -2\boldsymbol{\Omega}\times\boldsymbol{u}
 + \frac{\bfL}{\rho_m}
+\nu\nabla^2 \boldsymbol{u}, 
\label{eq:mom}
\\
\label{eq:theta}
\partial_t \vartheta =& -N^2\boldsymbol{u}\cdot
\boldsymbol{e}_R
+\kappa\nabla^2\vartheta,
\\
\partial_t \bb =&\nabla \times
\left(\boldsymbol{u}\times \boldsymbol{B}\right)
 +\eta\nabla^2 \bb, 
\label{eq:B}
\end{align}
where $\rho_m$ is the mean local density of the fluid, and $\boldsymbol{e}_R$ is the unit vector in the radial direction. The velocity field in the Boussinesq approximation is treated as incompressible,
\begin{equation}      
\nabla\cdot\boldsymbol{u}=0,
\end{equation}
and the buoyancy acceleration in the gravitational field $\boldsymbol{g}=-g\boldsymbol{e}_R$ is captured by the ``buoyancy variable'' $\vartheta =-g\rho'/\rho_m$.

The Brunt-V\"ais\"al\"a frequency $N$ is determined by the background stratification and enters the equation for $\vartheta$, in the coefficient of $u_R$.\footnote{This coefficient can be designated as $-N^2$ by definition, as it describes the appearance of density perturbation $\rho'$ in response to radial motions of the fluid $\boldsymbol{u}\cdot\be_R$ (resulting in the buoyancy force $-\rho'\boldsymbol{g}$).  For pure radial motions (and negligible $\kappa$), this supports oscillations with frequency $N$.} In the case of thermal stratification, $\rho'$ is related to the temperature perturbation $T'$ according to the equation of state $\rho'=\alpha T'$ where $\alpha=(\partial \rho/\partial T)_P$ is evaluated at constant pressure (the influence of pressure perturbation $P'$ on $\rho'$ is a second-order effect in the Boussinesq limit \citep{spiegel1960boussinesq}). Similarly, for compositional stratification, $\rho'$ is related to the chemical composition perturbation $C'$ by $\rho'=\beta C'$ where $\beta=(\partial \rho/\partial C)_P$.

Diffusion of $T'$ or $C'$ with a diffusion coefficient $\kappa$ directly translates to the diffusion of the buoyancy variable $\vartheta$ (\Eq~\ref{eq:theta}), with $\kappa$ being the thermal/compositional diffusivity in the case of thermal/compositional stratification. Extension to a mixed stratification will be discussed in an accompanying paper.

\subsection{Dispersion relation}
\label{sec:disp_relation}

\begin{figure}
	\includegraphics[width=0.9\linewidth]{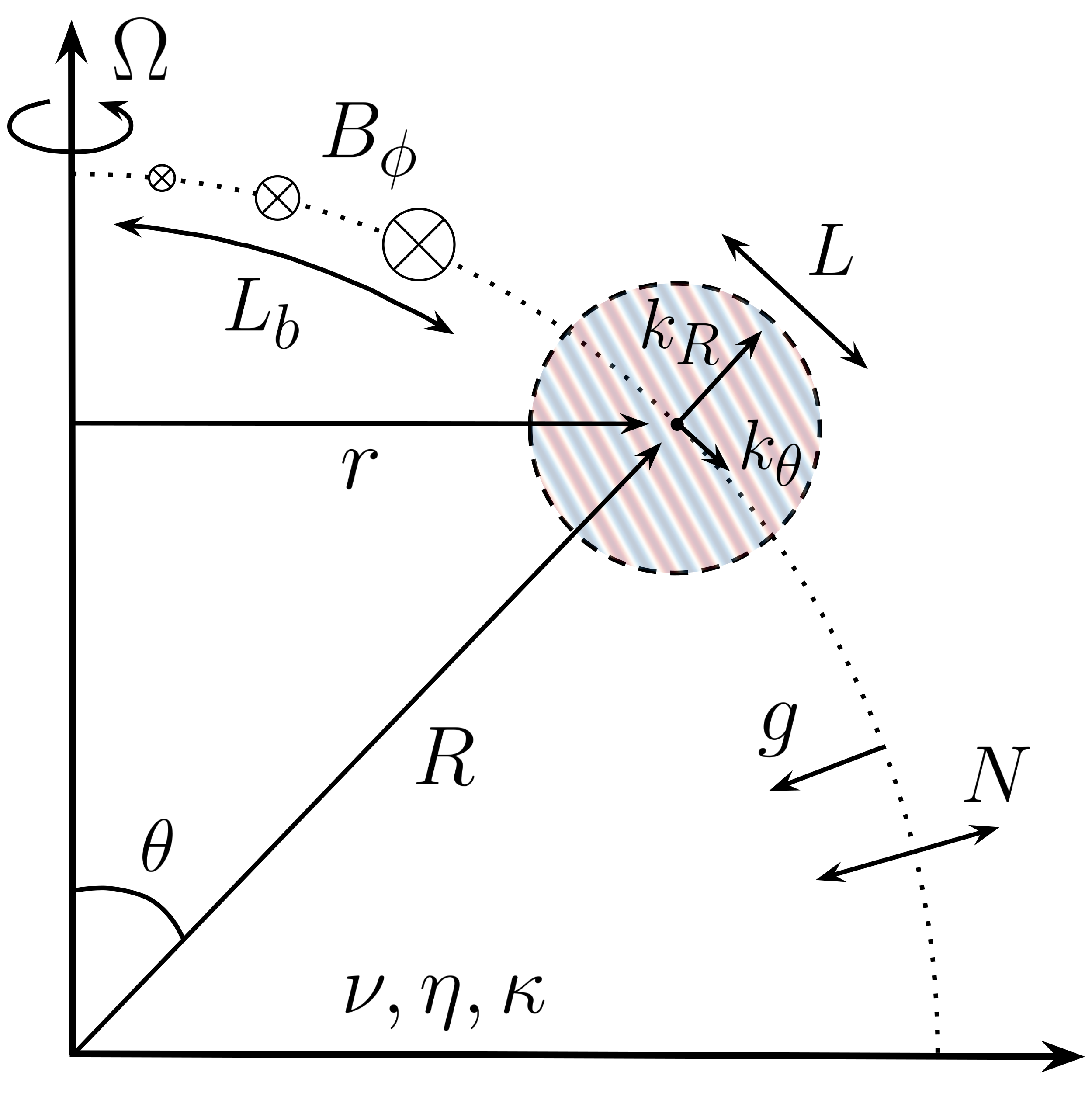}
    \caption{
Setup of the local stability analysis in spherical coordinates $(R,\theta,\phi)$. The star rotates with rate $\Omega$ about the (vertical) polar axis and has a stable radial stratification with a Brunt-V\"ais\"al\"a frequency $N$. The toroidal magnetic field $B_\phi$ varies on a length scale $L_b$. Circular region represents a local region of size $L\ll L_b$ where plane wave perturbations are considered with wavenumbers satisfying $k_R\gg k_\theta\gg L^{-1}$. The fluid has microphysical diffusivities $\nu$, $\eta$, and $\kappa$.
}
    \label{fig:Envelope}
\end{figure}

The most straightforward method to analyze the TI is to consider perturbations of the form 
\begin{equation}
\label{eq:pert}
 \boldsymbol{u},\bb, \vartheta
 \propto
  \exp[i(k_R R + l\theta + m\phi-\omega t)].
\end{equation}
The frequency is complex $\omega=\omega_r+i\gamma$ and instability corresponds to a positive imaginary component $\gamma>0$. The dependence on $R$ and $\theta$ corresponds to a local plane wave in the poloidal plane (the WKB approximation), as illustrated in Figure \ref{fig:Envelope}. 
The perturbation wavevector is
\beq
  \bk=(k_R,k_\theta,k_\phi), 
   \qquad k_\theta=\frac{l}{R}, 
  \quad k_\phi=\frac{m}{r}.
\eeq
While this approach may miss global modes and effects of boundary conditions, it should capture small-scale instabilities. Validity of the plane-wave solutions requires scale separation $k_{\rm pol}^{-1}\ll L\ll L_b$, where $k_{\rm pol}=\sqrt{k_R^2+k_\theta^2}$ is the poloidal wavenumber of the perturbation localized in a region of size $L$, and $L_b$ is the characteristic scale of the background gradients, 
$L_b\sim \min [(\nabla \ln B_\phi)^{-1},(\nabla \ln \Omega)^{-1}]$ \citep{ogilvie2007instabilities}. 

Since the velocity perturbation is solenoidal in the Boussinesq approximation, $\nabla\cdot\bu=0$, only solenoidal forces are relevant for the perturbation dynamics. Gradient forces (in particular $-\nabla P'$) are irrelevant and  can be removed by taking curl of the dynamical equation~(\ref{eq:mom}). Then, we obtain from \Eqs~(\ref{eq:mom})-(\ref{eq:B})
\begin{align}
\label{eq:curl}
   -i\omega_\nu \nabla\times \bu &=  \nabla\vartheta\times \be_R 
     +2\bOm\cdot\nabla\bu + \frac{\nabla\times \bfL}{\rho_m},
     \\
\label{eq:theta1}
    -i\omega_\kappa\vartheta &=-N^2u_R,  \\
\label{eq:B'}
    -i\omega_\eta\bb &=\bB\cdot\nabla \bu-\bu\cdot\nabla\bB, 
\end{align}
where we used $\nabla\cdot\bB=0$ and $\nabla\cdot\bu=0$, and defined
\beq
    \omega_\nu\equiv \omega+i\nu k^2,\;\;\,\omega_\eta \equiv \omega+i\eta k^2,\;\;\,\omega_\kappa\equiv \omega+i\kappa k^2.
\eeq

\Eqs~(\ref{eq:curl})-(\ref{eq:B'}) can be reduced to a single linear  equation for the fluid velocity $\bu$. The buoyancy variable $\vartheta$ is excluded using \Eq~(\ref{eq:theta1}), and the magnetic field perturbation $\bb$ is excluded using \Eq~(\ref{eq:B'}). The expression for $\bb$ should then be substituted into the Lorentz force perturbation $\bfL$, which we rewrite as 
\begin{equation}
\label{eq:fL}
   \bfL= \frac{\bB\cdot\nabla\bb + \bb\cdot\nabla\bB}{4\pi} -\nabla\left(\frac{\bB\cdot\bb}{4\pi}\right).
\end{equation}
The gradient term drops out from $\nabla\times\bfL$. 
Note that the relevant solenoidal part of $\bfL$ does not contain poloidal derivatives of $\bb$; it depends on the perturbation azimuthal wavenumber $m$ but not on the large poloidal wavevector $\bk_{\rm pol}$. For $m\sim 1$ the solenoidal part of the magnetic force is $\sim B b/r$. 

Using the expressions for $\vartheta$ and $\bb$ in terms of $\bu$, we obtain from \Eq~(\ref{eq:curl}) the linear differential equation for $\bu$. It simplifies in the WKB limit of $|k_j| \gg |\boldsymbol{e}_j\cdot \nabla\ln B_\phi|$, where $\boldsymbol{e}_j$ are the unit vectors for spherical ($j=R,\theta,\phi$) or cylindrical ($j=r,\phi,z$) coordinates.\footnote{For expected background fields with a moderate dependence on the distance to the polar axis, the WKB approximation is satisfied if $k_R r\gg 1$, $k_\theta r\gg 1$, and any $k_\phi$ (since $\partial_\phi\ln B_\phi=0$).} 
In particular, one can treat the coefficients of $\bu$ and $\partial_j\bu$ as constants and look for solutions in the form~(\ref{eq:pert}). This gives after some algebra 
\begin{align}
\nonumber
& \left( \omega_\nu - \frac{m^2\omA^2}{\omega_\eta} \right)  \bk \times \bu 
  = 
  \frac{N^2 u_R}{\omega_\kappa}  \, \bk \times \be_R + 2i (\bOm\cdot\bk)\,\bu \\
   & -
    \frac{2\omA^2}{\omega_\eta }  \left[ im u_r\,  \bk \times \be_\phi 
   - \left(im u_\phi - r\bu\cdot\nabla \ln \frac{B_\phi}{r}\right)\, \bk \times\be_r \right]. 
      \label{eq:u}
\end{align}
We have neglected $d\Omega/dR$ in \Eq~(\ref{eq:u}), which will limit our linear stability analysis to fluids in approximately uniform rotation. 

We will consider stably stratified regions of stars ($N^2>0$) where the Brunt-V\"ais\"sal\"a frequency $N\gg \Omega,\omA,|\omega|$ is the highest frequency in the system. For perturbations with $\bk$ not nearly aligned with $\be_R$, this ordering implies that the stratification term $\propto N^2$ dominates on the right side of \Eq~(\ref{eq:u}), and so the bulk of perturbations are ordinary internal gravity waves, with negligible effects of other forces. Hereafter, we focus on the more interesting modes with $\bk$ almost parallel to $\be_R$, for which the buoyancy term is reduced, allowing Coriolis and Lorentz forces to become important and make instability possible. The incompressibility condition $\nabla\cdot\bu=i\bk\cdot\bu=0$ then implies that the fluid velocity has a small radial component $u_R$, i.e. the fluid motions in the potentially unstable  modes are nearly horizontal. This fact is well known from previous work \citep{spruit1999differential}, which argued that nearly horizontal motions avoid the prohibitive energy penalty imposed by stratification. The modes of prime interest for instability have a low azimuthal number $m$, in particular $m=1$, as described in later sections. Therefore, the focus will be on wavevectors $\bk$  that satisfy
\begin{equation}
  k\approx k_R\gg k_\theta\gg  k_\phi.
\end{equation}
For such modes, $\bk\times\be_R\approx -k_\theta\be_\phi$ and $u_R\approx -k_\theta u_\theta/k$, so one can rewrite the stratification term in \Eq~(\ref{eq:u})
\beq
    \frac{N^2 u_R}{\omega_\kappa}  \, \bk \times \be_R  
    \approx  \frac{N^2}{\omega_\kappa}  \, \frac{k_\theta^2}{k}\,u_\theta \be_\phi.
\eeq
The non-zero misalignment of $\bk$ with $\be_R$ is important only in the stratification term where the small $\bk\times \be_R$ appears with the large prefactor $N^2$. In all other terms, one can take $\bk=k\be_R$, and then the condition $\bu\perp\bk$ leaves two unknowns in \Eq~(\ref{eq:u}), $u_\theta$ and $u_\phi$.

The $\theta$ and $\phi$ projections of \Eq~(\ref{eq:u}) give a system of two homogeneous linear equations for $u_\theta$ and $u_\phi$:
\begin{align}
\label{eq:u1}
 &  2i \cos\theta \left(\Omega+\frac{m\omA^2}{\omega_\eta}\right) u_\theta
   + \left(\omega_\nu-\frac{m^2\omA^2}{\omega_\eta} \right) u_\phi =0,\\
\label{eq:u2}
 & \left(\omega_\nu-\frac{m_\star^2\omA^2}{\omega_\eta} 
  - \frac{N^2k_\theta^2}{\omega_\kappa k^2} \right) u_\theta 
  = 2i \cos\theta \left(\Omega+\frac{m\omA^2}{\omega_\eta}\right) u_\phi,
\end{align}
where we have used $k_z/k\approx\cos\theta $ and defined
\beq
\label{eq:mstar}
  m_\star^2 \equiv  m^2 - 2\cos\theta\, \frac{r}{R}\,\partial_\theta \ln \frac{B_\phi}{r}
  = m^2 - 2 (p\cos^2\theta - 1).
\eeq
The determinant of this system must vanish for a non-zero solution, which gives the dispersion relation
\begin{align}
   \label{eq:SphericalFullIntro}
   D(\omega) =& \left(\omega_\nu\omega_\eta-m_\star^2\omA^2-\frac{k_\theta^2 N^2}{k^2}\frac{\omega_\eta}{\omega_\kappa}\right) \left(\omega_\nu\omega_\eta-m^2\omA^2\right)
    \nonumber\\
    &- 4\cos^2\! \theta \left(\Omega \omega_\eta+m\omA^2\right)^2=0.
\end{align}

The dispersion relation for MHD perturbations of the form~(\ref{eq:pert}) at low $m$ was first established by \cite{acheson1978instability} in cylindrical coordinates. In Appendix~\ref{ap:GeneralDispDer} we verify that the \cite{acheson1978instability} dispersion relation near the polar axis becomes \Eq~(\ref{eq:SphericalFullIntro}) after converting to spherical coordinates and taking the Boussinesq limit.
\Eq~(\ref{eq:SphericalFullIntro}) is also identical to the dispersion relations found in \cite{masada2007effect} and \cite{kagan2014role} for the case of $q=0$ and no poloidal field, and in \cite{ma2019angular} for the case of $\nu=0$ and $B_\phi\propto R\sin\theta$.

Note that \Eq~(\ref{eq:SphericalFullIntro}) assumes an approximately uniform rotation of the fluid, $\Omega\approx const$. The differential rotation parameter $q=d\ln\Omega/d\ln R$ is important for generating the toroidal fields and in general may be comparable to unity, however in our analysis of the TI we set $q=0$ for simplicity. A possible effect of $q\neq 0$ on the TI growth rate is left for future work.

In the next sections, we solve the dispersion relation~(\ref{eq:SphericalFullIntro}) for $\omega(k)$. We are particularly interested in the imaginary part $\gamma(k)=Im(\omega)$, which determines the stability of the mode, and in the wavenumbers  $k$ at which instability grows fastest. We investigate $\gamma(k)$ by deriving approximate analytical expressions in various regimes, which we have verified by direct numerical solution of Equation~(\ref{eq:SphericalFullIntro}). The analytic approach conveniently shows how the existence of instability and the characteristic unstable wavenumbers depend on the microphysical fluid parameters $\nu$, $\eta$, $\kappa$ and the configuration parameters $\omA$, $\Omega$, $N$, $R$, and $p$. We begin with an overview of relevant timescales, which will help interpret the subsequent results for $\gamma(k)$.

\subsection{Timescales}\label{sec:timescales}

Dynamics of MHD perturbations are controlled by several effects including the Coriolis force, buoyancy response, Lorentz force, viscosity and magnetic diffusivity. The contribution of each effect for a given mode with wavevector $\bk$ and frequency $\omega$ can be described by a characteristic timescale, which may significantly differ from $|\omega|^{-1}$.   

In particular, consider perturbations with $m=1$ in the polar region $\cos\theta\approx 1$. By Stokes theorem, the background configuration near the polar axis has electric current density $\bj= (c B_\phi/2\pi r)\boldsymbol{e}_z$ (where $\be_z$ is the unit vector along the axis), and so a finite $\bj$ on the axis implies $B_\phi\propto r$, i.e. $p=1$. Then the dynamical equation for $\nabla\times\bu$ (\Eq~\ref{eq:u}) simplifies to
\beq
\label{eq:mom2}
i (\partial_t + \nu k^2)\,\be_z\times \bu =  \frac{N^2}{\omega_\kappa} \frac{k_\theta^2}{k^2}\,u_\theta \be_\phi
 +2i \Omega\,\bu  + \frac{\omA^2}{\omega_\eta } \left( \be_z\times\bu + 2 i \bu \right). 
\eeq
Here one can see the contribution of each force to the fluid acceleration $\partial_t\bu$ and identify the corresponding timescales for buoyancy $\tb$, Coriolis force $\tiw$, and magnetic force $t_B$:
\beq
\label{eq:timescales}
   \tb=\frac{|\omega_\kappa|}{N^2} \frac{k^2}{k_\theta^2}, \qquad \tiw = \frac{1}{2\Omega},
   \qquad t_B=\frac{|\omega_\eta|}{\omA^2}.
\eeq
In addition, the timescales for viscosity and magnetic diffusion are:
\beq
   t_\nu\equiv\frac{1}{\nu k^2}, \qquad t_\eta\equiv\frac{1}{\eta k^2}.
\eeq

Note that the timescale $t_B$ for the magnetic force to affect fluid motions in a given perturbation mode depends on the mode frequency $\omega$. In general it differs from the Alfv\'en time,
\beq
   \tA\equiv\frac{1}{\omA}=\frac{r\sqrt{4\pi \rho}}{B_\phi}.
\eeq
One finds $t_B=\tA$ if the mode frequency $|\omega|$ equals $\omA$ and magnetic diffusion is slower than the mode oscillation, $\eta k^2\ll |\omega|$. In particular, $|\omega|=\omA$ can occur when rotation and stratification effects are negligible, so that the perturbation dynamics is governed by the magnetic force alone.

The timescale $\tb$ that characterizes buoyancy response in a given mode depends on the diffusivity $\kappa$,
\begin{eqnarray}
 \tb=\left\{ \begin{array}{lc}
 \displaystyle{t_N\equiv\frac{k^2|\omega|}{k_\theta^2 N^2}} & \quad\kappa k^2\ll |\omega| 
 \vspace*{1mm} \\
 \displaystyle{t_\kappa\equiv\frac{\kappa k^4}{k_\theta^2 N^2}} & \quad \kappa k^2\gg |\omega| 
 \end{array}\right.
\label{eq:tb}
\end{eqnarray}
The physical meaning of the timescales $t_N$ and $t_\kappa$ can be examined when buoyancy dominates the fluid response to the perturbation (and the Coriolis and magnetic forces are negligible). 
Then, in the low-$\kappa$ regime one finds the dispersion relation $\omega\approx k_\theta N/k=t_N^{-1}$, which describes the usual internal gravity waves. In the high-$\kappa$ regime one finds $\omega\approx -iN^2k_\theta^2/\kappa k^4$, which implies damping of the perturbation on the timescale $t_\kappa$.

Note that the effect of diffusivity $\kappa$ (which may be thermal or compositional) is more subtle than the usual diffusion of fluctuations due to viscosity or magnetic diffusivity, as $\kappa$ enters by qualitatively changing the buoyancy response to perturbations. The effect of high $\kappa$ is strong  at {\em small} $k$ (i.e. on large scales), since $t_\kappa\propto \kappa k^4$. The peculiar damping of large-scale perturbations  by the diffusive buoyancy response is known as the low Peclet regime \citep{lignieres1999small}, where internal gravity waves are overdamped and the usual buoyancy restoring force becomes a drag force on vertical motions with the timescale $t_\kappa$. This regime has been discussed in the context of dynamical shear instabilities and stably stratified turbulence in the low $Pr$ regions of stellar radiative zones \citep{lignieres1999small,lignieres2020turbulence,cope2020dynamics,garaud2021journey,skoutnev2023critical}.

\section{Tayler instability in non-rotating stars}
\label{sec:nonrotatingTI}

Before investigating the rapid rotation regime of primary interest, $\Omega\gg\omA$, we first examine the simpler non-rotating case $\Omega=0$. It shares a few key physical elements with the rapidly rotating case, and it will be shown that the growth rate curves $\gamma(k)$ in both regimes have some common features.

With no rotation ($\Omega=0$), instability proceeds via direct onset with a purely imaginary frequency $\omega=i\gamma$. The magnetic field can be thought of as nested and stacked loops, which develop the instability if they find a way to reduce the total magnetic energy by slipping past each other. The picture is particularly simple in the axisymmetric mode $m=0$, which turns out to be unstable in the non-rotating fluid for magnetic configurations with a steep gradient, $p>1$. The instability develops as interchange of loops of stronger fields further from the symmetry axis with loops of weaker fields closer to the axis thereby reducing magnetic energy while maintaining constant magnetic flux \citep{tayler1973adiabatic}.

The formal stability analysis of modes with $m\geq 0$ for any $p$ involves analyzing the five roots $\omega(k)$ of \Eq~(\ref{eq:SphericalFullIntro}) with $\Omega=0$,
\begin{align}\label{eq:NRfull}
    &\left(\omega_\nu\omega_\eta-
m_\star^2\omA^2-\frac{k_\theta^2 N^2}{k^2} \frac{\omega_\eta}{\omega_\kappa} \right)\left(\omega_\nu\omega_\eta-m^2\omA^2\right)
\nonumber\\
    &-\left(2m\omA^2\cos\theta\right)^2=0.
\end{align}
We look for roots that have $\gamma(k)>0$ (unstable modes). It is instructive to  examine first the case of ideal MHD ($\nu=\eta=\kappa=0$) and then include diffusive effects.

In the case of $\nu=\eta=\kappa=0$, the dispersion relation~(\ref{eq:NRfull}) becomes a quadratic in $\omega^2$ (the fifth root is  $\omega=0$),
\begin{align}
\label{eq:quadratic}
 \frac{\omega^4}{\omA^4}&-\left(m_\star^2+m^2+\frac{k_N^2}{k^2}\right)\frac{\omega^2}{\omA^2}\nonumber\\
 &+m^2\left(m_\star^2+\frac{k_N^2}{k^2}-4\cos^2\theta\right)=0,
\end{align}
where
\begin{equation}
    k_N\equiv\frac{k_\theta N}{\omA}.
\end{equation}
For axisymmetric modes $m=0$ this gives $ \omega^2/\omA^2=2-2p\cos^2\theta+k_N^2/k^2$, and the instability condition $\omega^2<0$ is reduced to
\begin{equation}
\label{eq:instability_zero_m}
    p\cos^2\theta>1+\frac{k_N^2}{2k^2} \qquad (m=0).
\end{equation}
For non-axisymmetric modes, \Eq~(\ref{eq:quadratic}) gives two roots for $\omega^2$.
One of the roots satisfies $\omega^2<0$ (i.e. there is an instability) if $m^2[m_\star^2+(k_N^2/k^2)-4\cos^2\theta]<0$, which is reduced to
\begin{equation}
\label{eq:instability_positive_m}
  (2+p)\cos^2\theta-\frac{m^2}{2}>1+\frac{k_N^2}{2k^2} \qquad (m>0).
\end{equation}
This condition requires $m^2<2+2p$.

In particular, in the case of main interest $p=1$, only modes with $0<m<2$ can have $\omega^2<0$, i.e. instability requires $m=1$. Moreover, instability develops only in the polar region $\cos^2\theta>1/2$. The solution for $\omega^2$ in the case of $p=1$ for the  modes with $m=1$ is
\begin{equation}
\label{eq:ideal_p1}
  \frac{\omega^2}{\omA^2} = 1+\sin^2\theta + \frac{k_N^2}{2k^2}
     \pm \sqrt{\left(\sin^2\theta + \frac{k_N^2}{2k^2} \right)^2+4 \cos^2\theta}.
\end{equation} 
The instability is fastest near the polar axis $\cos^2\theta\approx 1$ and at high wavenumbers $k\gg k_N$; then the unstable root is $\omega^2/\omA^2\approx -1$, i.e. the growth rate is $\gamma\approx\omA$. Perturbations with lower wavenumbers $k\ll k_N$ have $\omega^2>0$ and so are stable. Note that the instability condition $k\gtrsim k_N$ implies $k/k_\theta\gtrsim N/\omA\gg 1$. This highlights the fact that stratification allows TI only for perturbations with $k_R\gg k_\theta$.

The simple regime of ideal MHD in a non-rotating star exhibits a general feature of the TI: instability develops near the polar axis, and magnetic configurations with $p=1$ are unstable only in the $m=1$ modes. This feature can be verified analytically or by scanning the numerical solutions of the dispersion relation at various parameters (including the rapidly rotating case described in Section~\ref{sec:RotatingTI} below). Therefore, hereafter our description of the TI mainly focuses on the polar regions $\cos^2\theta\approx 1$ and $m=1$ . Note that these parameters imply $m_\star^2=3-2p$, which gives  $m_\star=1$ for $p=1$ and $m_\star^2>0$ for $p<3/2$.

\begin{figure}
	\includegraphics[width=0.95\linewidth]{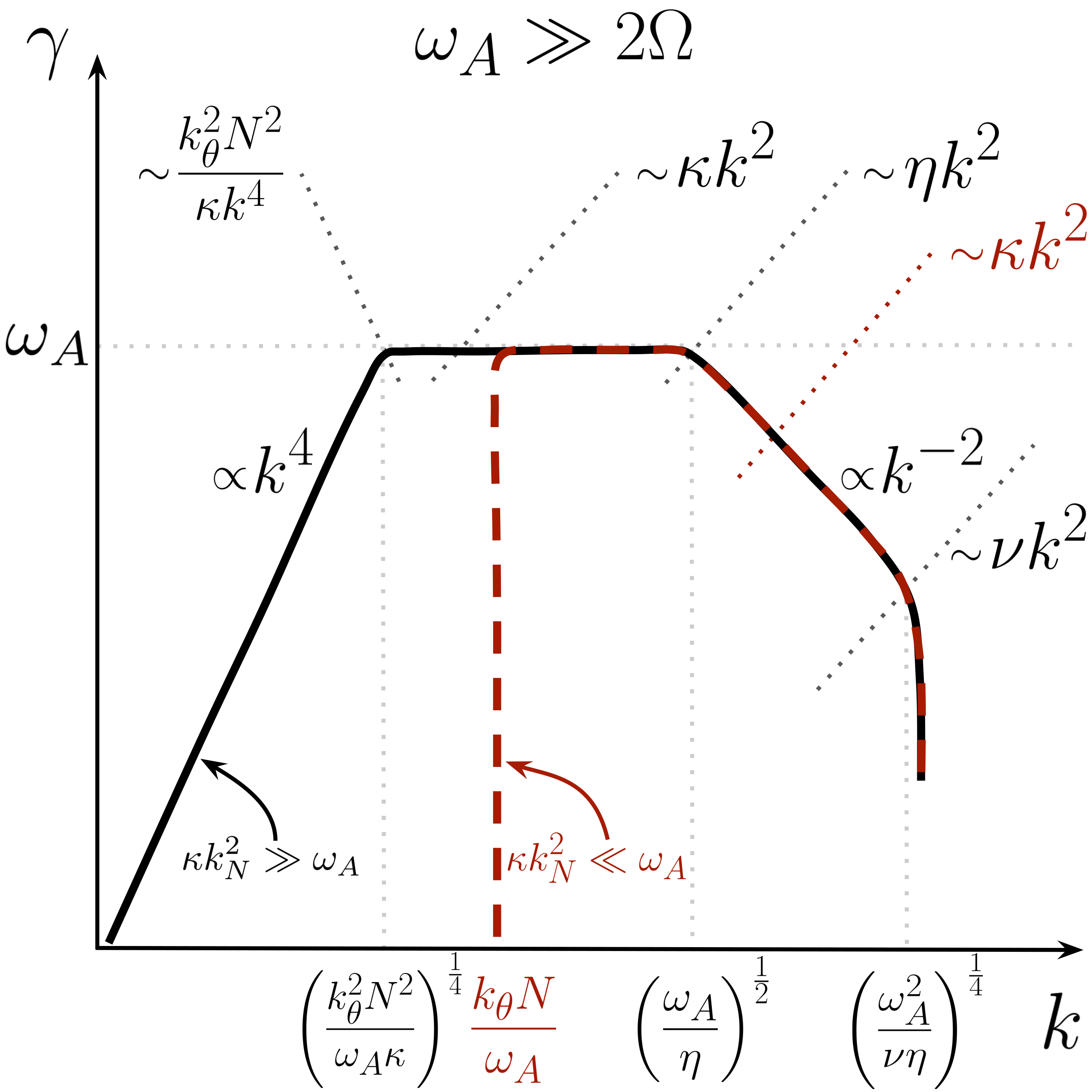}
    \caption{
    Growth rate $\gamma$ versus wavenumber $k\approx k_R\gg k_\theta$ of the Tayler instability for the $p=1$ magnetic configuration in a non-rotating star. Both vertical and horizontal axes are on a logarithmic scale. The diagram assumes $Pm=\nu/\eta\ll1$, however the dispersion relation with $\Omega=0$ is symmetric with respect to $\nu$ and $\eta$ and so a similar diagram for $Pm\gg 1$ is obtained by the simple change $\nu\leftrightarrow\eta$. Instability is suppressed at high $k$ by viscous/magnetic diffusion. At low $k$, instability is limited by buoyancy forces due to stratification. The solid black and dashed red curves correspond to the cases when buoyancy forces are in a highly diffusive or non-diffusive regime, respectively (see text).
    \\
    }  
\label{fig:NonRotatingTI}
\end{figure}
 
We now derive $\gamma(k)$ in non-ideal MHD; the result for the unstable root is summarized in Figure~\ref{fig:NonRotatingTI}. First, consider the high wavenumbers $k$ where buoyancy effects are negligible, $\tb\gg \tA=\omega_A^{-1}$. Then, the term $\propto k_\theta^2N^2/k^2$ is negligible in the dispersion relation~(\ref{eq:NRfull}), and it is reduced to two quadratic equations that can be immediately solved:
\begin{align}
\label{eq:omegaNR_diff}
  \omega=&-\frac{i(\nu+\eta)k^2}{2}\nonumber\\
  &+ \epsilon_1\sqrt{-\frac{(\nu+\eta)^2k^4}{4} + ( 2\epsilon_2 +1)\omA^2+\nu\eta k^4}, 
\end{align}
where $\epsilon_1=\pm1$ and $\epsilon_2=\pm1$ give the four roots. If all diffusive and buoyancy effects are negligible on the instability timescale ($\tA\ll t_\eta,t_\nu,\tb$), then the root with $\epsilon_1=1$ and $\epsilon_2=-1$ gives $\gamma\approx \omA$, recovering the ideal MHD result.
At sufficiently high wavenumbers where effects of viscous or magnetic diffusion become important, the unstable solution in \Eq~(\ref{eq:omegaNR_diff}) becomes

\begin{equation}
\label{eq:gamma_visc_NR}
    \gamma\approx \frac{k_{\rm A}^2}{k^2}\omA-\frac{\nu\eta}{\nu+\eta} k^2,\qquad 
   \kA\ll k < k_{\nu\eta},
\end{equation}
where
\begin{equation}
    k_{\rm A}^2 \equiv \frac{\omA}{\nu+\eta},
    \qquad  k_{\nu\eta}^2\equiv\frac{\omA}{\sqrt{\nu\eta}}. 
\end{equation}
The growth rate $\gamma(k)$ is reduced below $\omA$ at $k>\kA$ 
because the slipping magnetic loops experience diffusion (magnetic or viscous, whichever is dominant) on the timescale $\min\{t_\eta,t_\nu\}<\tA$. The instability is completely suppressed ($\gamma<0$) at $k>k_{\nu\eta}$ where both $t_\eta$ and $t_\nu$ are shorter than $\gamma^{-1}$.

Buoyancy effects become important at sufficiently low wavenumbers $k$, so that the term $\propto k_\theta^2N^2/k^2$ is no longer negligible, and the dispersion relation becomes a fifth-order polynomial with five roots. Four of them are extensions of the four roots found in ideal MHD, modified by diffusivities as described below. The remaining fifth root $\omega_5$ can be non-zero when stratification effects ($N\neq0$) combine with the effects of $\eta\neq0$ or $\kappa\neq0$. However, we find that any instability in the fifth branch has a growth rate $\gamma_5\ll\omA$, small compared with the main branch of instability described below. Therefore, we omit the analysis of $\gamma_5$ for non-rotating stars. The fifth branch becomes more interesting in stars with fast rotation (Section~\ref{sec:RotatingTI}) where it can become the only unstable root at some parameters.

At low wavenumbers, where $\tb\ll\tA\ll t_\nu,t_\eta$, buoyancy effects are strong, and so stratification suppresses instability. The onset of this effect, where $\omega^2$ begins to deviate from $-\omA^2$, can be examined by looking for the solution in the form $\omega=i\omA(1+\varepsilon)$ with unknown $\varepsilon\ll 1$. Assuming that stratification becomes important at $k$ where effects of viscous and magnetic diffusion are negligible, we set $\omega_\nu=\omega_\eta=\omega$ in Equation~(\ref{eq:NRfull}) and find $\omega=i\gamma$ with

\begin{align}
\label{eq:NRgamma_stratpert}
    &\gamma\approx\omA \left[ 1-\frac{k_N^2}{4k^2}\frac{\omA}{\omA+\kappa k^2}\right],\quad \tA \ll \tb \ll t_\nu,t_\eta.
\end{align}
One can see two distinct behaviors of $\gamma(k)$ when $\kappa k_N^2 \ll \omA$ and $\kappa k_N^2 \gg \omA$ (shown by the red and black curves in Figure~\ref{fig:NonRotatingTI}).

In particular, when the role of diffusivity $\kappa$ in the stratification response is negligible, $\kappa k_N^2 \ll \omA$, \Eq~(\ref{eq:NRgamma_stratpert}) becomes
\begin{align}
\label{eq:gN}
    \gamma\approx \omA\left(1-\frac{k_N^2}{4k^2}\right)
    =\omA\left(1-\frac{1}{4t_N^2\omA^2}\right),
    \quad 
    k_N<k\ll k_{\rm A}.
\end{align}
It reproduces the ideal MHD solution expanded at $k\gg k_N$. The ideal MHD solution at any $k/k_N$ was given in Equation~(\ref{eq:ideal_p1}), and the instability criterion (Equation~\ref{eq:instability_positive_m}) implies stability ($\gamma<0$) at $k<k_N/\sqrt{3}$. Note that $k\lesssim k_N$ corresponds to $t_N\lesssim t_A$, i.e. stratification acts on the perturbations faster than the instability could develop. The suppression of instability at $k\lesssim k_N$ also admits a simple interpretation using the MHD energy principle: the potential energy required to overcome stratification exceeds the released magnetic energy for perturbations with $k\ll k_N$ \citep{tayler1973adiabatic}. 

In the regime of fast diffusivity,
$\kappa k_N^2\gg \omA$, \Eq~(\ref{eq:NRgamma_stratpert}) becomes
\begin{align}
\label{eq:gkappa_NR}
    \gamma\approx \omA\left(1-\frac{k_{\rm d}^4}{4 k^4}\right)
    =\omA\left(1-\frac{1}{4t_\kappa\omA}\right),
\;\; k_{\rm d}\ll k \ll k_{\rm A},
\end{align}

\begin{equation}
     k_{\rm d}\equiv \left(\frac{k_\theta^2N^2}{\omA \kappa}\right)^{1/4}.
\end{equation}
Note that $k_{\rm d}/k_N=(\omA/\kappa k_N^2)^{1/4}\ll 1$, so the high-$\kappa$ regime gives $\gamma\approx\omA$ in a greater interval of $k$ compared with the low-$\kappa$ regime (compare \Eqs~(\ref{eq:gN}) and (\ref{eq:gkappa_NR})). This occurs because the effects of stratification are reduced by the fast $\kappa$-diffusivity, and the perturbation experiences damping on the longer timescale $\tb=t_\kappa\gg t_N$ (\Eq~\ref{eq:tb}).

Lastly, at wavenumbers $k\ll k_{\rm d}$ the damping time is short $t_\kappa\ll\tA$, and we find that the instability persists, but with a diminished growth rate. When $t_\kappa\ll\tA$, the dispersion relation~(\ref{eq:NRfull}) can be satisfied only for small $\omega=i\gamma\ll i\omA$, implying an approximate balance between the stratification term $\propto N^2$ (with $\omega_\kappa\approx i\kappa k^2$) and the Alfv\'enic terms $\propto \omA^2$. The instability growth rate is then given by 
\begin{equation}
\label{eq:low_k}
      \gamma\approx 3\omA^2 t_\kappa=3\frac{k^4}{k_{\rm d}^4}\omA, \quad t_\kappa\ll \tA,\; k\ll k_{\rm d}.
\end{equation}
The growth rate is reduced at these low wavenumbers because slipping magnetic loops feel a significant effective drag force from the diffusive
buoyancy response (which resembles the effects of viscous/magnetic diffusion at high wavenumbers). When considering the low $k$ one should also keep in mind that \Eq~(\ref{eq:low_k}) is derived assuming  $k\approx k_R\gg k_\theta>r^{-1}$, which sets a lower bound on $k$.

In summary, the TI of magnetic configurations with $p=1$ occurs only for modes with $m=1$ and develops near the polar axis. Its growth rate in a non-rotating star is $\gamma\approx \omega_A$ for a range of wavenumbers bounded below by $k_N$ (when $\kappa k_N^2\ll\omA$) or $k_{\rm d}$ (when $\kappa k_N^2\gg\omA$), and bounded above by $k_{\rm A}$. In ideal MHD ($\nu=\eta=\kappa=0$), all perturbations with wavenumbers $k\gg k_N$ grow with rate $\gamma=\omA$.

\section{Tayler instability in rotating stars}
\label{sec:RotatingTI}

Rotation creates a Coriolis acceleration $-2\boldsymbol{\Omega}\times\boldsymbol{u}$ in response to velocity perturbations $\bu$. It qualitatively changes the fluid response to perturbations and reduces the parameter space for TI. We will investigate below the fast rotation regime, $\Omega\gg\omA$, where the Coriolis force acts much faster than the instability could develop, promoting stability. In this regime, the fluid is known to support two types of oscillations enabled by the Coriolis force: inertial waves (IW) and magnetostrophic waves (MW). Unlike the non-rotating case, a direct onset of instability with a pure imaginary $\omega=i\gamma$ becomes impossible. However, TI may still develop via overstability of the waves. As shown below, this can happen with both IW and MW, depending on the fluid parameters. In addition, there is a separate branch of the dispersion relation $\omega_5(k)$ which can also become unstable, although with a smaller growth rate. This ``fifth branch'' will be described in the end of this section.

One way to understand the rotating regime is to again start with the simple case of ideal MHD ($\nu=\eta=\kappa=0$). We use the ideal case as a brief introduction to IW and MW in Section \ref{sec:ideal_rot_MHD}, where we demonstrate the stability of IW in ideal MHD configurations with any $p=d\ln B_\phi/d\ln r$ and stability of MW with $p<3/2$. In Section~\ref{sec:effects_of_nu_eta} we examine the destabilizing role of viscosity and magnetic diffusion for waves at high wavenumbers where stratification effects are negligible. After this initial analysis, we describe the full stability problem, including stratification. Most of our analytical results will be given for the most relevant case of $p=1$. The results for IW (Section~\ref{sec:effects_of_strat_IW}) are similar for any $p$, and the results for MW (Section~\ref{sec:effects_ofkappa_onMW}) are similar for any $p<3/2$. MW instability at $p>3/2$ (including stratification and non-ideal effects) is described separately in Section~\ref{sec:strongfieldgradient}.

\subsection{Stability in ideal MHD}
\label{sec:ideal_rot_MHD}

\bigskip

With $\nu=\eta=\kappa=0$ the dispersion relation~(\ref{eq:SphericalFullIntro}) becomes
\begin{align}
\label{eq:disp_ideal}
   \left( \omega^2 - m_\star^2\omA^2 - \frac{k_\theta^2 N^2}{k^2} \right)&
  \left( \omega^2 - m^2\omA^2 \right)\nonumber\\
  &= 4\cos^2\!\theta \left( \Omega\omega + m\omA^2 \right)^2 .
\end{align}
The small parameter $\omA/\Omega\ll 1$ ($m$ turns out to be order unity for any unstable modes) leads to a large separation of two pairs of the roots $\omega(k)$ of this equation. It has two roots with $\omega\gg\omA$,
\begin{align}
\label{eq:IW_ideal}
   &\omega_{\rm IW}\approx\pm\sqrt{4\Omega^2\cos^2\theta+\omA^2\left(\frac{k_N^2}{k^2}+m_\star^2+m^2\pm 4m|\cos\theta|\right)},   
\end{align}
where we have kept terms up to order $(\omA/\Omega)^2$; we did not use any expansion in $k_N/k$ and only assumed $k_N/k\ll (\Omega/\omA)^2$. The other two roots with $\omega\ll\omA$ are
\begin{equation}
\label{eq:MW_ideal}
\omega_{\rm MW}\approx\frac{m\omA^2}{\om}\left(-2 \pm \frac{1}{|\cos\theta|}\sqrt{m_\star^2+\frac{k_N^2}{k^2}} \right).
\end{equation}

Near the polar axis ($|\cos\theta|\approx1$) \Eq~(\ref{eq:IW_ideal}) describes the IW with frequency $\omega_{\rm IW}\approx \pm 2\Omega$ (when stratification is negligible, i.e. when $t_N\gg t_\Omega$), and \Eq~(\ref{eq:MW_ideal}) describes the MW with frequency $\omega_{\rm MW}\propto \omA^2/2\Omega$. One can see that in ideal MHD the IW are always stable while MW are stable if $m_\star^2+k_N^2/k^2>0$. 

The physical pictures of the two waves types are quite different. In the IW, the fluid acceleration is dominated by the Coriolis force, and other forces are negligible, so the IW oscillates on the Coriolis timescale, $\omega_{\rm IW}\approx t_\Omega^{-1}$. Its instability may only occur in non-ideal MHD and will require a fast diffusive process on a timescale comparable to $\tiw$ (Section~\ref{sec:effects_of_nu_eta}). By contrast, the MW oscillates with a far lower frequency, so that the inertial term in the equation of motion is negligible, and the Lorentz force is in approximate balance with the Coriolis force. Note also that the MW with $\omega\neq 0$ are supported only for non-axisymmetric perturbations $m>0$ that invoke magnetic tension forces. 

Substituting the solutions $\omega(k)$ into the dynamical equations (\ref{eq:u1}) and (\ref{eq:u2}) reveals the polarization of the wave modes. Recall that we only consider wavevectors $\boldsymbol{k}$ dominated by the radial component $k_R$, because they are most prone to instability. In particular, waves near the polar axis have $\boldsymbol{k}$ nearly parallel to $\boldsymbol{\Omega}$. In both IW and MW, the velocity $\boldsymbol{u}$ and the corresponding magnetic field perturbation $\bb$ are nearly confined to the plane perpendicular to $\bOm$. In IW, $\bb$ has a small amplitude in the sense that $b/\sqrt{4\pi \rho}$ is much smaller than $u$:  $b/\sqrt{4\pi \rho}u\sim \omA/2\Omega\ll 1$. As a result, the Lorentz force is suppressed by the factor $\sim\left(\omA/2\Omega\right)^2$ compared to the Coriolis force. By contrast, MW have $b/\sqrt{4\pi \rho}u\sim(\omA/2\Omega)^{-1}\gg 1$. The MW oscillate with a small velocity (which allows the Coriolis force to be balanced by the Lorentz force), building up a large displacement and hence a large magnetic perturbations $b$ during the long wave period.

Since only MW may become unstable in ideal MHD, the stability analysis is reduced to examining \Eq~(\ref{eq:MW_ideal}). One can see that stratification promotes stability (this will not be true in non-ideal MHD with diffusivity $\kappa$, as discussed in Section~\ref{sec:effects_ofkappa_onMW}). For modes with $k\gg k_N /|m_\star^2|$ the stratification effects are negligible and the instability condition is $m_\star^2<0$. This condition is easiest to satisfy near the rotation axis, where it takes the form $p>m^2/2+1$. Since the lowest $m$ for MW is $m=1$, the instability condition in ideal MHD becomes $p>3/2$. This result is known from previous work \citep{spruit1999differential,zahn2007magnetic}. 

Magnetic configurations of primary interest have $p=1$ near the axis, including configurations created by differential rotation winding a dipole field into a strong $B_\phi$ (Section~\ref{sec:backgroundstate}). One can see that such configurations are stable in ideal MHD and are fully stabilized by rotation alone, with no need to invoke stratification.

\subsection{Effects of viscosity and magnetic diffusion}
\label{sec:effects_of_nu_eta}

We first examine the role of viscosity $\nu$ and magnetic diffusivity $\eta$ in the absence of stratification effects, which corresponds to dropping the stratification term $\propto k_\theta^2 N^2/k^2$ in the dispersion relation. This approximation holds for perturbations at sufficiently large $k$, which have $\tb\gg |\omega|^{-1},\omA^{-1}$. Calculations of $\gamma(k)$ for arbitrary $p$ and $\theta$ are given in Appendices \ref{ap:IW} and \ref{ap:MW}. In the main text below, we will focus on the configurations of prime interest $p=1$ and consider perturbations near the polar axis $\cos^2\!\theta\approx 1$ where instability is fastest (away from the axis, we find that modes $m=1$ can be unstable only at $|\cos \theta|> 1/\sqrt{2}$ for MW and at $|\cos \theta|> \sqrt{3}-1$ for IW).

For $p=1$ and $\cos^2\!\theta\approx 1$ we have $m_\star\approx m$. Then, the dispersion relation~(\ref{eq:SphericalFullIntro}) simplifies to
\begin{equation}
\label{eq:disp1}
   \left(\omega_\nu\omega_\eta-m^2\omA^2\right)^2
   - \left(2\Omega\omega_\eta + 2 m\omA^2\right)^2=0.
\end{equation}
It is reduced to two quadratic equations for $\omega$, whose solutions are
\begin{align} 
  \omega =& \frac{-i(\eta+\nu)k^2\pm \om}{2}\nonumber\\
  &+ \epsilon \sqrt{\left(\frac{i(\eta-\nu)k^2\pm\om}{2}\right)^2+m(m\pm 2)\omA^2 },
\end{align}
where $\epsilon=\pm 1$ corresponds to the IW and MW branches. At $\eta=\nu=0$ and $\omA\ll\Omega$ the solutions with $\epsilon=1$ and $\epsilon=-1$ are reduced to equations~(\ref{eq:IW_ideal}) and (\ref{eq:MW_ideal}), respectively.

\begin{figure}
	\includegraphics[width=\linewidth]{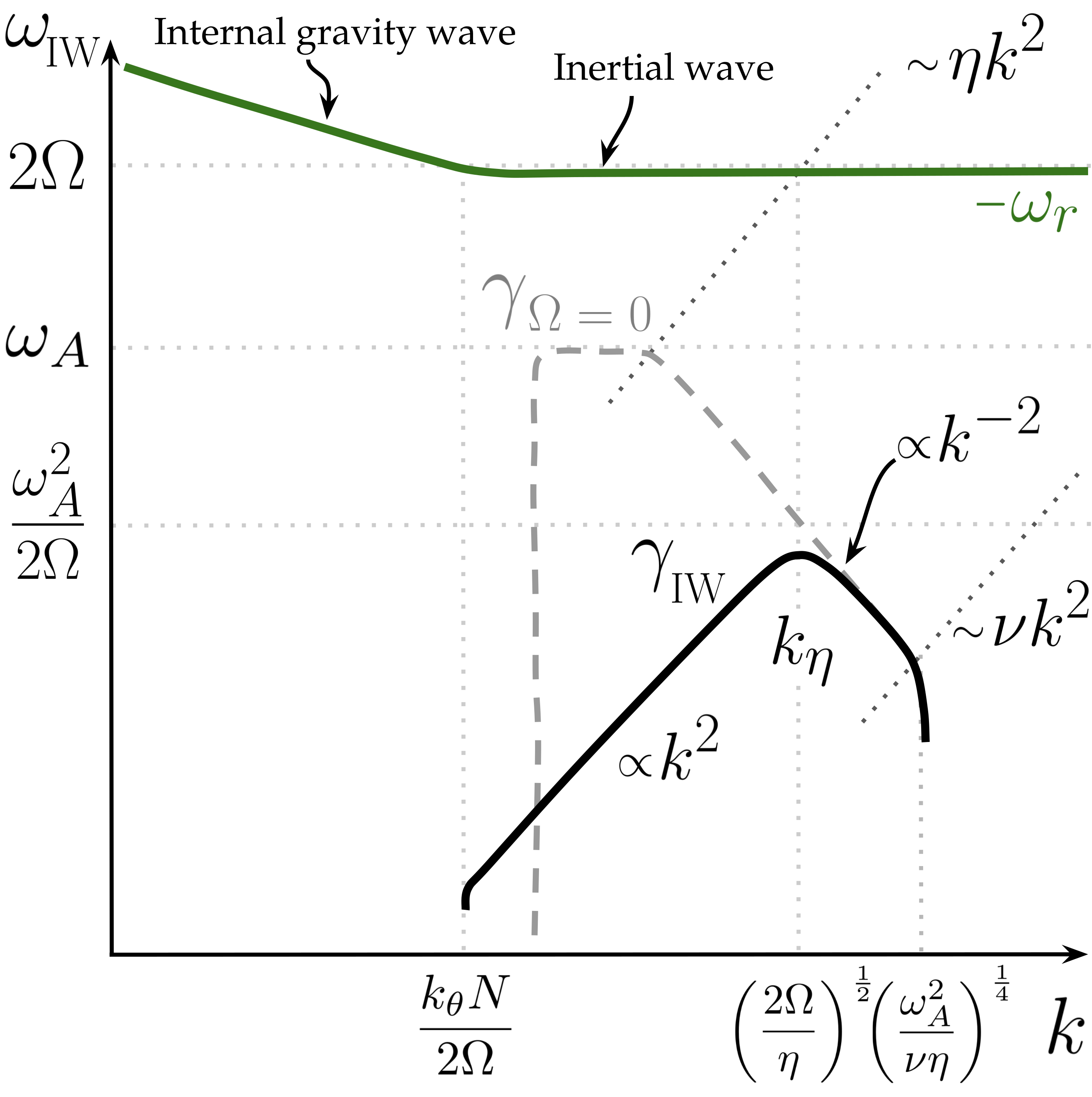}
    \caption{Complex frequency $\omega=\omega_r+i\gamma$ versus wavenumber $k$ for the IW branch of the TI in the rapidly rotating case $\Omega\gg\omA$. Both axes are on a logarithmic scale. Green curve shows $-\omega_r(k)$, and black curve shows $\gamma(k)$. For comparison, dashed grey curve shows $\gamma_{\Omega=0}(k)$ for the TI at $\Omega=0$, with otherwise the same parameters. The diagram is based on the obtained analytical expressions for $\omega(k)$ for the $m=1$ mode near the polar axis ($|\cos\theta|\approx 1$) in the  regime of a low magnetic Prandtl number $Pm\ll1$ and negligible buoyancy diffusion $\kappa k_\eta^2\ll\omA^2/\om$. The maximum growth rate $\gamma_{\max}\approx \omA^2/4\Omega$ is attained at $k_\eta$ where $ t_{\Omega}\sim t_\eta$. The IW growth rate curve is insensitive to $p$.
}

    \label{fig:IW_branch}
\end{figure}

First note that modes with large $m$ satisfying
$m(m\pm 2)\omA^2 \gg |i(\eta-\nu)k^2/2\pm\Omega|^2$ are stable: this
limit gives $\omega_r\approx \pm m\omA$ with $Im(\omega)<0$. The large-$m$ limit corresponds to small-scale Alfv\'en waves propagating along the magnetic loops. These stable waves are supported by the local magnetic tension $B_\phi^2/4\pi$ while the global hoop stress plays no role (or, more formally, the term $\propto 2m$ in the dispersion relation is negligible). 

The global hoop stress is essential for the TI, and this makes perturbations with small $m$ most relevant. Modes of main interest obey the condition $|m(m\pm 2)|\omA^2\ll |i(\eta-\nu)k^2/2\pm\Omega|^2$ (which is easily satisfied since $\omA\ll\Omega$). Then, one can expand the square root in \Eq~(\ref{eq:disp1}), and the solutions with $\epsilon=+1$ (IW) and $\epsilon=-1$ (MW) simplify to
\begin{equation}
\label{eq:om_IW}
  \omega_{\rm IW} \approx \pm \om - i\nu k^2 + \frac{m(m\pm 2)\omA^2 }{i(\eta-\nu)k^2\pm \om},
 \end{equation}
 \begin{equation}
 \label{eq:om_MW}
   \omega_{\rm MW} \approx - i\eta k^2 - \frac{m(m\pm 2)\omA^2 }{i(\eta-\nu)k^2\pm \om}.
\end{equation}
It is easy to show that for either wave mode $\gamma=Im(\omega)>0$ is possible only if $m(m\pm 2)<0$. The only possible negative value of $m(m\pm 2)$ is $-1$, which happens at $m=1$. This proves that configurations with $p=1$ can become unstable only in the $m=1$ mode.

The growth rate for the unstable $m=1$ mode of IW (see Figure~\ref{fig:IW_branch}) is found from the imaginary part of $\omega_{\rm IW}$,
\begin{equation}\label{eq:gammaIW_viscresis}
 \gamma_{\rm IW}(k) \approx \frac{\omA^2(\eta-\nu)k^2}{4\Omega^2+(\eta-\nu)^2k^4}-\nu k^2.
\end{equation}
One can see that $\gamma_{\rm IW}>0$ occurs if $\omA/2\Omega>(\nu/\eta)^{1/2}=Pm^{1/2}$. This requires $Pm\ll 1$ and implies $\eta-\nu\approx \eta$. The growth rate reaches the peak of 
\begin{equation}
\label{eq:IW_max}
   \gamma_{\rm IW}^{\max}
   \approx \frac{\omA^2}{4\Omega} \quad {\rm at} \quad k\approx k_\eta \equiv \left(\frac{2\Omega}{\eta}\right)^{1/2}
   \quad \left(Pm\ll \frac{\omA^2}{4\Omega^2}\right).
\end{equation}
Note that at $k\gg k_\eta$ rotational effects disappear and the instability occurs similarly to the $\Omega=0$ case (compare \Eq~(\ref{eq:gammaIW_viscresis}) 
with 
\Eq~(\ref{eq:gamma_visc_NR}) for $Pm\ll1$).

\begin{figure}
	\includegraphics[width=\linewidth]{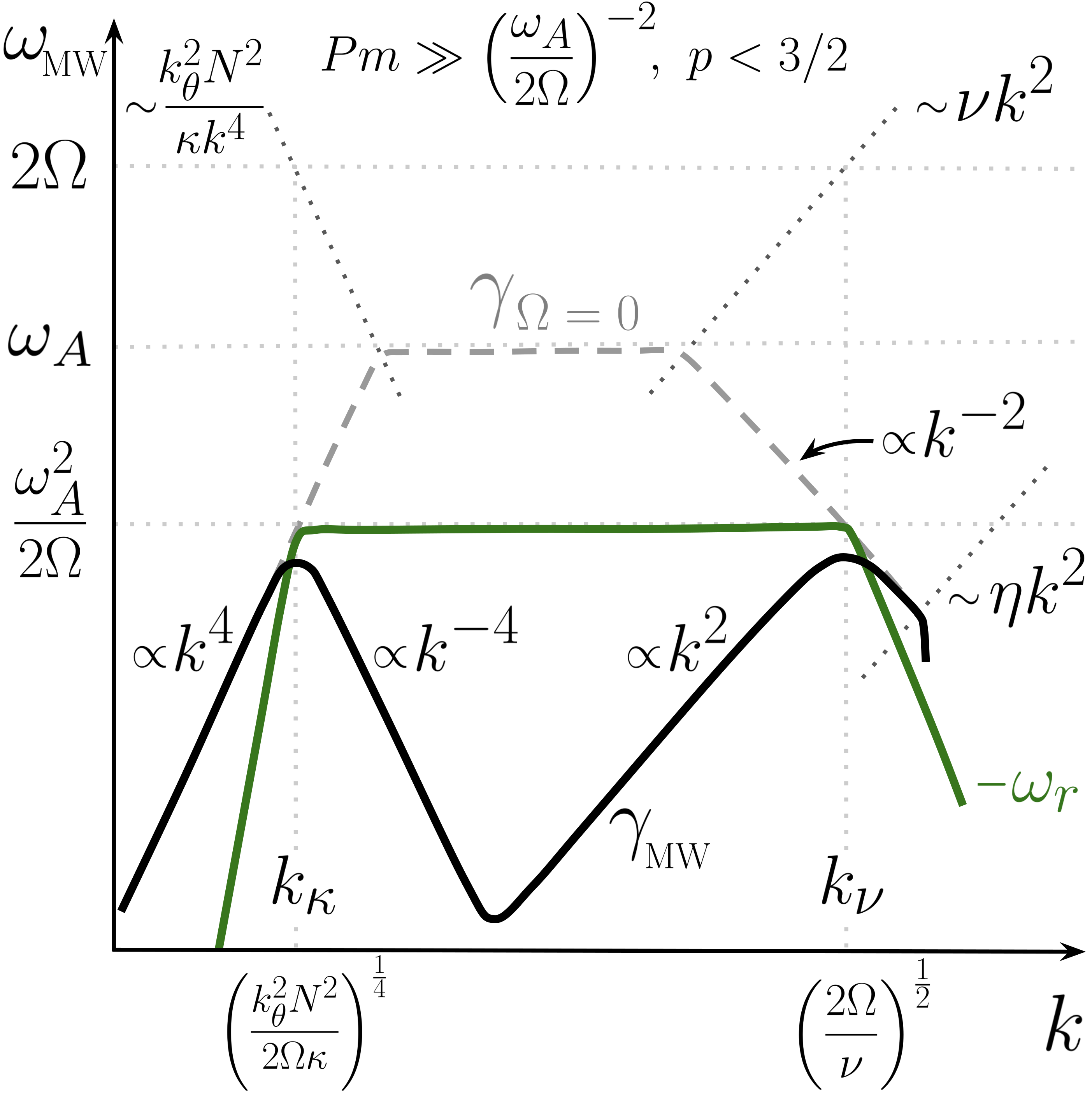}
    \caption{
    Complex frequency $\omega=\omega_r+i\gamma$ versus wavenumber $k$ of the MW branch of the TI in the rapidly rotating case $\Omega\gg \omA$ for magnetic configurations with a moderate gradient $p<3/2$. The results are shown in the regime of high $Pm\gg(\omA/\om)^{-2}$ and fast buoyancy diffusion $\kappa k_N^2\gg\omA^2/\om$ ($k_\kappa\ll k_N$) for the $m=1$ mode near the polar axis ($|\cos\theta|\approx 1$). Both the horizontal and vertical axes in the figure are on a logarithmic scale. Green curve shows $-\omega_r(k)$, black curve shows the growth rate $\gamma(k)$, and dashed grey curve shows $\gamma_{\Omega=0}(k)$ for the non-rotating TI with otherwise the same parameters. The maximum growth rate $\gamma_{\max}\approx \omA^2/4\Omega$ is attained at both $k_\kappa$  (where $t_\Omega\sim t_\kappa$) and $k_\nu$ (where $t_\Omega\sim t_\nu$) as long as $k_\kappa\ll k_\nu$. The shown result for $\omega_{\rm MW}(k)$ remains similar at different $p$ and $\theta$ as long as they give $m_\star^2>0$ (\Eq~\ref{eq:mstar}).
    } \label{fig:RotatingTI_MS_weakfieldgrad_highPm}
\end{figure}

Similarly, for the unstable $m=1$ mode of MW (see Figure~\ref{fig:RotatingTI_MS_weakfieldgrad_highPm}) we find,
\begin{equation}\label{eq:gammaMW_viscresis}
 \gamma_{\rm MW}(k) \approx \frac{\omA^2(\nu-\eta)k^2}{4\Omega^2+(\nu-\eta)^2k^4} - \eta k^2
\end{equation}
Here, $\gamma_{\rm MW}>0$ occurs if $\omA/2\Omega>Pm^{-1/2}$, which requires $Pm\gg 1$ and implies $\nu-\eta\approx \nu$. The instability reaches the peak of 
\begin{equation}
    \label{eq:MWmax}
   \gamma_{\rm MW}^{\max}
   \approx \frac{\omA^2}{4\Omega} \quad {\rm at} \quad k\approx k_\nu \equiv \left(\frac{2\Omega}{\nu}\right)^{1/2}
   \quad \left(Pm\gg \frac{4\Omega^2}{\omA^2}\right).
\end{equation}
At $k\gg k_\nu$ rotational effects disappear and the instability occurs similarly to the $\Omega=0$ case (compare \Eq~(\ref{eq:gammaMW_viscresis}) with \Eq~(\ref{eq:gamma_visc_NR}) for $Pm\gg1$).

The peaked shape of the growth rate $\gamma(k)$ for both IW and MW implies that the magnetic configuration is most unstable at characteristic wavenumbers where a single diffusive process and the Coriolis force operate on the same timescale: $\tiw\sim t_\eta$ for IW and $\tiw\sim t_\nu$ for MW. The growth rate is reduced for smaller $k$  ($k<k_\eta$ for IW and $k<k_\nu$ for MW) because the perturbations approach the ideal MHD limit where Coriolis force suppresses instability. On the other hand, perturbations with larger $k$ ($k>k_\eta$ for IW and $k>k_\nu$ for MW) experience too much diffusion and instability proceeds similarly to the non-rotating case $\Omega=0$ as the diffusion timescale becomes the shortest one. Thus, for both IW and MW, the growth rate $\gamma(k)$ peaks where it approaches  $\gamma_{\Omega=0}(k)$ given by \Eq~(\ref{eq:gamma_visc_NR}). One can verify that $\gamma_{\Omega=0}(k_\eta)\approx\gamma_{\rm \Omega=0}(k_\nu)\approx \omA^2/\om$.

\subsection{Effects of stratification on inertial waves}
\label{sec:effects_of_strat_IW}

When the ideal MHD picture of perturbations is extended to include the effects of stratification (while keeping $\eta=\nu=0$), the IW remain stable. IW instability occurs when $\eta\neq 0$ and $Pm\ll1$, and stratification changes the growth rate $\gamma_{\rm IW}(k)$. This change is described by \Eq~(\ref{eq:gamma_IW2}) derived in Appendix~\ref{ap:IW}. Below we apply it to perturbations with $m=1$ in the polar regions $\cos^2\theta\approx 1$ of the magnetic configuration with $p=1$.

In particular, when viscous effects are negligible at $k_\eta$ ($Pm\ll\omA^2/4\Omega^2$), the peak of IW instability (\Eq~\ref{eq:IW_max}) is changed by stratification as follows
\beq
\label{eq:IW_max_N}
 \gamma_{\rm IW}^{\max}=\frac{\omA^2}{4\Omega}-\frac{\kappa k_\theta^2N^2}{8\Omega^2(1+\kappa^2/\eta^2)} -\frac{\omA^2\eta k_\theta^2 N^2}{32\Omega^4(1+\kappa^2/\eta^2)}.
\eeq
The peak is weakly changed as long as both negative terms $\propto N^2$ are small compared to $\omA^2/4\Omega$. In this case, stratification can affect the instability growth rate only at small $k\ll k_\eta$, far from the peak. Expanding  \Eq~(\ref{eq:gamma_IW2}) at $k\ll k_\eta$, we find
\beq   
\label{eq:LowKIW}
    \gamma_{\rm IW}\approx \frac{\omA^2}{2\Omega}\frac{k^2}{k_\eta^2}\left[1 - \frac{k_\theta^2N^2}{2 k^2(4\Omega^2+\kappa^2k^4)}\left(3+\frac{\kappa}{\eta}\frac{4\Omega^2}{\omA^2}\right)\right].
\eeq
This expression can be used to estimate the wavenumber $k_1$ below which stratification suppresses instability. For instance, when $\kappa/\eta\ll \omA^2/\Omega^2$ one finds $k_1\sim k_\theta N/\Omega$.
Perturbations with wavenumbers $k\ll k_\theta N/\Omega$ occur with $t_N\ll\tiw$ and their real frequency shifts from $\omega_r\sim -\tiw^{-1}$ to $\omega_r\sim -t_N^{-1}$, so they become internal gravity waves, which are stable (Appendix~\ref{ap:IGW}). This behavior of $\omega_r(k)$ and $\gamma_{\rm IW}(k)$ is shown in Figure~\ref{fig:IW_branch}.

\begin{figure}
	\includegraphics[width=\linewidth]{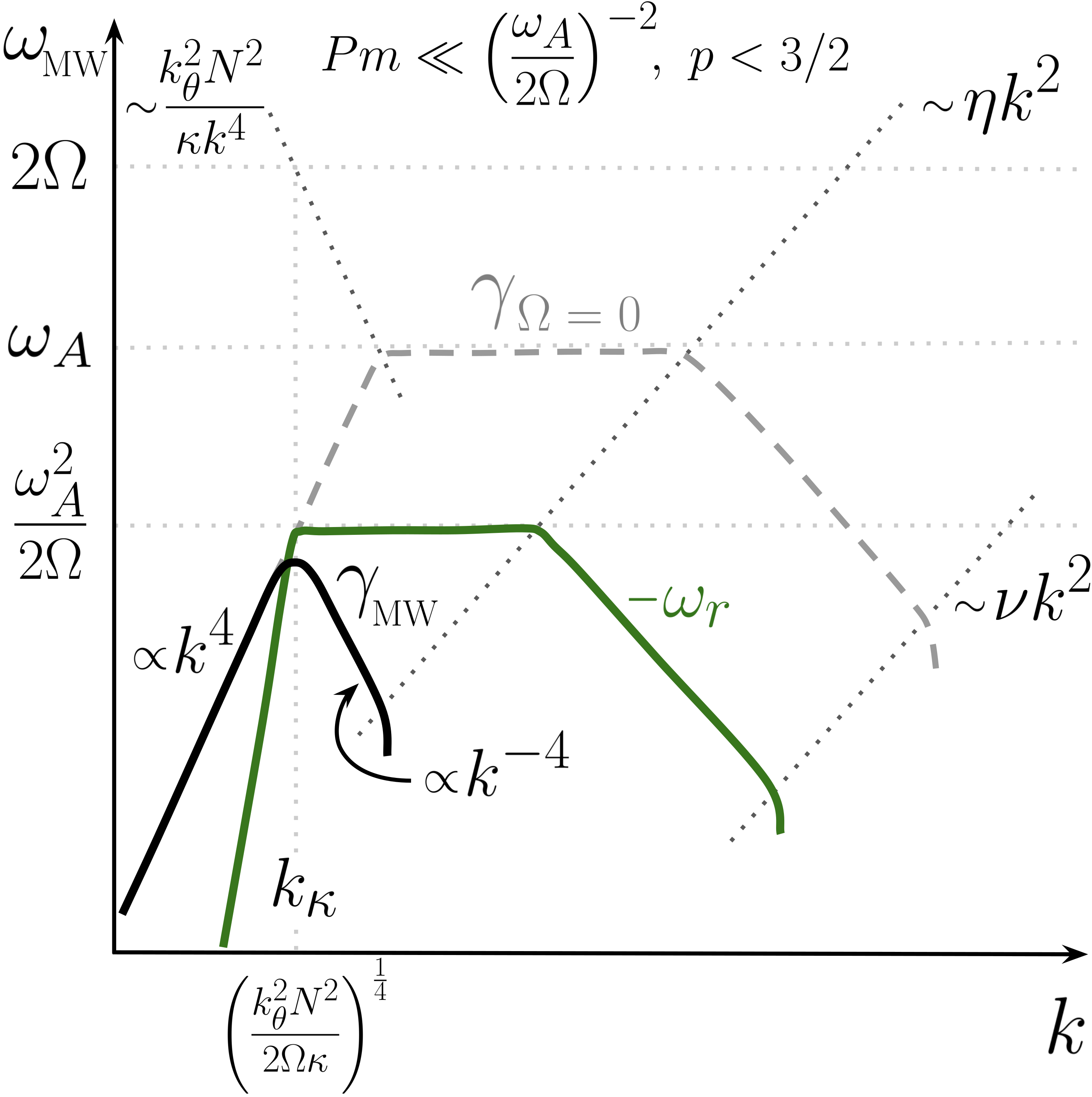}
    \caption{Complex frequency $\omega=\omega_r+i\gamma$ versus wavenumber $k$ of the MW branch of the TI in the regime of $Pm\ll(\omA/\om)^{-2}$ (which may be satisfied with $Pm\ll 1$ or $Pm\gg 1$; in the figure we assumed $Pm\ll 1$). Other parameters are the same as in Figure~\ref{fig:RotatingTI_MS_weakfieldgrad_highPm}. Magnetic diffusion suppresses instability at $k_\nu$ and the maximum growth rate $\gamma_{\max}\approx \omA^2/4\Omega$ is only attained at $k_\kappa$.
    }   \label{fig:RotatingTI_MS_weakfieldgrad_lowPm}
\end{figure}
\subsection{Effects of stratification on magnetostrophic waves}\label{sec:effects_ofkappa_onMW}

\subsubsection{Effects of stratification with $\nu=\eta=0$}
\label{sec:kappa_rot}
MW instability in magnetic configurations with $m_\star^2>0$ requires at least one of the diffusivities $\nu$, $\eta$, $\kappa$ to be non-zero. For perturbations with sufficiently small $k$, the effects of viscous and magnetic diffusion are negligible while the stratification effects, with a diffusive buoyancy response, can be significant. Therefore, we now consider the instability enabled by $\kappa\neq 0$ while neglecting terms with $\nu k^2$ and $\eta k^2$, which is equivalent to setting $\nu=\eta=0$ in the dispersion relation. In Appendix~\ref{ap:MW}, we verify that for $p=1$ only the $m=1$ modes are unstable and instability occurs in the polar regions $\cos^2\theta>1/2$. 

The low frequency of MW, $|\omega|\ll\omA\ll\Omega$, allows one to further simplify the dispersion relation by dropping the terms with high powers of $\omega$. Then, the dispersion relation~(\ref{eq:SphericalFullIntro}) becomes
\begin{align}\label{eq:D_MWthermalbranch_full}
    \left(1+\frac{k_N^2}{k^2}\frac{\omega}{\omega_\kappa}\right)\omA^4-\left(2\Omega  \omega+2\omA^2\right)^2=0.
\end{align}
It differs from the ideal MHD limit of MW only in the factor $\omega/\omega_\kappa\neq 1$ due to $\kappa\neq 0$. The stratification term $\propto k_N^2/k^2$ is significant if it is at least comparable to unity, which requires $k\lesssim k_N$. The onset of stratification effects  at $k\sim k_N$ can be examined in two limits: $\kappa k_N^2\ll |\omega|$ and $\kappa k_N^2\gg|\omega|$ where $|\omega(k_N)|\approx \omA^2/\om$. The case of $\kappa k_N^2\ll|\omega|$ corresponds to ideal MHD, which is stable. In the opposite limit, $\kappa k_N^2\gg|\omega|$, the stratification term for $k\lesssim k_N$ is

\begin{align}
    &\frac{k_N^2}{k^2}\frac{\omega}{\omega_\kappa}\approx -\frac{i\om\omega}{\omA^2}\frac{k_\kappa^4}{k^4},
  \quad\qquad   \kappa k_N^2\gg \frac{\omA^2}{\om},
\\
    &k_\kappa\equiv\left(\frac{k_\theta^2N^2}{\om \kappa}\right)^{1/4}.
\end{align}

The dispersion relation~(\ref{eq:D_MWthermalbranch_full}) then becomes quadratic in $\omega$:
\beq \label{eq:D_MWthermalbranch_quad_form}
    4\Omega^2 \omega^2+\left(i\frac{k_\kappa^4}{k^4}+4 \right)\om\omA^2\omega+3\omA^4=0.
\eeq
One of its roots $\omega_{\rm MW}$ has the imaginary part $\gamma_{\rm MW}>0$ and describes unstable perturbations:
\beq
  \gamma_{\rm MW}=\frac{\omA^2}{4\Omega}\left[ \sqrt{\frac{x^2}{2}-2+\sqrt{\frac{x^4}{4}+14x^2+4}} -x\right], 
\eeq
where $x\equiv k_\kappa^4/k^4$. The instability growth rate has a peak of $\gamma_{\max}\approx 0.21\omA^2/\Omega$ at $k\approx 0.78 k_\kappa$. Far from the wavenumber $k_\kappa$ the solution of \Eq~(\ref{eq:D_MWthermalbranch_quad_form}) simplifies to
\begin{eqnarray}
\label{eq:omega_kappa_MW}
    \omega_{\rm MW}\approx \frac{\omA^2}{2\Omega}\times \left\{\begin{array}{lr}
\displaystyle{ -1 + i\,  \frac{k_\kappa^4}{2k^4}, } & \quad k\gg k_\kappa
\vspace*{1mm}
\\
\displaystyle{ -12\frac{k^8}{k_\kappa^8}+3i\,\frac{ k^4}{k_\kappa^4}, } &  \quad k\ll k_\kappa
     \end{array}\right.
\end{eqnarray}
Note that the condition $\kappa k^2\gg|\omega_{\rm MW}|$ remains valid at $k\ll k_\kappa$ when it holds at $k\sim k_\kappa$, i.e. when $\kappa k_N^2\gg \omA^2/\om$. This condition for the diffusive buoyancy response is equivalent to $k_\kappa\ll k_N$:
\begin{equation}
\label{eq:fast_kappa}
 \frac{k_\kappa^4}{k_N^4}= \frac{\omA^2}{2\Omega \, \kappa k_N^2} \ll 1.
\end{equation}
At $k\ll k_\kappa$, the solution $\omega_{\rm MW}\approx 3i\omA^2 \kappa k^4/k_\theta^2 N^2$ becomes independent of $\Omega$ and reproduces $\gamma_{\Omega=0}(k)$ found in the non-rotating case at $k\ll k_{\rm d}$ (\Eq~\ref{eq:gkappa_NR}). This is expected, since $t_\Omega\gg t_\kappa$ at $k\ll k_\kappa$, so the Coriolis acceleration is negligible in the perturbation dynamics dominated by the diffusive buoyancy response. By contrast, at wavenumbers $k\gg k_\kappa$ the Coriolis effect dominates, $t_\Omega\ll t_\kappa$, and here the instability growth rate is reduced as $\gamma_{\rm MW}\propto k_\kappa^4/k^4$. The peak $\gamma_{\max}$ corresponds to $t_\kappa\sim t_\Omega$: the timescales for the Coriolis acceleration and the diffusive buoyancy response approximately match for the most unstable perturbations with $k\approx k_\kappa$.

Recall that our derivation assumed negligible effects of viscosity and magnetic diffusion at the relevant wavenumbers around $k_\kappa$. This description holds when  $\nu k_\kappa^2\ll 2\Omega$ and $\eta k_\kappa^2\ll \omA^2/2\Omega$, which is equivalent to $k_\kappa\ll k_\nu$ and $\omA/2\Omega\gg(N/2\Omega)^{1/2}(\kappa k_\theta^2/2\Omega)^{1/4}(\eta/\kappa)^{1/2}$.

\subsubsection{Combined effects of viscosity, magnetic diffusion, and stratification}
\label{sec:MW_alldiffusivities}

The full picture of MW instability in magnetic configurations with $p=1$ can be dissected into four cases of $k_\nu\gtrless k_\kappa$ and $Pm\gtrless 1$.

(1) $k_\nu\gg k_\kappa$ and $Pm\gg 1$.
In this case, the MW instability reaches the growth rate $\gamma_{\max} \approx \omA^2/4\Omega$ at $k_\nu$ if $\omA/\om>Pm^{-1/2}$ (Sections~\ref{sec:effects_of_nu_eta}) and also at $k_\kappa$ if $k_N>k_\kappa$ (Section~\ref{sec:kappa_rot}). The growth rate $\gamma_{\rm MW}(k)$ can be further studied between the two peaks, in the range of intermediate wavenumbers $k_\kappa\ll k \ll k_\nu$ where interplay between the stratification and viscous effects can occur. More generally, one can consider the interval $\min(k_\kappa,k_N)\ll k\ll k_\nu$ (this allows the case of $k_\kappa>k_N$, in which the instability peak at $k_\kappa$ is absent). In this interval of $k$ the stratification effects amount to a small correction to $\omega$, which can be found with a perturbative approach (Appendix~\ref{ap:MW}). The result for the imaginary part of $\omega$ is
\begin{align}\label{eq:lowMWnu}
    \gamma_{\rm MW} &\approx \frac{\omA^2}{\om}\,\frac{k^2}{k_\nu^2}\left[1-\frac{k_\theta^2 N^2}{2k^2|\omega_\kappa|^2}\left(\frac{\omA^2}{4\Omega^2}-\frac{\kappa}{\nu}\right)\right],
\end{align}
while the real part of $\omega$ remains close to $\omega_r\approx-\omA^2/\om$. 

The effect of diffusivity $\kappa$ is negligible when $\kappa\ll \nu (\omA/\om)^2$
(which gives $\kappa k_\nu^2\ll|\omega|$). In this regime we find
\begin{align}\label{eq:gammaMW_stratnokappa}
    \gamma_{\rm MW} &\approx \frac{\omA^2}{\om}\,\frac{k^2}{k_\nu^2}\left(1-\frac{k_N^2}{2k^2}\right),
    \;\, \frac{\kappa}{\nu}\ll\frac{\omA^2}{4\Omega^2},\; k_N\ll k\ll k_\nu.
\end{align}
Then, no instability peak is present at $k_\kappa$. The growth rate $\gamma_{\rm MW}(k)$ has a single peak at $k_\nu$, and the stratification effects cut off the instability curve at $k\lesssim k_N$ where $\gamma_{\rm MW}(k)$ becomes negative.

In the opposite regime $\kappa\gg\nu (\omA/\om)^2$, 
\Eq~(\ref{eq:lowMWnu}) becomes
\begin{align}\label{eq:MSthem_strat}
    \gamma_{\rm MW} &\approx \frac{\omA^2}{\om}\left(\frac{\nu k^2}{\om}+\frac{\kappa k_\theta^2 N^2}{4\Omega|\omega_\kappa|^2}\right)\nonumber\\
  &\approx \frac{\omA^2}{\om}\left[\frac{k^2}{k_\nu^2}+\frac{1}{2}\left(\frac{k_\kappa^4}{k_N^4}+\frac{k^4}{k_\kappa^4}\right)^{-1}\right].
\end{align}
The instability peak at $k_\kappa$ is present if $k_\kappa\ll k_N$ (\Eq~\ref{eq:fast_kappa}), and then we find
\begin{align}\label{eq:gamma_MS_viscous_and_thermal}
    \gamma_{\rm MW} &\approx \frac{\omA^2}{\om}\left(\frac{ k^2}{k_\nu^2}+\frac{k_\kappa^4}{2 k^4} \right), \; k_\kappa\ll k\ll  k_\nu
    \; 
    \left(\frac{\omA}{2\Omega}\gg Pm^{1/2}\right).
\end{align}
This expression describes the instability curve $\gamma_{\rm MW}(k)$ between the two peaks at $k_\kappa$ and $k_\nu$ (Figure~\ref{fig:RotatingTI_MS_weakfieldgrad_highPm}).

\medskip

(2) $k_\nu\gg k_\kappa$ and $Pm\ll 1$.
In this case, the growth rate of MW instability can only have the peak at $k_\kappa$. The instability peak at $k_\nu$ is removed by the dominant magnetic diffusivity $\eta>\nu$ (Section~\ref{sec:effects_of_nu_eta}). Including the effects of magnetic diffusion (with neglected viscosity) simply amounts to reinstating  $\omega \rightarrow \omega_\eta$ in \Eq~(\ref{eq:D_MWthermalbranch_full}), and then the solution~(\ref{eq:omega_kappa_MW}) at $k\gg k_\kappa$ changes to
\begin{align}\label{eq:omega_kkappa_highk_pluseta}
    &\gamma_{\rm MW}\approx  \frac{\omA^2}{\om}\frac{k_\kappa^4}{2k^4}-\eta k^2, \quad k\gg k_\kappa 
    \quad (Pm\ll 1).
\end{align}
The peak of $\gamma_{\rm MW}\approx 0.2 \omA^2/\Omega$ at $k\approx 0.78 k_\kappa$ is not degraded if $\eta k_\kappa^2\ll  \gamma_{\rm MW}$, i.e. magnetic diffusion at $k_\kappa$ is slow compared to the instability growth. When combined with condition~(\ref{eq:fast_kappa}), this requires $\kappa>\eta$. The behavior of $\gamma_{\rm MW}(k)$, including its suppression at high $k$ by magnetic diffusivity, is summarized in Figure~\ref{fig:RotatingTI_MS_weakfieldgrad_lowPm}. \cite{spruit2002dynamo} gave a correct heuristic argument that $k_\kappa$ is a minimum characteristic wavenumber for the TI in the regime of $\kappa\gg\eta\gg\nu$ and estimated $\gamma\sim\omA^2/\Omega$, which differs only by a numerical factor of $5$ from the actual maximum $\gamma_{\rm MW}\approx 0.2 \omA^2/\Omega$. \cite{spruit2002dynamo} incorrectly stated that $\gamma(k)\sim \omA^2/\Omega$ holds with increasing $k>k_\kappa$ as long as magnetic diffusivity effects are small. In fact, $\gamma(k)$ steeply decreases at $k\gg k_\kappa$.

Here another disagreement with previous work should be mentioned. Note that $\kappa\neq 0$ is required for the MW instability of the $p=1$ configurations with $Pm<1$. In particular, MW are stable when $\kappa=\nu=0$ and $\eta\neq0$. By contrast, the TI analysis in \cite{spruit1999differential}, \cite{zahn2007magnetic}, and \cite{ma2019angular} suggests that instability exists when $\kappa=\nu=0$ with the growth rate $\gamma\sim\omA^2/\Omega$ in an extended range of wavenumbers $k_N<k<(\omA^2/\om\eta)^{1/2}$. In the limit of $\eta\rightarrow 0$ this comes in contradiction with the known absence of TI in ideal MHD configurations with $p=1$. The contradiction is explained in Appendix~\ref{ap:PreviousProofs}: the analysis in the previous three works implicitly switched to a different branch of the dispersion relation. This ``fifth branch'' is described in Section~\ref{sec:fifth_branch} below. It is indeed unstable in a broad range of $k$ at small $\eta$. However, its growth rate has a narrow peak at $k=k_N/\sqrt{3}$ with $\gamma_{\max}\propto \eta^{1/2}$ and vanishes at $\eta\rightarrow 0$.

\medskip

(3) $k_\nu\ll k_\kappa$ and $Pm\gg 1$. 
In this case, the instability peak at $k_\kappa$ disappears --- it is killed by viscosity. Numerical exploration shows that $\gamma_{\rm MW}(k)$ has a single peak $\gamma_{\rm MW}(k_\nu)\approx \omA^2/4\Omega$ as long as $k_N\ll k_\nu$, so that stratification does not suppress the instability at $k_\nu$. The conditions $k_N\ll k_\nu\ll k_\kappa$ imply $k_N\ll k_\kappa$ and $\kappa k_N^2\ll |\omega_{\rm MW}|$, so the onset of stratification effects with decreasing $k\ll k_\nu$ occurs with negligible diffusivity $\kappa$ and the suppression of $\gamma_{\rm MW}$ at $k\lesssim k_N$ is described by \Eq~(\ref{eq:gammaMW_stratnokappa}). For $k>k_\nu$, stratification effects are unimportant and $\gamma_{\rm MW}(k)$ is well described by \Eq~(\ref{eq:gammaMW_viscresis}).

\medskip

(4) $k_\nu\ll k_\kappa$ and $Pm\ll 1$. In this case, there is no instability. The instability peak at $k_\kappa$ is removed by viscosity, and the instability peak at $k_\nu$ is removed by magnetic diffusion (Section~\ref{sec:effects_of_nu_eta}).

\begin{figure}
	\includegraphics[width=\linewidth]{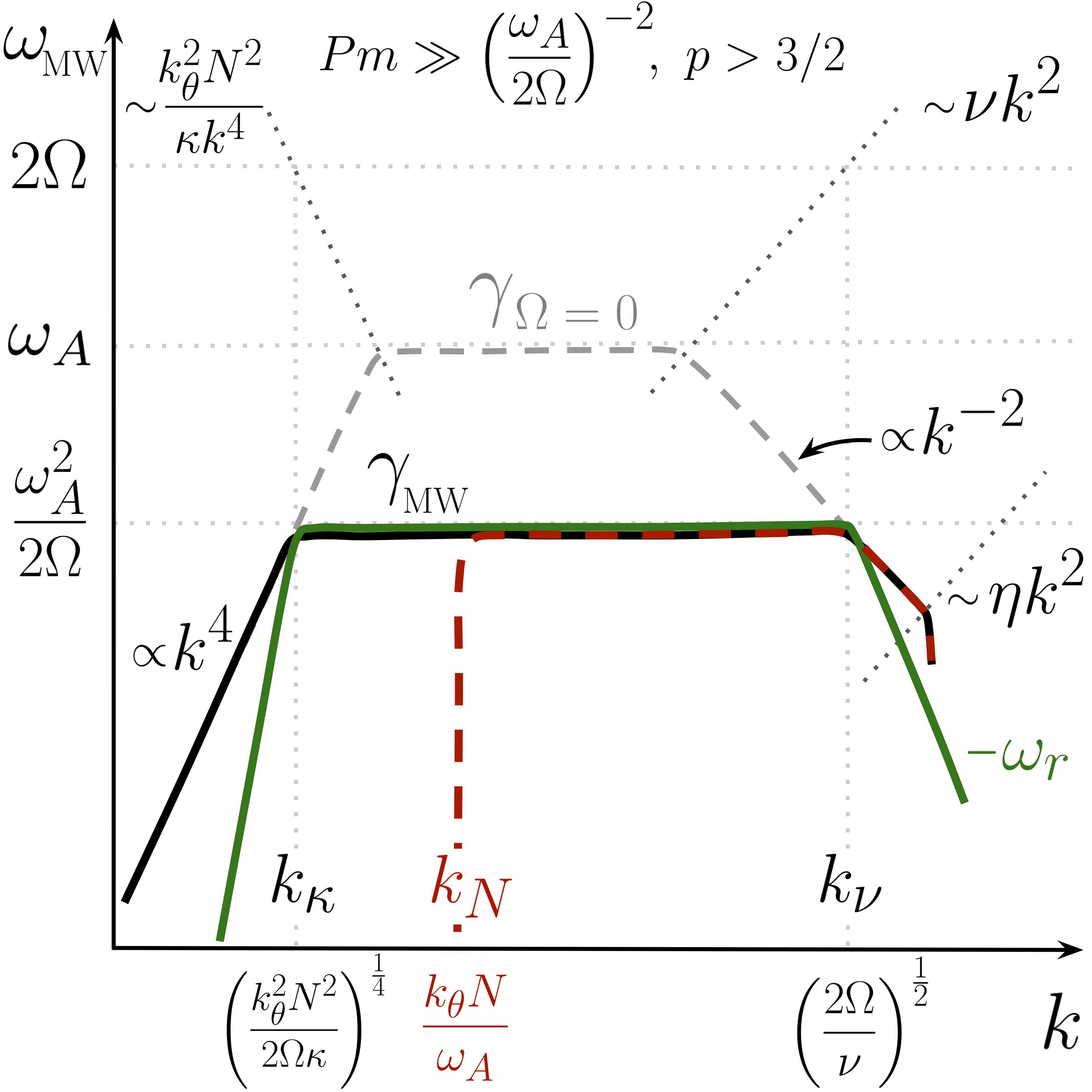}
    \caption{Complex frequency $\omega=\omega_r+i\gamma$ versus wavenumber $k$ of the MW branch of the TI with rapid rotation, a strong gradient $p>3/2$ and $Pm\gg(\omA/\om)^{-2}$ for the $m=1$ mode near the polar axis. The instability growth rate $\gamma_{\rm MW}(k)$ is shown for two cases: when buoyancy diffusion is slow ($\kappa k_N^2\ll |\omega|$, dashed red curve) and fast (solid black curve). The real part of the MW frequency is shown by the green curve. A similar behavior of $\omega_{\rm MW}$ is found for different values of $p$ and different $\theta$ near the polar axis, as long as $m_\star^2<0$. For comparison, the dashed grey curve shows $\gamma_{\Omega=0}(k)$ for the non-rotating TI in otherwise the same parameter regime (and assuming fast buoyancy diffusion).
    }
    \label{fig:RotatingTI_MS_strongfieldgrad_highPm}
\end{figure}
\begin{figure}
	\includegraphics[width=\linewidth]{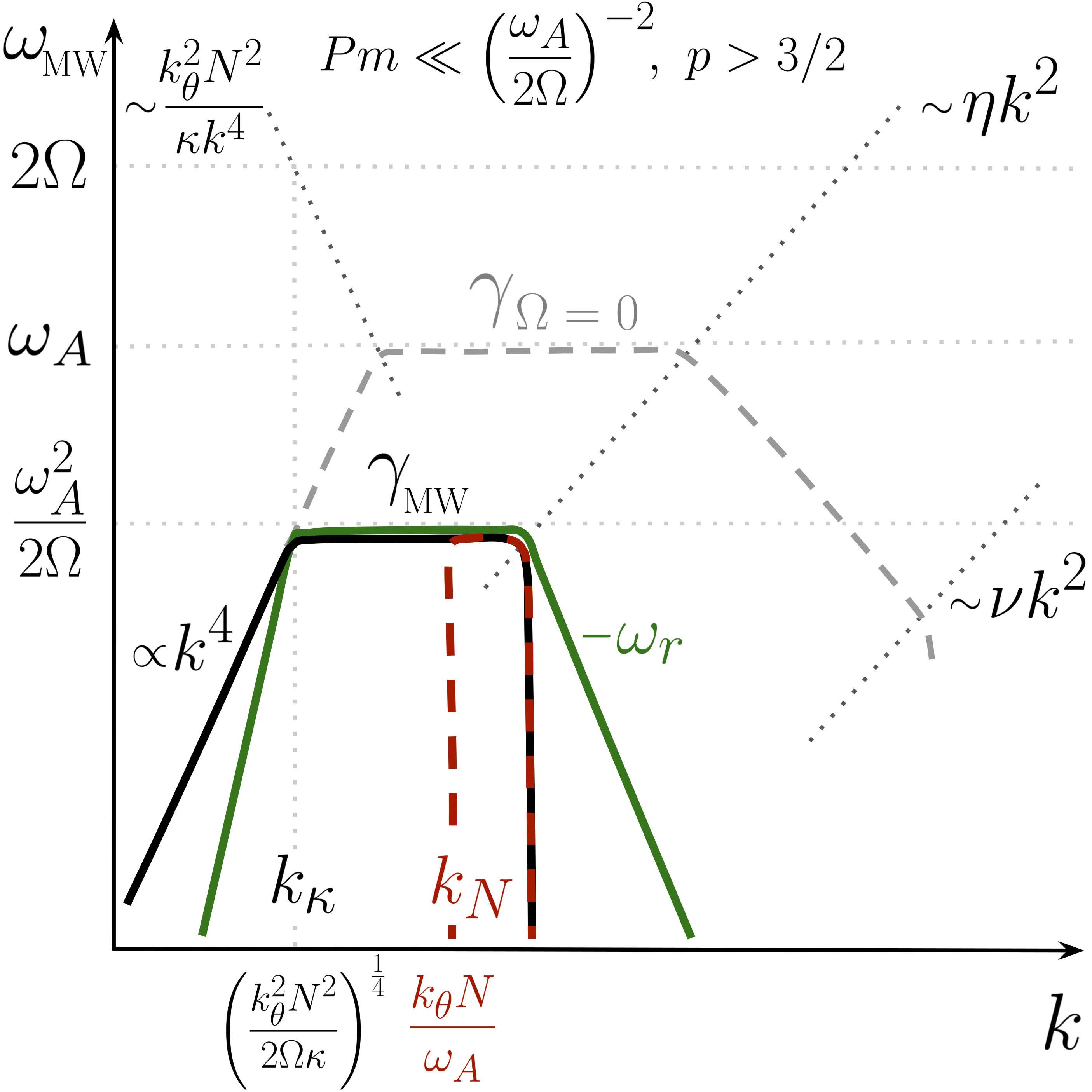}
    \caption{
    Same as Figure~\ref{fig:RotatingTI_MS_strongfieldgrad_highPm} but for $Pm\ll (\omA/\om)^{-2}$ (which may be satisfied with $Pm\ll 1$ or $Pm\gg 1$; in the figure we assumed $Pm\ll 1$). The highest wavenumber of perturbations growing with $\gamma_{\max}\approx \omA^2/\om$ is now set by magnetic diffusion (instead of the wavenumber $k_\nu$  shown in Figure~\ref{fig:RotatingTI_MS_strongfieldgrad_highPm}).
    }
    \label{fig:RotatingTI_MS_strongfieldgrad_lowPm}
\end{figure}

\subsection{Magnetostrophic waves with large gradients of $B_\phi$}
\label{sec:strongfieldgradient}

It remains to examine the stability of MW in magnetic configurations with large gradients of the toroidal field, $p=\partial \ln B_\phi/\partial\ln r$, when perturbations exist with $m_\star^2\equiv m^2-2(p\cos^2\theta-1)<0$. This condition is easiest to satisfy at $m=1$ (the lowest $m>0$) and $\cos^2\theta=1$. Hence, perturbations with $m_\star^2<0$ exist if $p>3/2$. Then, MW instability appears in ideal MHD at $k>k_N$ (\Eq~\ref{eq:MW_ideal}).

The presence of diffusive processes reduces the range of $k$ where instability occurs as predicted by ideal MHD. The  growth rate $\gamma_{\rm MW}(k)$ in non-ideal MHD is derived in Appendix~\ref{ap:MW}, and the results are summarized in Figures~\ref{fig:RotatingTI_MS_strongfieldgrad_highPm} and \ref{fig:RotatingTI_MS_strongfieldgrad_lowPm}. The shape of $\gamma_{\rm MW}(k)$ is similar to the non-rotating case where the growth rate was constant across an interval $k_1<k<k_2$ (cf. Figure~\ref{fig:NonRotatingTI}) except that it takes on a value of $\gamma_{\rm MW}\approx\omA^2/\om$ instead of $\gamma_{\Omega=0}\approx \omA$. In particular, perturbations with $m=1$ at $\cos^2\theta\approx 1$ follow the solution
\beq
\label{eq:strong_gradient}
  \omega_{\rm MW} \approx \frac{\omA^2}{2\Omega}\left(-2+i\sqrt{3p-2}\right)
\eeq 
in the range $k_1<k<k_2$
where
\beq
\label{eq:k1k2}
    k_1 = \min\Big\{k_\kappa,k_N\Big\}, \quad 
    k_2 = \min\Big\{k_\nu,k_\eta\frac{\omA}{\om}\Big\}.
\eeq

Figure~\ref{fig:RotatingTI_MS_strongfieldgrad_highPm} shows $\omega_{\rm MW}$ when
$Pm\gg(\omA/\om)^{-2}$. In this case, $k_2=k_\nu$ is set by viscosity. At wavenumbers $k>k_\nu$, the fast viscous diffusion ($t_\nu<t_B$) reduces both the instability growth rate $\gamma_{\rm MW}$ and the real part of $\omega_{\rm MW}$. Note also that the rotation effects become subdominant at $k>k_\nu$ ($t_\Omega>t_\nu$) and $\gamma_{\rm MW}$ follows the solution $\gamma_{\Omega=0}(k)$ found in the non-rotating case. By contrast, when $Pm\ll(\omA/\om)^{-2}$, $k_2$ is set by magnetic diffusivity: $k_2\sim (\omA^2/\om \eta)^{1/2}$ corresponds to $t_\eta^{-1}\sim \omA^2/\om$. In this case, the growth rate $\gamma_{\rm MW}$ does not approach $\gamma_{\Omega=0}$ at high $k$ (Figure~\ref{fig:RotatingTI_MS_strongfieldgrad_lowPm}).

Stratification impacts $\omega_{\rm MW}$ at wavenumbers below $k_1$ where the timescales for the buoyancy response and the magnetic force become comparable, $\tb\sim t_B$ (see Section~\ref{sec:timescales} for the discussion of timescales). The basic effect is similar to that in the non-rotating case, although the timescale $t_B$ is different, since it scales with the mode frequency $|\omega|$ (\Eq~\ref{eq:timescales}): $t_B=|\omega_{\rm MW}|/\omA^2 \sim (2\Omega)^{-1}$. If diffusivity $\kappa$ is negligible, the buoyancy timescale $\tb$ also scales with $|\omega|$, leading to $k_1=k_N$, same as in the non-rotating case. The buoyancy response is diffusive if $\kappa k_N^2>|\omega|$, which corresponds to $k_\kappa<k_N$, and in this case $k_1$ is reduced from $k_N$ to $k_\kappa$. Suppression of the instability growth rate at $k\ll k_\kappa$ occurs with $\tb\ll t_\Omega$, so here rotation effects are negligible and $\gamma_{\rm MW}=\gamma_{\Omega=0}$ (Figures~\ref{fig:RotatingTI_MS_strongfieldgrad_highPm} and \ref{fig:RotatingTI_MS_strongfieldgrad_lowPm}).


\subsection{The fifth branch}
\label{sec:fifth_branch}
\begin{figure}
	\includegraphics[width=\linewidth]{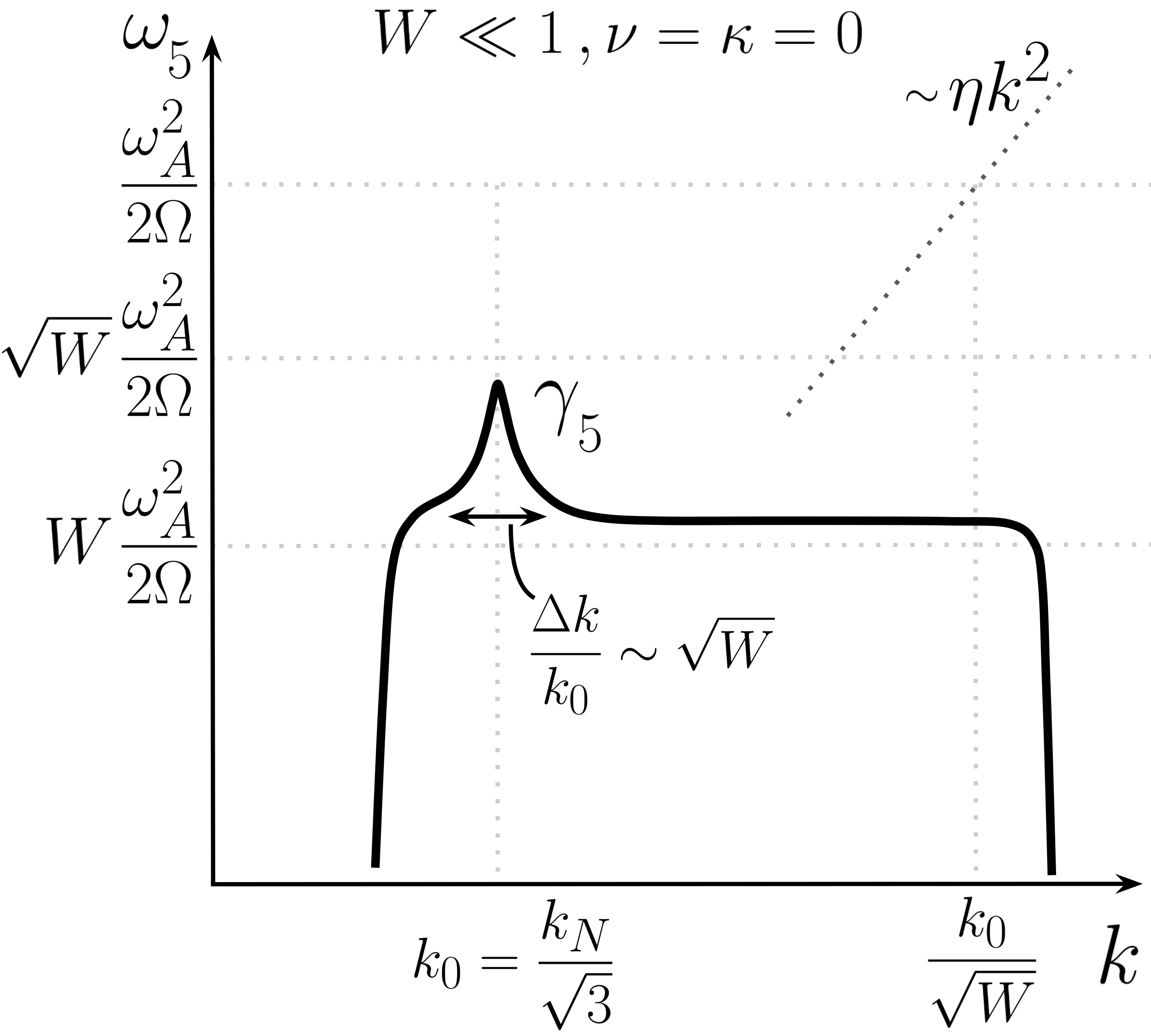}
    \caption{Growth rate $\gamma_5(k)$ for the fifth branch when $W\ll 1$. The growth rate has a narrow peak at $k=k_0$ with $\gamma_5^{\max}\approx \sqrt{W/2}\,\omA^2/4\Omega$. Instability in the fifth branch exists when $\kappa\ll\eta$, $k_\nu\gg k_N$, and $W<11$.
    }
    \label{fig:GammaFive}
\end{figure}

The full dispersion relation (\Eq~\ref{eq:SphericalFullIntro}) may be written in the form $F(\omega)\equiv \omega_\kappa D(\omega)=0$, where $F(\omega)$ is a fifth-order polynomial. It has two pairs of roots $\omega_{\rm IW}=\omega_{1,2}\approx \pm 2\Omega+i\gamma_{1,2}$ and $\omega_{\rm MW}=\omega_{3,4}\approx (-2\pm1)\omA^2/2\Omega+i\gamma_{3,4}$, with the roots $\omega_2$ and $\omega_4$ giving instability. In addition, $F(\omega)$ has a fifth root with a low frequency $\omega_5$.

The solution for the fifth branch $\omega_5(k)$ is derived in Appendix~\ref{ap:fifth_root}. We find that this branch can be unstable, $\gamma_5=Im(\omega_5)>0$, when the effects of viscosity $\nu$ and diffusivity $\kappa$ are small; this requires $k_N< k_\nu$ and $\kappa<\eta$, respectively. The instability is driven by magnetic diffusivity $\eta$ and peaks at a wavenumber $k_0$ near $k_N=k_\theta N/\omA$. It develops for non-axisymmetric perturbations $m\neq 0$ and requires $4\cos^2\theta>m_\star^2$. For magnetic configurations with $p=1$ this condition is satisfied only for $m=1$ and at $\cos^2\theta>1/2$. Instability also requires
\beq\label{eq:fifthbranch_instcriteria}
  W\equiv \frac{2\Omega\eta k_N^2}{\omA^2}<11 \qquad ({\rm at~} |\cos\theta|\approx1).
\eeq

The shape of the instability peak $\gamma_5(k)$ can be derived using an expansion of $F(\omega)$ in $W$ and $\omega$ (Appendix~\ref{ap:fifth_root}).
For $p=1$, $m=1$, and $|\cos\theta|\approx 1$, we find
\beq
\label{eq:om5}
  \omega_5(k)\approx\frac{\omA^2}{16\Omega}\left(3-\frac{k_N^2}{k^2}\right)\left[\sqrt{1+\frac{16iW}{(3-k_N^2/k^2)^2}}-1\right].
\eeq
Its imaginary part $\gamma_5$  has a sharp peak at $k_0=k_N/\sqrt{3}$ of width $\Delta k/k_0\sim \sqrt{W}$, reaching 
\beq 
\label{eq:gam5_max}
  \gamma_5^{\max}\approx \frac{\omA^2}{4\Omega}\sqrt{\frac{W}{2}}=\sqrt{ \frac{\eta k_\theta^2 N^2 }{16\Omega}} \qquad (W<1).
\eeq
Perturbations are unstable, $\gamma_5(k)>0$, in a broad interval $k_1<k<k_2$, whose boundaries are given by the marginal stability analysis in Appendix~\ref{ap:PreviousProofs}. In particular, when $W\ll 1$ the boundaries are $k_1^2=k_N^2/(4\sqrt{3}+6)$ and $k_2^2=k_N^2\sqrt{3}/W$.
The growth rate $\gamma_5$ is significantly below $\gamma_5^{\max}$ outside the narrow peak at $k_0$. In the extended interval of $k_0\ll k<k_2$ we find $\gamma_5\approx \eta k_N^2/3\approx \sqrt{W}\gamma_5^{\max}$. The shape of $\gamma_5(k)$ is shown in Figure~\ref{fig:GammaFive}.

The fifth branch becomes important when it is the only unstable branch. This occurs in the parameter space of $\nu$, $\kappa$, $\eta$ where $\kappa\ll\eta$ (necessary for $\gamma_5>0)$ and where the fifth branch has the lowest $\omA$-threshold for instability, so that it can develop in the absence of other (faster) instabilities. Note that the condition $\kappa\ll\eta$ implies absence of MW instability at $k_\kappa$, and the remaining competing instabilities occur at $k_\nu$ for the MW branch and $k_\eta$ for the IW branch. The threshold for the fifth-branch instability is given by \Eq~(\ref{eq:fifthbranch_instcriteria}): $\omA/2\Omega\gg (\eta k_\theta^2 k_N^2/8\Omega^3)^{1/4}$. We compare it with the condition $\omA/2\Omega\gg Pm^{1/2}$ required for the $k_\eta$ instability peak (\Eq~\ref{eq:IW_max}) and $\omA^2/2\Omega\gg Pm^{-1/2}$ required for the $k_\nu$ instability peak (\Eq~\ref{eq:MWmax}). Thus, we find that both IW and MW instabilities have a higher threshold if 
\beq
  \frac{\nu}{\eta}+\frac{\eta}{\nu}\gg \left(\frac{\eta k_\theta^2 N^2}{8\Omega^3}\right)^{1/2}.
\eeq
Then, the fifth-branch instability can develop without competition from the other, faster instabilities. Its  growth rate saturates at $\gamma_5^{\max}\approx (\eta k_\theta^2 N^2/16\Omega)^{1/2}$ above the threshold (\Eq~\ref{eq:gam5_max}), so  $\gamma_5^{\max}$ does not increase as $\propto\omA^2$, in contrast to the MW and IW instabilities, which grow with $\gamma_{\max}\approx \omA^2/4\Omega$ above their thresholds.


\section{Conditions for instability}
\label{sec:CondInst}

It is convenient to have a criterion for  TI formulated as a condition on the field strength $B_\phi$ or the corresponding $\omA$ in a region of a star with given configuration parameters $\Omega$, $N$, $p$, $R$, and diffusivities $\nu$, $\eta$, $\kappa$. Consider increasing the field strength from $\omA=0$. In non-ideal MHD, any configuration will be stable ($\gamma<0$ at all $k$) for a sufficiently low $\omA$, and so there is a threshold for the  onset of instability. For $\omA$ sufficiently above the threshold, the instability growth rate $\gamma(k)$ can approach its canonical maximum value,
\beq
  \gamma_{\max}\approx\left\{\begin{array}{rl}
  \omA, & \Omega\ll\omA 
    \vspace*{1mm} \\
  \displaystyle{\frac{\omA^2}{4\Omega}}, & \Omega\gg\omA
  \end{array} \right.
\eeq
Similar to \cite{spruit1999differential,spruit2002dynamo} we will  focus on the conditions for the TI with $\gamma\approx\gamma_{\max}$ (rather than $\gamma>0$) because the fully developed instability is of main interest for the Tayler-Spruit dynamo and the associated angular momentum transport in stars. 

We will first summarize the non-rotating case in Section~\ref{sec:CondInstNR}, and then describe the rotating case, with the IW branch discussed in Section~\ref{sec:CondInstIW} and the MW branch in Section \ref{sec:CondInstMW}. 
In addition, instability can occur in the fifth branch when $\eta\gg \kappa$, $k_\nu\gg k_N$ (Section~\ref{sec:fifth_branch}); its growth rate approaches $\gamma\sim \omA^2/4\Omega$ only if $\omA$ is near a special value such that $\omA^2/2\Omega\approx\eta k_N^2$.

\subsection{Non-rotating case ($\Omega\ll\omA$)}
\label{sec:CondInstNR}

Magnetic configurations with $p=1$ in weakly rotating stars develop the TI with the growth rate $\gamma_{\Omega=0}\approx\omA$ in a range of wavenumbers $\min\{k_N,k_{\rm d}\}<k<\kA$ (Section~\ref{sec:nonrotatingTI}). This range of $k$ opens up when $\omA$ exceeds a critical value at which $\min\{k_N,k_{\rm d}\}=\kA$. This transition, i.e. the onset of $\gamma_{\Omega=0}=\omA$, occurs with negligible $\kappa$-diffusivity if $\kappa\kA^2\ll \omA$, which is equivalent to $\kappa\ll \nu+\eta$. In this regime, $\min\{k_N,k_{\rm d}\}=k_N$ and so $\gamma_{\Omega=0}(k)\approx \omA$ exists if $k_N<\kA$, which requires
\begin{equation}
\label{eq:NR_TI_cond1}
    \omA^3>(\nu+\eta)k_\theta^2N^2, \qquad \kappa\ll \nu+\eta.
\end{equation}
In the opposite case of $\kappa\gg \nu+\eta$,  the onset of $\gamma_{\Omega=0}\approx\omA$ occurs with $\min\{k_N,k_{\rm d}\}=\kd$ 
(the regime of diffusive buoyancy response). Then, $\gamma_{\Omega=0}(k)\approx\omA$ exists if $\kd<\kA$, which requires
\begin{equation}
\label{eq:NR_TI_cond2}
    \omA^3>\frac{(\nu+\eta)^2}{\kappa}\,k_\theta^2N^2, \qquad \kappa\gg \nu+\eta.
\end{equation}
The conditions~(\ref{eq:NR_TI_cond1}) and (\ref{eq:NR_TI_cond2}) can be combined into a single approximate condition:
\begin{equation}
\label{eq:NR_TI_cond}
    \gamma_{\Omega=0}\approx\omA \quad {\rm when} \quad \omA>\left[\frac{(\nu+\eta)^2k_\theta^2 N^2}{\kappa+\nu+\eta}\,\right]^{1/3}.
\end{equation}

\subsection{Inertial wave branch ($\Omega\gg \omA$)}
\label{sec:CondInstIW}

Magnetic configurations with $p=1$ in stars with $\Omega\gg\omA$ develop the TI with $\gamma_{\rm IW}\approx\gamma_{\max}\approx \omA^2/4\Omega$ near wavenumber $k_\eta$ on the IW branch and near wavenumbers $k_\kappa$, $k_\nu$ on the MW branch (Section~\ref{sec:RotatingTI}). We first review the IW instability. Its growth rate is
\begin{equation}
    \gamma_{\rm IW}\approx \frac{\omA^2}{2\Omega}
 \times\left\{\begin{array}{lr}
 \displaystyle{\frac{k^2}{k_\eta^2}\left[1-\frac{k_\theta^2 N^2(3+2\Omega\kappa k_\eta^2/\omA^2)}{2k^2(4\Omega^2+\kappa^2k^4)}\right]} & k<k_\eta 
\vspace*{5mm}
 \\
 \displaystyle{\frac{k_\eta^2}{k^2}\left(1-\frac{k^4}{k_{\nu\eta}^4}\right)} & k>k_\eta
    \end{array}\right.
\end{equation}
where $k_\eta=(2\Omega/\eta)^{1/2}$ and $k_{\nu\eta}=(\omA^2/\nu\eta)^{1/4}$.
This result was derived assuming that the corrections due to  viscosity (the term $\propto k^4/k_{\nu\eta}^4$) and stratification (the term $\propto N^2$) are both small at $k\sim k_\eta$. Both conditions are required for the existence of the peak $\gamma_{\rm IW}(k_\eta)\approx \omA^2/4\Omega$. The viscous suppression is small at $k\sim k_\eta$ if $k_{\nu\eta}\gg k_\eta$, i.e.
\begin{equation}
\label{eq:IW_cond1}
    \frac{\omA}{2\Omega}\gg 
    Pm^{1/2}.
\end{equation}

The correction to the instability peak due to stratification is given by \Eq~(\ref{eq:IW_max_N}). Stratification does not suppress the peak if
\begin{equation}
 \frac{\eta k_\theta^2 N^2}{8\Omega^3(1+\kappa^2/\eta^2)}
 +\frac{\kappa k_\theta^2 N^2 }{2\Omega\omA^2 (1+\kappa^2/\eta^2)}\ll 1.
\end{equation}
This gives two conditions, one on the Brunt-V\"ais\"al\"a frequency and one on the magnetic field strength:
\begin{align}
    \label{eq:IW_cond2}
    &\frac{N}{2\Omega}\ll\left(\frac{\eta k_\theta^2}{2\Omega}\right)^{-1/2}\left(1+\frac{\kappa^2}{\eta^2}\right)^{1/2},\\
    \label{eq:IW_cond3}
    &
    \frac{\omA}{2\Omega}\gg \frac{N}{2\Omega}\left(\frac{\kappa k_\theta^2}{2\Omega}\right)^{1/2}\left(1+\frac{\kappa^2}{\eta^2}\right)^{-1/2}.
\end{align}

The development of IW instability with rate $\gamma_{\rm IW}\approx\omA^2/4\Omega$ occurs if all three conditions~(\ref{eq:IW_cond1}), (\ref{eq:IW_cond2}), and (\ref{eq:IW_cond3}) are satisfied. They remain unchanged for magnetic configurations with different gradients $p$.

\subsection{Magnetostrophic wave branch ($\Omega\gg \omA$)}
\label{sec:CondInstMW}

\subsubsection{Moderate toroidal field gradients}

Next, we obtain necessary and sufficient conditions for instability of the MW branch with the growth rate $\gamma_{\max}\approx\omA^2/4\Omega$. We here again focus on the case of $p=1$, the primary example of magnetic configurations with moderate gradients $p<3/2$ ($m_\star^2>0$). 

As shown in Section~\ref{sec:RotatingTI}, the MW instability can develop with $\gamma_{\rm MW}\approx\omA^2/4\Omega$ at wavenumbers $k_\kappa$ and/or $k_\nu$, depending on the parameters of the star. The instability is easiest to develop when $k_\nu\gg k_\kappa$, so we begin with this case.

(1) $k_\nu\gg k_\kappa$. As shown in Section~\ref{sec:MW_alldiffusivities},
instability at $k_\kappa$ with $\gamma_{\rm MW}(k_\kappa)\approx\omA^2/4\Omega$ occurs only in the regime of diffusive buoyancy $\kappa  k_N^2\gg \omA^2/2\Omega$, which is equivalent to $k_\kappa\ll k_N$ and requires
\begin{align}
\label{eq:MWNS1}
    \frac{\omA}{\om}\ll\left(\frac{N}{\om}\right)^{1/2}\left(\frac{\kappa k_\theta^2}{\om}\right)^{1/4}
     \qquad  (\mathrm{TI\; at\;} k_\kappa).
\end{align}
The condition $k_\nu\gg k_\kappa$ implies negligible viscosity effects at $k\sim k_\kappa$, and the only remaining requirement for the successful TI at $k_\kappa$ is the negligible suppression by magnetic diffusion, $\eta k_\kappa^2\ll \gamma_{\rm MW}(k_\kappa)$ (see \Eq~(\ref{eq:omega_kkappa_highk_pluseta})). It is satisfied if
\begin{align}
\label{eq:MWNS2}
    \frac{\omA}{\om}\gg \left(\frac{N}{ \om}\right)^{1/2}\left(\frac{\kappa k_\theta^2}{\om}\right)^{1/4}\left(\frac{\eta}{\kappa}\right)^{1/2} 
   \quad (\mathrm{TI\; at\;} k_\kappa).
\end{align}
\cite{spruit1999differential} found a similar condition for the TI using a marginal stability analysis (by setting $\gamma=0$ in the dispersion relation) with $\nu=0$ and $\kappa\gg\eta$. However, \cite{spruit1999differential} missed the condition~(\ref{eq:MWNS1}). Taken together, the conditions (\ref{eq:MWNS1}) and (\ref{eq:MWNS2}) are necessary and sufficient for instability at $k_\kappa$ when $k_\kappa\ll k_\nu$. Note that they require $\eta\ll \kappa$.

Consider now the instability at $k_\nu$. The condition $k_\nu\gg k_\kappa$ implies negligible stratification effects at $k\sim k_\nu$ for both regimes of buoyancy response, diffusive ($k_\kappa<k_N$) or non-diffusive ($k_\kappa>k_N$). The instability peak $\gamma_{\rm MW}(k_\nu)\approx\omA^2/4\Omega$ may only be suppressed by magnetic diffusion. The condition for negligible suppression $\eta k_\nu^2\ll \gamma_{\rm MW}(k_\nu)$ requires

\begin{align}\label{eq:MWNS3}
    &\frac{\omA}{2\Omega}\gg
    Pm^{-1/2} \qquad (\mathrm{TI\; at\;}  k_\nu).
\end{align}
A similar condition $\omA/2\Omega\gtrsim Pm^{-1/2}$ describes the marginal stability threshold $\gamma_{\rm MW}(k_\nu)>0$, as we verified numerically.

(2) $k_\nu\ll k_\kappa$. In this case, the MW instability with $\gamma_{\rm MW}\approx \omA^2/4\Omega$ is possible only at $k\approx k_\nu$ and only if the stratification effects are weak at $k_\nu$ (Section~\ref{sec:MW_alldiffusivities}).  This requires $k_N\ll k_\nu$, which implies non-diffusive buoyancy response ($k_\kappa\gg k_N$). 
The condition $k_N\ll k_\nu$ sets a lower limit of $\omA\gg k_\theta N/k_\nu$. In addition, the peak $\gamma_{\rm MW}(k_\nu)\approx\omA^2/4\Omega$
requires negligible effects of magnetic diffusion, $\eta k_\nu^2\ll \gamma_{\rm MW}(k_\nu)$. Combining the two constraints, we find
\beq
    \frac{\omA}{2\Omega} \gg \max\left\{\frac{N}{2\Omega} \left(\frac{\nu k_\theta^2}{2\Omega} \right)^{1/2}, \, Pm^{-1/2}\right\} \quad  (\mathrm{TI\; at\; } k_\nu).
\eeq

The instability conditions are summarized in Table~\ref{tab:SummaryMFG}, for both IW and MW. The main parameter
differentiating the possible cases for MW instability is given by
\beq
\label{eq:kkappa_knu}
   \frac{k_\kappa}{k_\nu}=\left(\frac{k_\theta^2 N^2 \nu^2}{8\Omega^3\kappa}\right)^{1/4}.
\eeq
Unstable modes with the largest scales, i.e. with smallest $k$, are usually of main interest. Therefore, the lowest allowed value of $k_\theta$ will be most important for the $k_\kappa$ peak of instability. Our analysis is valid only for $k_\theta r\gg 1$, and so the lowest allowed $k_\theta$ has a wavelength that is a fraction of the distance to the polar axis, $r$. This typically gives $k_\kappa\ll k_\nu$ in regions of stellar radiative zones dominated by thermal stratification.

\begin{deluxetable*}{cccc}
\tablewidth{\linewidth}

 \tablecaption{ Summary of instability criteria for magnetic configurations with $p=1$ near the polar axis in a rotating star, $\omA\ll\Omega$.   \label{tab:SummaryMFG}}

 \tablehead{
 \colhead{Unstable wave branch} & \colhead{Peak wavenumber} &  \colhead{Sub-case} & \colhead{Interval of $\omA$ that gives instability with 
        $\gamma\approx \displaystyle\frac{\omA^2}{4\Omega}$}
 }
 \startdata 
         Inertial  & $\displaystyle {k_\eta=\left(\frac{\om}{\eta}\right)^{1/2}}$ & $\displaystyle{\frac{N}{2\Omega}\ll\left(\frac{\eta k_\theta^2}{2\Omega}\right)^{-1/2}\left(1+\frac{\kappa^2}{\eta^2}\right)^{1/2}}$ &  $\displaystyle{\frac{\omA}{\om}\gg\max\left\{\frac{N}{\om}\left(\frac{\kappa k_\theta^2}{\om}\right)^{1/2}\left(1+\frac{\kappa^2}{\eta^2}\right)^{-1/2},
         \, Pm^{1/2}\right\}}$ \\[3mm]
         & & $\displaystyle{\frac{N}{2\Omega}\gg\left(\frac{\eta k_\theta^2}{2\Omega}\right)^{-1/2}\left(1+\frac{\kappa^2}{\eta^2}\right)^{1/2}}$ & No instability with $\gamma_{\rm IW}(k_\eta)\approx \displaystyle{\frac{\omA^2}{4\Omega}}$ \\[3mm]
        \hline
        Magnetostrophic & $\displaystyle{k_\kappa=\left(\frac{k_\theta^2 N^2}{ \om\kappa}\right)^{1/4}}$  & $k_\kappa\ll k_\nu$ &  $\displaystyle{\left(\frac{N}{\om}         \right)^{1/2}\left(\frac{\kappa k_\theta^2}{\om}\right)^{1/4}\gg\frac{\omA}{\om}\gg \left(\frac{N}{\om} \frac{\eta}{\kappa}\right)^{1/2}\left(\frac{\kappa k_\theta^2}{\om}\right)^{1/4}}$
        \\[3mm]
         & &$\displaystyle{k_\kappa\gg k_\nu}$ &   No instability with $\gamma_{\rm MW}(k_\kappa)\approx \displaystyle{\frac{\omA^2}{4\Omega}}$ \\[3mm]
        \hline
        Magnetostrophic& $\displaystyle{k_\nu=\left(\frac{\om}{\nu}\right)^{1/2}}$ & $k_\kappa\ll k_\nu$& $\displaystyle{\frac{\omA}{\om}\gg Pm^{-1/2}}$\\[3mm]
         & &  $k_\kappa\gg k_\nu$ & $\displaystyle{\frac{\omA}{\om}\gg\max\left\{\frac{N}{\om}\left(\frac{\nu k_\theta^2}{\om}\right)^{1/2}, \, Pm^{-1/2}\right\}}$ \\[4mm]
        \hline
        Fifth branch& $\displaystyle{\frac{k_N}{\sqrt{3}}=\frac{k_\theta N}{\sqrt{3}\,\omA}}$ & $k_N\ll k_\nu$ and $\kappa<\eta$ & Only near $\displaystyle{\frac{\omA}{\om}\approx\left(\frac{N}{\om}\right)^{1/2}\left(\frac{\eta k_\theta^2}{\om}\right)^{1/4}}$\\[3mm]
         & &  $k_N\gg k_\nu$ or $\kappa>\eta$ & No instability\\[1mm]
 \enddata

 \tablecomments{
 The maximum growth rate $\gamma_{\max}\approx\omA^2/4\Omega$ is attained at the characteristic wavenumbers given in the second column for intervals of $\omA$ given in the fourth column. Values of $\omA$ outside the listed intervals give either stability or a slow instability with $\gamma\ll \omA^2/4\Omega$. For IW, the instability criteria remain practically unchanged for more general values of $p$ and $\theta$ around the polar axis. For MW, the instability criteria are weakly changed as long as $m_\star^2>0$, which corresponds to $p<3/2$ for $m=1$ modes in polar regions.}

\end{deluxetable*}

\subsubsection{Large toroidal field gradients}

Perturbations in magnetic configurations with large gradients $p$ can have $m_\star^2<0$ and should be considered separately (Section~\ref{sec:strongfieldgradient}). In particular, the $m=1$ modes near the polar axis have $m_\star^2<0$ if $p>3/2$. Such perturbations grow with the maximum rate $\gamma_{\rm MW }\approx \omA^2/\om$ in an interval of wavenumbers $k_1<k<k_2$ (\Eq~\ref{eq:k1k2}). The TI with the maximum growth rate exists if $k_2>k_1$, which we wish to express as a condition on $\omA$.

The dependencies of $k_1$ and $k_2$ on $\omA$ may be approximated as 
\begin{align}
   k_1 &\approx \frac{k_\kappa}{1+\omA/\omega_1},  \qquad  
   \omega_1 \equiv (2\Omega\kappa)^{1/4}(k_\theta N)^{1/2}, \\
   k_2 &\approx \frac{k_\nu}{1+\omega_2/\omA},  \qquad  
   \omega_2 \equiv 2\Omega\,Pm^{-1/2},
\end{align}
which captures the change $k_1=k_\kappa\rightarrow k_1=k_N$ when $\omA$ exceeds $\omega_1$, and $k_2=k_\nu\, Pm^{1/2}\omA/2\Omega\rightarrow k_2=k_\nu$ when $\omA$ exceeds $\omega_2$. Note that $k_2/k_1$ is monotonically increasing with $\omA$. The interval of $k_1<k<k_2$ opens up when $k_2=k_1$. This condition gives a quadratic equation for $\omA$,
\beq
   \left(\frac{\omA}{\omega_1}\right)^2-\left(\frac{k_\kappa}{k_\nu}-1\right)\frac{\omA}{\omega_1}-\left(\frac{\eta}{\kappa}\right)^{1/2}=0.
\eeq
Its positive root determines the minimum value of $\omA$ required for $k_2>k_1$. So, the TI with $\gamma_{\rm MW}\approx \omA^2/2\Omega$ occurs if
\beq
\label{eq:InstCond_SF}
   \omA \gg \frac{\omega_1}{2} \left[ \frac{k_\kappa}{k_\nu} - 1 
   + \sqrt{ \left(\frac{k_\kappa}{k_\nu}-1\right)^2+4\left(\frac{\eta}{\kappa}\right)^{1/2} } \right],
\eeq
where $k_\kappa/k_\nu$ is given in \Eq~(\ref{eq:kkappa_knu}). The obtained condition can be further simplified in the limits of $k_\kappa/k_\nu\ll 1$ and $k_\kappa/k_\nu\ll 1$, 
\begin{eqnarray}
\label{eq:cond_sf}
  \frac{\omA}{2\Omega}\gg \frac{k_\kappa}{k_\nu}\left(\frac{\eta\kappa}{\nu^2}\right)^{1/4} \times \left\{\begin{array}{lr}
  \displaystyle{ \left(1+\frac{\kappa}{\eta}\right)^{-1/4} }
  \vspace*{1mm} & k_\kappa\ll k_\nu \\
  \displaystyle{ \left(1+\frac{k_\kappa^4}{k_\nu^4}\frac{\kappa}{\eta}\right)^{1/4} }  & k_\kappa\gg k_\nu
  \end{array}\right.
\end{eqnarray}
This result is summarized in Table~\ref{tab:SummarySFG}.

\begin{deluxetable*}{cc}
\tablewidth{\linewidth}

 \tablecaption{Summary of MW instability criteria for $m=1$ modes of magnetic configurations with $p>3/2$ near the polar axis in a rotating star $\omA\ll\Omega$. \label{tab:SummarySFG}}

 \tablehead{
 \colhead{Case} & \colhead{Interval of $\omA$ that gives instability with 
        $\gamma\approx \displaystyle\frac{\omA^2}{2\Omega}$}
 }
 \startdata 
    $k_\kappa\ll k_\nu$ & $\displaystyle{\frac{\omA}{\om}\gg \left(\frac{N}{\om}\right)^{1/2}\left(\frac{\eta k_\theta^2}{\om}\right)^{1/4} \min\left\{1,\left(\frac{\eta}{\kappa}\right)^{1/4}\right\}}$\\[3mm]
    \hline
    $k_\kappa\gg k_\nu$&  $\displaystyle{\frac{\omA}{\om}\gg \max\left\{\frac{N}{\om}\left(\frac{\nu k_\theta^2}{\om}\right)^{1/2},\left(\frac{N}{\om}\right)^{1/2}\left(\frac{\eta k_\theta^2}{\om}\right)^{1/4}\right\}}$\\[3mm]
 \enddata

 \tablecomments{
 Instability occurs with the maximum growth rate $\gamma_{\rm MW}\approx\omA^2/4\Omega$ when the magnetic field strength (or the corresponding $\omA$) exceeds the critical value given in the second column. The instability criteria remain unchanged to within factors of order unity for other values of $p$ and $\theta$ as long as $m_\star^2<0$.}

\end{deluxetable*}

\subsection{Role of the poloidal magnetic field}
\label{sec:radialfield}

The TI is driven by the hoop stress of the toroidal magnetic field $B_\phi$, and our calculations focused on pure toroidal field configurations, neglecting the poloidal field that is present in real stars. It is well known that the presence of both $B_\phi$ and $\mathbf{B}_{\mathrm{pol}}=(B_R,B_\theta,0)$ can stabilize the configuration (see e.g. \cite{braithwaite2009axisymmetric} for a detailed discussion). The following simple estimate summarizes the effect of the poloidal field and when it can be neglected.

Recall that perturbations of a pure toroidal configuration generate perturbations of the Lorentz force $\bfL$ (\Eq~\ref{eq:fL}) whose solenoidal part is independent of the poloidal wavevector $\bk_{\rm pol}$ and depends only on $k_\phi=m/r$. By contrast, in the presence of $B_R\neq 0$ the large radial component of the perturbation wavevector, $k_R\approx k$, appears in the solenoidal magnetic force, since the term $\bB\cdot\nabla\bb$ now contains a radial gradient $\sim B_R\partial_R \bb$. Tension of the magnetic field lines now resists the large radial shear created by the horizontal fluid motions involved in the TI. This effect arrests the TI when the tension force due to the radial background field $\sim k_R B_R \bb/4\pi$ becomes comparable to the tension force $\sim m B_\phi\bb/4\pi r$ that drives the TI. Using $k_R\approx k$ and $m=1$ one can see that the stabilizing effect of $B_R$ may be neglected if
\beq
\label{eq:B_R_B_phi}
   \frac{B_R}{B_\phi}\ll \frac{1}{kr}.
\eeq
Similarly, $B_\theta$ provides a tension force $\sim k_\theta B_\theta \bb/4\pi$. Its stabilizing effect is  negligible compared to that of $B_R$ when $k_\theta B_\theta\ll k_R B_R$ (which is normally satisfied, since the perturbations relevant for TI have $k_\theta\ll k_R\approx k$).

This gives an additional necessary condition for the instability of a perturbation with a given $k$. In particular, for a magnetic configuration with fast rotation $\Omega\gg\omA$ and moderate field gradients $p$ the TI 
can peak at the following wavenumbers: $k_\eta$ in the IW branch, $k_\nu$ and $k_\kappa$ in the MW branch, and $k_0\sim k_N$ in the fifth branch. For each of them, the condition~(\ref{eq:B_R_B_phi}) may be stated 
\begin{align}\label{eq:radialfieldcondition}
    \omA\gg (k_s r)\omA^R,  \qquad s\in\{\eta,\nu,\kappa,N\},
\end{align}
where $\omA^R\equiv B_R/(4\pi \rho r^2)^{1/2}$ is defined for $B_R$ similarly to the definition of $\omA$ for $B_\phi$. The TI can be suppressed even by a weak radial field since the most unstable perturbations have $k\approx k_R\gg r^{-1}$ and invoke a large radial shear. With increasing $B_R$ the instability peak with highest $k_s$ will be suppressed first and the peak with lowest $k_s$ will be suppressed last. In the case of configurations with strong gradients ($m_\star^2<0$), the TI with the canonical growth rate $\gamma=\omA^2/2\Omega$ will be suppressed in the MW branch if  $B_R/B_\phi>(k_1 r)^{-1}$ (where $k_1=\min\{k_N,k_\kappa\}$) and in the IW branch if $B_R/B_\phi>(k_\eta r)^{-1}$.

\section{Discussion}\label{sec:discussion}
\subsection{Summary of main results}
Our results clarify the physical picture of TI in stratified, rapidly rotating stars ($N\gg\Omega\gg\omA$) and give instability criteria for any toroidal magnetic configurations and fluid microphysical parameters. For plausible stellar magnetic configurations with moderate gradients of $B_\phi$, the instability is enabled by diffusive processes which cause overstability of eigenmodes supported by the rotating star. Modes of main interest are inertial and magnetostrophic waves, and we have shown that both can become unstable with the maximum growth rate $\gamma\approx \omA^2/4\Omega$. In addition, instability can occur in another branch of the dispersion relation (which we called the fifth branch) with a smaller growth rate.

Remarkably, the peak growth rate $\gamma\approx \omA^2/4\Omega$ is reached at three characteristic wavenumbers: $\kTI=k_\eta$, $k_\nu$, and $k_\kappa$. Each peak is enabled by a single diffusive process --- magnetic diffusivity $\eta$, viscosity $\nu$, and thermal/chemical diffusivity $\kappa$, respectively. The instability at $k_\eta$ develops only in the IW branch while the instabilities at $k_\nu$ and $k_\kappa$ develop only in the MW branch. Note that both IW and MW modes exist because the perturbed fluid experiences the Coriolis acceleration $-2\bOm\times\bu$ in the rotating star. The mode overstability in each peak results from matching between the Coriolis timescale $t_\Omega=(2\Omega)^{-1}$ and a diffusive timescale: $t_\Omega$ matches the magnetic diffusion timescale $t_\eta$ in IW at $k_\eta$, the viscous timescale $t_\nu$ in MW at $k_\nu$, and the timescale for diffusive buoyancy response $t_\kappa$ in MW at $k_\kappa$. The existence of each instability peak with growth rate $\gamma\approx \omA^2/4\Omega$ depends on the fluid parameters and the magnetic field strength $B_\phi$. Table~\ref{tab:SummaryMFG} gives the complete (necessary and sufficient) conditions for the existence of instability peaks at $k_\eta$, $k_\nu$, and $k_\kappa$.

We have also examined (perhaps more exotic) magnetic configurations with a large gradient of $B_\phi$, which develop the TI differently and more easily, with no assistance from a diffusive process. Such configurations develop instability with the maximum $\gamma\approx \omA^2/\om$ in an extended range of wavenumbers. The presence of diffusive processes tends to reduce this range and may even eliminate it if $B_\phi$ is weak; therefore the TI still has a finite threshold for $B_\phi$ (Table~\ref{tab:SummarySFG}).

The linear stability analysis presented in this paper will have implications for Tayler-Spruit dynamo, which is broadly used in stellar evolution models as a means of angular momentum transport. The dynamo models will need to be reevaluated because our results imply that (1) the TI has a new set of characteristic wavenumbers: $k_{\mathrm{TI}}=k_\eta$, $k_\nu$, $k_\kappa$, $k_N$, and (2) the TI occurs in three distinct branches: inertial $\omega_{\rm IW}(k)$, magnetostrophic $\omega_{\rm MW}(k)$, and the fifth branch $\omega_5(k)$. The instability criteria obtained in this paper significantly differ from previous work \citep{spruit1999differential,zahn2007magnetic,ma2019angular}, which did not realize the presence of three unstable branches with distinct instability peaks (Appendix~\ref{ap:PreviousProofs}). The dependence of the TI regime on the microphysical fluid properties $(\nu,\eta,\kappa)$ and the configuration parameters $(p,\Omega,N,R)$ indicates that there may be no universal prescription for the Tayler-Spruit dynamo. 

One limitation of our results is the neglect of differential rotation $q=d \ln \Omega/d \ln R$ in the TI analysis. Keeping this usual approximation facilitates comparison with previous work. Note however that $q\neq 0$ is responsible for creating a magnetic configuration dominated by toroidal field $B_\phi$ in the first place. The timescale to wind a sufficiently strong $B_\phi$ and trigger TI will scale as $q^{-1}$. The presented analysis holds for sufficiently low $q$, which correspond to long winding timescales.

\subsection{Implications for numerical simulations}
Our results offer a clearer picture of existing numerical simulations of the TI and help inform future numerical work. Below, we examine three simulation studies that have investigated the TI in the presence of rapid rotation, stable stratification, and explicit diffusivities, so direct comparison can be made with predictions of our linear stability analysis.  

Simulations in \cite{ji2023magnetohydrodynamic} examine an initial toroidal magnetic field in the presence of both rotation and stratification with $N/\om\approx 1$, $\nu/\Omega r^2\approx10^{-5}$, $Pr\equiv \nu/\kappa=1$, and $Pm\equiv\nu/\eta=0.5$. Their choice of $p=2$ for the background field profile in a cylindrical geometry put the simulation into the strong field gradient regime $m_\star^2<0$ for $m=1$. For these parameters, we find $k_\kappa/k_\nu\approx 0.12 (k_\theta r/2\pi)^{1/2}$ (\Eq~\ref{eq:kkappa_knu}), and TI is expected in the MW branch if $\omA/2\Omega\gg  0.13(k_\theta r/2\pi)^{1/2}$ (\Eq~\ref{eq:cond_sf}).
One can see that the instability should develop for perturbations with the lowest $k_\theta\sim 2\pi/r$ (marginally consistent with our WKB approximation) if $\omA/2\Omega\gg 0.1$. This condition is satisfied in the simulations of \cite{ji2023magnetohydrodynamic}, as they have $\omA/\om\gtrsim 0.25$. Thus, the TI observed in their simulations is consistent with our linear stability analysis. The predicted lowest wavenumber $k_1$ of unstable MW is close to $k_N$. For their simulation setup IW can also become unstable, but only for strong fields, when $\omA$ approaches $2\Omega$. Note also that if their setup was changed to $p=1$, we would predict that the MW branch should become stable, and the large magnetic diffusivity $\eta>\kappa=\nu$ could enable instability in the fifth branch.

Two recent works simulated spherical Couette flow with shear $q<0$ \citep{petitdemange2023spin}
and $q>0$ \citep{barrere2023numerical}, and reported evidence for TI and nonlinear dynamo. The setup with $q>0$ should be free of MRI and appears more suitable for TI studies. However, the growth of magnetic energy observed in the simulations is inconsistent with Tayler-Spruit dynamo: the observed growth is much too fast. In general, dynamo cannot be faster than the instability that drives it: $\gamma_{\rm dyn} < \gamma_{\max}^{\rm TI}$. For instance, \cite{barrere2022} argue that $\gamma_{\rm dyn}$ is smaller than $\gamma_{\max}^{\rm TI}$ by a factor of $\sim (q\omA/N)^{0.5}\ll1$,
\begin{align}
    \gamma_{\rm dyn}\sim \left(\frac{q\omA}{N}\right)^{\frac{1}{2}}\frac{\omA^2}{\Omega}\ll \gamma_{\max}^{\rm TI}.
\end{align}
By contrast, the simulations of \cite{petitdemange2023spin}
and \cite{barrere2023numerical} show $\gamma_{\rm dyn}\gg\gamma_{\max}^{\rm TI}$, as detailed below. This primary issue indicates that the dynamo in their simulations may not be driven by the TI and could be enabled by another mechanism, perhaps specific to the Couette flow setup.\footnote{ Couette flow simulations are complicated by the presence of a ``weak'' dynamo driven by the local shear and small-scale velocity/magnetic fluctuations in unstable Stewartson layers. The weak dynamo may be of two types: $\alpha$-$\Omega$ \citep{brandenburg2005astrophysical,tobias2021turbulent,rincon2019dynamo} or shear-current \citep{rogachevskii2003electromotive,squire2015generation,skoutnev2022large}.
}

The linear stability analysis can be used to predict the expected onset of TI in the simulations as follows. Simulations of \cite{petitdemange2023spin} have fiducial parameters $N/\om\approx 1$, $Ek=\nu/\Omega r^2=10^{-5}$, $Pr=0.1$, $Pm=1$, and a dynamically changing $\omega_A(t)\ll\Omega$. In this case, we find $k_\kappa/k_\nu\approx 6.7\times 10^{-2} (k_\theta r/2\pi)^{1/2}$. These parameters should give MW instability for a sufficiently strong $B_\phi$ (with any gradient $p$) while IW should be stable. However, we find that neither the instability threshold nor the growth rate $\gamma_{\rm dyn}$ observed in their simulations match the linear theory. TI is expected to  develop when $\omA/2\Omega\gg(N/\om)^{1/2}(\kappa k_\theta^2/\om)^{1/4}(\eta/\kappa)^{1/2}$ (Table~\ref{tab:SummaryMFG}). For a direct comparison, one can express this condition in terms of Elsasser number $\Lambda\equiv(\omA/\Omega)^2(\Omega r^2/\eta)$, as \cite{petitdemange2023spin} use $\Lambda$ to parameterize the field strength; it can also be expressed as $\Lambda=4\gamma_{\max}^{\rm TI}T_\eta$ where $T_\eta\equiv r^2/\eta$ is the global magnetic diffusion timescale.
Then, the TI condition becomes $\Lambda>\Lambda^c$ where $\Lambda^c=(N/\Omega) (\Omega r^2/\kappa)^{1/2}$. For the parameters used in \cite{petitdemange2023spin} we find $\Lambda^c\sim 10^2$. However, they report the development of instability at a much lower $\Lambda\sim 1$ (and $\Lambda\sim 10^2$ is barely reached in the simulations). Furthermore, they observe a growth rate $\gamma_{\rm dyn}$ far exceeding $\gamma_{\max}^{\rm TI}$ in the rapidly rotating regime $\Omega\gg\omA$. In particular, the fiducial simulation with $Pm=1$ presented in their Figure~1 shows amplification of the magnetic energy by a factor $E_f/E_i\sim 10$ over a growth period of duration $\Delta t\sim 0.05 T_\eta$ once $\Lambda\gtrsim 2$ and $\omA/\Omega=(\Lambda\, Ek)^{0.5}\gtrsim 10^{-2.5}$. Comparing

\begin{align}
    \gamma_{\rm dyn}=\frac{\ln\left(\frac{E_f}{E_i}\right)}{2\Delta t}\approx \frac{20}{T_\eta},\quad \gamma_{\max}^{\rm TI}=\frac{\Lambda}{4T_\eta}\gtrsim \frac{0.5}{T_\eta},
\end{align}
one can see that the observed growth rate exceeds the maximum TI growth rate by a factor of $\sim 40$. Hence, the observed growth is unrelated to TI.

Simulations in \cite{barrere2023numerical} have similar parameters, but with weaker stratification $N/\Omega=0.1$, $Ek=\nu/\Omega r^2\approx 1.8\times 10^{-5}$, $Pr=0.1$, $Pm=1$. We find that the instability threshold observed in their simulations is consistent with linear theory while the growth rate is not. The TI threshold $\Lambda^c\sim 7$ is reached around a turbulent resistive time $\overline{\tau}_\eta\approx 0.1T_\eta$ in the simulation shown in their Figure 4 (their estimate $\Lambda^c\sim 3$ is slightly lower due to geometrical factors). The magnetic field energy then grows by a factor $E_f/E_i\sim10^2$ in a time $\Delta t\approx 0.3\overline{\tau}_\eta\approx 0.03T_\eta$. Again comparing
\begin{align}
    \gamma_{\rm dyn}=\frac{\ln\left(\frac{E_f}{E_i}\right)}{2\Delta t}\approx \frac{80}{T_\eta},\quad \gamma_{\max}^{\rm TI}=\frac{\Lambda}{4T_\eta}\gtrsim \frac{2}{T_\eta}
\end{align}
one can see that the observed dynamo growth is much faster than the TI. 

Future simulations should give more attention to magnetic field configurations with moderate gradient $p=1$, which is most likely in stellar applications. Then, the TI development relies on diffusivities, in contrast to the more unstable configurations with $p>3/2$. Particular care needs to be taken in the choice of the Prandtl numbers, $Pr$ and $Pm$, to simulate the TI regime relevant for astrophysical systems of interest. Note that simulations with a common choice of $Pm=1$ (or if dissipation is controlled by numerical diffusion) automatically cannot observe the $k_\eta$ peak of the IW instability and the $k_\nu$ peak of the MW instability.

\begin{acknowledgments}
A.M.B. is supported by NSF grants AST-2009453 and PHY-2206609, NASA grant 21-ATP21-0056, and Simons Foundation award No. 446228.
\end{acknowledgments}

%

\vspace{5mm}





\appendix

\section{Comparison with Acheson dispersion relation}
\label{ap:GeneralDispDer}

The general linear stability of a differentially rotating, stratified, non-ideal MHD fluid with an axisymmetric toroidal field $B_\phi$ in the WKB limit was first established in \cite{acheson1978instability} using cylindrical coordinates $(r,z,\phi)$. Since many background variables in stably stratified regions depend primarily on the spherical radius, here we convert Acheson's dispersion relation to spherical coordinates $(R,\theta,\phi)$. Then we simplify it in the Boussinesq limit to compare with the dispersion relation derived in Section~\ref{sec:disp_relation}.

We begin with the dispersion relation from \cite{acheson1978instability} stated in cylindrical coordinates:

\begin{align}\label{eq:AchesonFull}
    &v_A^2\left[\frac{2\Omega m}{r}+\omega_\nu\left\{\frac{2}{r}-\frac{G}{c_s^2}\left(\frac{\omega_\kappa}{\omega+i\kappa k^2/\gamma}\right)\right\}\right]\bigg[\frac{m}{\omega_\eta}\frac{\partial \Omega}{\partial h}+\frac{\partial F}{\partial h}-\frac{\omega}{\omega \gamma+i\kappa k^2}\frac{\partial E}{\partial h}\bigg]\nonumber\\
    &+\left[\frac{k^2}{k_z^2}\left(\omega_\nu-\frac{m^2v_A^2/r^2}{\omega_\eta}\right)-\frac{G}{\omega\gamma+i\kappa k^2}\frac{\partial E}{\partial h}\right]\left[\omega_\nu\omega_\eta-\frac{m^2v_A^2}{r^2}+\frac{v_A^2}{c_s^2}\left(\frac{\omega_\kappa}{\omega+i\kappa k^2/\gamma}\right)\omega \omega_\nu\right]\nonumber\\
    &-\left[\frac{\partial}{\partial h}\left(\Omega r^2\right)+\frac{mv_A^2}{\omega_\eta}\frac{\partial Q}{\partial h}\right]\Bigg[\frac{2\Omega}{r}\left\{\omega_\eta+
    \frac{v_A^2}{c_s^2}\left(\frac{\omega_\kappa}{\omega+i\kappa k^2/\gamma}\right)\omega\right\}+\frac{m v_A^2}{r^2}\left\{\frac{2}{r}-\frac{G}{c_s^2}\left(\frac{\omega_\kappa}{\omega+i\kappa k^2/\gamma}\right)\right\}\Bigg]=0,\\ 
    &\frac{\partial}{\partial h}=\frac{\partial}{\partial r}-\frac{k_r}{k_z}\frac{\partial}{\partial z},
\end{align}
where $v_A=B_\phi/\sqrt{4\pi \rho}$ is the Alfv\'en speed, $c_s=\sqrt{\gamma P/\rho}$ is the sound speed, $G=g_r-g_z(k_r/k_z)$ is related to the local gravitational acceleration, $E=\ln (p/\rho^\gamma)$ is the entropy, and $F=\ln (B_\phi /\rho r)$ and $Q=\ln (B_\phi r)$ are related to the magnetic field profile. The derivation of this dispersion relation assumed $m/r\ll k_r,k_z$, allowing global modes in the $\phi$ direction (low $m$) and using the WKB approximation in the poloidal plane.

Converting to spherical coordinates requires expressing $k_r$, $k_z$ in terms of $k_R$ and $k_\theta$, 
\beq
    k_z=k_R\cos\theta-k_\theta\sin\theta, \qquad k_r=k_R\sin\theta+k_\theta\cos\theta,
\eeq
and expressing the derivative operator $\partial/\partial h$ in spherical coordinates,
\begin{align}
    &\frac{\partial}{\partial h}=-\frac{k_\theta}{k_z}\frac{\partial}{\partial R}+\frac{k_R}{k_z}\frac{\partial}{R\partial \theta}.
\end{align}
We consider stably stratified regions with $N\gg\Omega$ and assume that the rotation rate $\Omega$, gravitational acceleration $\boldsymbol{g}$, and stratification function $E$ all depend only on the spherical radius $R$. Then several derivatives $\partial/\partial h$ simplify:
\beq
    \frac{\partial E}{\partial h}=-\frac{k_\theta}{k_z}\frac{\gamma N^2}{ g},\qquad 
    N^2=\frac{g}{\gamma}\frac{\partial}{\partial R} \ln \left(\frac{p}{\rho^\gamma}\right),
   \qquad 
   R\frac{\partial \Omega}{\partial h}=-\frac{k_\theta}{k_z}q\Omega,\qquad 
    q\equiv\frac{d \ln \Omega}{d \ln R}.
\eeq
The gravitational function $G$ is
\beq
  G=-\frac{k_\theta}{k_z}g, \qquad g=|\boldsymbol{g}|.
\eeq

Further simplification can be made by ordering out small terms in the relevant limit of large sound speed $c_s$ and large background scale heights. In particular, the condition $c_s\gg v_{\rm A}$ implies that the two terms of order $O(v_A^2/c_s^2)$ in the third and fourth lines of \Eq~(\ref{eq:AchesonFull}) can be dropped. The remaining term $G/c_s^2=-k_\theta/k_zH$ depends on the pressure scale height $H=c_s^2/g$, which is a fraction of $R$. For the modes of main interest, $k_\theta\ll k_R$, the term $G/c_s^2$ is non-negligible compared to $2/r$ only near the equatorial plane, $\theta\approx \pi/2$. However, the equatorial region is not interesting for TI, since the instability is known to be quenched at $\theta=\pi/2$. Therefore, we will neglect the term $G/c_s^2$ in the dispersion relation.

Lastly, we substitute $B_\phi\propto R^p\sin^p\theta \cos\theta$, which gives a sufficiently general class of magnetic configurations, as described in Section~\ref{sec:backgroundstate}. Then,
\begin{align}
    \label{eq:Fder}
    &\frac{\partial F}{\partial h}=-\frac{k_\theta}{k_z}\left(\frac{p-1}{R}-\frac{\partial \ln \rho}{\partial R}\right)+\frac{k_R}{k_z}\left(\frac{(p-1)\cos\theta}{R\sin\theta}-\frac{\tan\theta}{R}\right) \approx\frac{p-1-\tan^2\theta}{r},\\
    \label{eq:Qder}
    &\frac{\partial Q}{\partial h}=-\frac{k_\theta}{k_z}\frac{p+1}{R}+\frac{k_R}{k_z}\left(\frac{(p+1)\cos\theta}{R\sin\theta}-\frac{\tan\theta}{R}\right) \approx\frac{p+1-\tan^2\theta}{r},
\end{align} 
where the approximations assume $k_\theta/k_R\ll1$ and $\theta$ away from the equatorial plane.
Note that the dependence of $B_\phi$ on $R$ turns out to be unimportant near the polar axis --- the terms coming from the derivative $\partial/\partial R$ in \Eq~(\ref{eq:Fder}) and (\ref{eq:Qder}) are negligible near the axis. Therefore, choosing $B_\phi$ with a more general dependence on $R$ would not change the results near the axis.

Applying all of the above approximations, one finds that \Eq~(\ref{eq:AchesonFull}) becomes
\begin{align}
   \label{eq:Spherical_with_q}
   D(\omega) =& \left(\omega_\nu\omega_\eta-m_\star^2\omA^2-\frac{k_\theta^2 N^2}{k^2}\frac{\omega_\eta}{\omega_\kappa}\right) \left(\omega_\nu\omega_\eta-m^2\omA^2\right)- 4\frac{k_z^2}{k^2}\left(\Omega \omega_\eta+m\omA^2\right)^2\nonumber\\
    & +2q\sin\theta \frac{k_\theta k_z}{k^2} \Omega^2\left(\omega_\eta^2 -m^2\omega_A^2+2m\frac{\omega_A^2}{2\Omega}(\omega_\eta-\omega_\nu)\right)=0.
\end{align}
A similar dispersion relation is found in \cite{kagan2014role} in the case of a pure toroidal field. When $q=0$, \Eq~(\ref{eq:Spherical_with_q}) reduces to the dispersion relation given in \Eq~(\ref{eq:SphericalFullIntro}), which is derived independently in Section~\ref{sec:disp_relation}.

\section{Instability of inertial waves}
\label{ap:IW}

\subsection{Stable IW in a non-magnetized star}

First, let us consider the case of an unmagnetized star, $\omA=0$. Then, the full dispersion relation $D(\omega)=0$ (\Eq~\ref{eq:SphericalFullIntro}) simplifies to
\beq
\label{eq:D0_IW}
    D_0(\omega)\equiv \omega_\eta^2\left(\omega_\nu^2-4\Omega^2\mu^2-\frac{\omega_N^2\omega_\nu}{\omega_\kappa}\right) = 0,   \qquad  \mu\equiv \cos\theta, \qquad \omega_N\equiv \frac{k_\theta N}{k}.
\eeq
IW are supported at  wavenumbers $k$ where buoyancy effects are weak, i.e. the buoyancy term $\propto\omega_N^2$ is small compared to the Coriolis term $4\Omega^2\mu^2$. This condition is satisfied for sufficiently large wavenumbers:
\beq
\label{eq:IW_appr}
   \left|\frac{\omega_N^2\omega_\nu}{4\Omega^2\mu^2\omega_\kappa}\right|
  \approx \frac{k_\theta^2 N^2}{4\Omega^2|\mu|^2 k^2} \left|[1\pm \frac{i(\kappa-\nu)k^2}{2\Omega|\mu|}\right|^{-1} \ll 1
   \quad {\rm for} \quad
  k\gg k_{\rm b}\equiv \frac{k_\theta N}{2\Omega|\mu|}\left(1+\frac{k_\theta^2 N^2|\kappa-\nu|}{8\Omega^3|\mu|^3}\right)^{-1/4}.
\eeq
 
For wavenumbers $k\gg k_b$, the IW solutions of $D_0(\omega)=0$ can be expanded in the small parameter $\propto \omega_N^2$:
\beq
\label{eq:IW_non-mag}
   \omega_\nu = \omnu +\frac{\omega_N^2}{2\omkap} \qquad   (k\gg k_{\rm b}),
\eeq
where we neglected the higher order terms $\propto\omega_N^4$. The frequency $\omm$ describes IW in the absence of stratification ($\omega=\omm$ when $\omega_N=0$):
\beq
   \omm=\omnu-i\nu k^2, \qquad \omnu = \pm 2\Omega\mu,   \qquad \ometa = \omnu +i(\eta-\nu)k^2, 
   \qquad \omkap = \omnu + i(\kappa-\nu)k^2.
\eeq

\subsection{Unstable IW in a magnetized star}

When the star is magnetized, $\omA\neq 0$, IW can become unstable.
Let us find the correction to the wave frequency $\omega$ due to $\omA\ll\Omega$.
The full dispersion relation $D(\omega)=0$ (\Eq~\ref{eq:SphericalFullIntro}) can be written in the form 
\beq
   D(\omega) = D_0(\omega)-D_1(\omega)=0, \qquad   
   D_1(\omega)\equiv \omA^2\omega_\eta \left[ (m_\star^2+m^2)\,\omega_\nu + 8m\Omega\mu^2 
   -\frac{m^2\omega_N^2}{\omega_\kappa}\right].
\eeq
Let $\omega_0$ be the solution of $D_0(\omega)=0$ given by \Eq~(\ref{eq:IW_non-mag}). 
The solution of $D(\omega)=0$ differs from $\omega_0$ by a small correction $\omega-\omega_0\propto D_1\propto \omA^2$ in the leading order of $\omA^2/\Omega^2\ll 1$. It can be found by expanding $D(\omega)$ about $\omega_0$:
\beq
\label{eq:expand}
  D(\omega)=D(\omega_0) + (\omega-\omega_0) \left.\frac{dD}{d\omega}\right|_{\omega_0} 
  + {\cal O}\left[ (\omega-\omega_0)^2 \right] = 0 \quad \Rightarrow \quad
  \omega-\omega_0 = \frac{D_1(\omega_0)}{D_0'(\omega_0)}
   + {\cal O}\left[ (\omega-\omega_0)^2 \right],
\eeq
where we used $D(\omega_0)=-D_1(\omega_0)$ and $D'\equiv dD/d\omega= D'_0[1+{\cal O}(\omA^2/\Omega^2)]$. A straitghtforward calculation gives
\begin{align}
   D'_0(\omega_0) &= \omega_\eta^2\omega_\nu\left(2+\frac{\omega_N^2}{\omega_\kappa^2}-\frac{\omega_N^2}{\omega_\nu\omega_\kappa}\right), \\
   D_1(\omega_0) &= \omA^2\omega_\nu\omega_\eta \left[ - 2f_\pm 
   - \frac{m\omega_N^2 }{\Omega\omega_\kappa} 
   -\frac{m^2\omega_N^2}{\omega_\nu\omega_\kappa} +{\cal O}\left(\frac{\omega_N^4}{\Omega^4}\right) \right], 
   \qquad f_\pm\equiv -\frac{1}{2}(m^2+m_\star^2\pm 4m|\mu|).
\end{align}
In all terms $\propto\omega_N^2$ one can use $\omega\approx\omm$, since we neglect high-order corrections $\propto\omega_N^4/\Omega^4$. This gives
\beq
   \frac{D_1(\omega_0)}{D_0'(\omega_0)} = -\frac{f_\pm\omA^2}{\omega_\eta}
    -\frac{\omA^2\omega_N^2}{2\omnu\ometa\omkap}\left[m^2\pm 2m|\mu|+f_\pm\left(1-\frac{\omnu}{\omkap}\right)\right].
\eeq
In the leading-order term $\propto f_\pm$ we use $\omega_\eta^{-1}=(\ometa+\omega_N^2/2\omkap)^{-1}\approx \ometa(1-\omega_N^2/2\ometa\omkap)$. Then, we find from \Eq~(\ref{eq:expand}) the solution for the IW frequency $\omega=\omega_{\rm IW}$:
\beq
\label{eq:om_IW2}
   \omega_{\rm IW}(k)=\pm 2\Omega|\mu| -i\nu k^2 +\frac{\omega_N^2}{2\omkap}
   -\frac{f_\pm\omA^2}{\ometa} + \frac{f_\pm\omA^2\omega_N^2}{2\ometa^2\omkap}
    \mp \frac{\omA^2\omega_N^2}{4\Omega|\mu|\ometa\omkap}\left[m^2\pm 2m|\mu|+f_\pm\frac{i(\kappa-\nu)k^2}{\omkap}\right].
\eeq

The obtained result significantly simplifies in the absence of stratification, $\omega_N=0$:
\beq
\label{eq:om_IW0}
   \omega_{\rm IW}(k)=\pm 2\Omega|\mu| -i \nu k^2 
   - \frac{f_\pm\omA^2}{\pm 2\Omega |\mu|+i(\eta-\nu) k^2} 
   \qquad (N=0).
\eeq
If $|\mu|=1$ and $p=1$ (which gives $m_\star^2=m^2$), the solution $\omega_{\rm MW}(k)$ reproduces \Eq~(\ref{eq:om_IW}). The ``$-$'' root ($\omnu=-2\Omega|\mu|$) is of main interest, since it can give $\gamma_{\rm IW}>0$: 
\beq
\label{eq:gamma_IW0}
   \gamma_{\rm IW}(k) = -\nu k^2 + \frac{f_- \omA^2(\eta-\nu)k^2 }{4\Omega^2\mu^2+(\eta-\nu)^2 k^4} 
  \qquad (N=0).
\eeq
Note that instability exists only when $\nu\ll \eta\omA^2/\Omega^2\ll\eta$, so $\nu$ may be neglected in the term $\propto\omA^2$. In the limit of negligible viscosity, \Eq~(\ref{eq:gamma_IW0}) gives the following maximum instability growth rate, 
\beq
\label{eq:IW_peak}
   \gamma_{\rm IW}^{\max} = \frac{f_-\,\omA^2}{4\Omega|\mu|}
    \;\;\; {\rm at} \;\;\; k = k_\eta \equiv \sqrt{\frac{2\Omega|\mu|}{\eta}}  
    \qquad \left( \nu k_\eta^2\ll \gamma_{\rm IW}^{\max} \right).
\eeq
Note also that $\gamma_{\rm IW}>0$ requires $f_->0$, i.e.
\beq
\label{eq:condition_IW0}
   4m|\mu|-m^2-m_\star^2>0 \qquad \Leftrightarrow \qquad 
   p>\frac{m^2+1-2m|\mu|}{\mu^2}, \qquad |\mu|>\frac{1}{p}\left(\sqrt{m^2+p(m^2+1)}-m\right).
\eeq
Here we used the definition of $m_\star^2\equiv m^2-2(p\mu^2-1)$, where $p\equiv \partial \ln B_\phi/\partial\ln r$.
Condition~(\ref{eq:condition_IW0}) is easiest to satisfy in the polar regions $|\mu|=1$, and one can see that the IW instability develops if $p>(m-1)^2$. In the case of main interest, $p=1$, the IW instability is enabled only for $m=1$ and develops at $|\mu|>\sqrt{3}-1$. Magnetic configurations with stronger gradients $1<p<4$ enable instability for $m=0,1,2$.

Next, let us evaluate how the instability is affected by stratification, i.e. include corrections $\propto\omega_N^2\neq 0$. Taking the imaginary part of \Eq~(\ref{eq:om_IW2}) we find  
\beq
\label{eq:gamma_IW2}
   \gamma_{\rm IW}(k) = -\nu k^2 + \frac{f_- \omA^2\eta k^2 }{|\ometa|^2}
   - \frac{\omega_N^2\kappa k^2 }{2|\omkap|^2} 
   - \frac{\omA^2\omega_N^2 \eta k^2}{2|\ometa|^2|\omkap|^2}
     \left(\frac{8\Omega^2\mu^2 f_-}{|\ometa|^2}+2m|\mu|-m^2\right) + \gamma_\nu+\gamma_\kappa,
\eeq
where $\gamma_\nu$ and $\gamma_\kappa$ collect small terms $\propto \nu k^2$ and $\propto \kappa k^2$ that are negligible compared to $-\nu k^2$ and $(\omega_N^2/|\omkap|^2)\kappa k^2$,  so $\gamma_\nu$ and $\gamma_\kappa$ may be omitted. If $|\mu|=1$, $m=1$, $p=1$, and viscosity is neglected ($\nu=0$), the above solution for $\gamma_{\rm IW}(k)$ gives \Eq~(\ref{eq:LowKIW}) at wavenumbers $k\ll k_\eta$.
Note that both corrections $\propto\omega_N^2$ are negative in the main case of interest ($m=1$ and $f_->0$), reducing the growth rate $\gamma_{\rm IW}$. Instability occurs when both corrections are smaller than the main positive term $f_-\omA^2\eta k^2/|\ometa|^2$:
\beq
   \frac{\omega_N^2\kappa}{|\omkap|^2} <  \frac{f_- \omA^2\eta}{|\ometa|^2},  \qquad
    \frac{\omega_N^2}{2|\omkap|^2} \left(\frac{8\Omega^2\mu^2 f_-}{|\ometa|^2}+2m|\mu|-m^2\right) < f_-. 
\eeq
Existence of the instability peak $\gamma_{\rm IW}^{\max} = f_-\,\omA^2/4\Omega|\mu|$ requires that both of these conditions be satisfied at $k=k_\eta$. In addition, the instability requires $k_\eta\gg k_{\rm b}$.

\subsection{Transition to internal gravity waves at small wavenumbers}
\label{ap:IGW}

Perturbations with $k<k_{\rm b}$ are different from IW. Then,
the buoyancy term $\propto\omega_N^2$ in the dispersion relation $D_0(\omega)=0$ exceeds $\Omega^2$, so there is a transition from IW to a different type of wave. The effects of magnetic field and viscosity are small at the transition, and it is well described by the dispersion relation (\Eq~\ref{eq:SphericalFullIntro}) with neglected $\nu$ and $\omA^2$. It gives
\beq
\label{eq:IGW}
   \omega^2-4\Omega^2\mu^2 -\frac{\omega_N^2\omega}{\omega_\kappa} =0 
   \quad \Rightarrow \quad 
   \omega = \left\{ \begin{array}{lr}
    \displaystyle{ \pm 2\Omega|\mu| \sqrt{1+\frac{k_{\rm IGW}^2}{k^2}} }  
    & \quad\kappa k^2\ll |\omega|  \\ [4mm]
    \displaystyle{ \pm 2\Omega|\mu| \left( \sqrt{1 - \frac{k_\kappa^8}{4k^8} } \mp \frac{i k_\kappa^4}{2k^4} \right) }  &    \quad \kappa k^2\gg |\omega|
                              \end{array}\right\} 
     \qquad \begin{array}{l}
     \displaystyle{ k_{\rm IGW}\equiv \frac{k_\theta N}{2\Omega|\mu|} } \\ [4mm]
     \displaystyle{ k_\kappa\equiv \left(\frac{k_\theta^2N^2}{2\Omega|\mu| \kappa}\right)^{1/4} }
                 \end{array}
\eeq
This dispersion relation gives IW at $k\gg k_{\rm b}=\min\{k_{\rm IGW},k_\kappa\}$, with neglected effects of viscosity and magnetic field.
Behavior of perturbations with $k<k_{\rm b}$ is controlled by buoyancy response, which depends on diffusivity $\kappa$. 
Note that $k_{\rm IGW}/k_\kappa=(\kappa/\kappa_0)^{1/4}$ where $\kappa_0\equiv 8\Omega^3|\mu|^3/k_\theta^2N^2$. 
If $\kappa<\kappa_0$, perturbations with $k<k_{\rm IGW}$ are the usual internal gravity waves (IGW) with frequency $\omega=k_\theta N/k$. If $\kappa>\kappa_0$, fast buoyancy diffusion results in overdamped IGW with a pure imaginary frequency $\omega$.

\section{Instability of magnetostrophic waves}
\label{ap:MW}

The MW branch has a low frequency $|\omega|\sim\omA^2/\om\ll \omA\ll\Omega$. This ordering allows one to simplify the full dispersion relation $D(\omega)=0$ (\Eq~\ref{eq:SphericalFullIntro}) by dropping terms proportional to $\omega^4$, $\omega^3$, and $\omega^2\omA^2$. 
The resulting dispersion relation $D_{\rm MW}(\omega)=0$ can be easily solved for $\omega$.

The solution depends on viscosity, magnetic diffusivity, and stratification. 
It is helpful to examine three limits where only one of the parameters $\nu$, $\eta$, or $N$ is non-zero, so that each effect is viewed in isolation. The dispersion relation $D_{\rm MW}(\omega)=0$ in each limit will be reduced to a quadratic equation. We will retain relevant roots $\omega_{\rm MW}$ that match the known ideal MHD limit for MW (\Eq~\ref{eq:MW_ideal}) when $\nu=0$, $\eta=0$, and $N=0$.
\medskip

(1) $\nu\neq 0$, $\eta=0$, $N=0$: 
\beq
\label{eq:DMW_nu}
    D_{\rm MW}(\omega)=
  \left(4\Omega^2\mu^2+\nu^2k^4\right)\omega^2 + \omA^2 \left[ i\nu k^2 
       (m_\star^2+m^2)+8m\Omega\mu^2\right] \omega + m^2\omA^4(4\mu^2 -m_\star^2) = 0,
\eeq
where $\mu\equiv \cos\theta$. The solution $\omega_{\rm MW}(k)$ depends on the characteristic wavenumber $k_\nu=(2\Omega|\mu|/\nu)^{1/2}$. For instance, for $m=0$,
 \beq
    \omega_{\rm MW}=-\frac{i\nu k^2 m_\star^2\omA^2}{4\Omega^2\mu^2+\nu^2k^4}
    =-\frac{m_\star^2\omA^2}{2\Omega|\mu|}\,\frac{ik^2/k_\nu^2}{1+k^4/k_\nu^4} \qquad (m=0).
 \eeq
For $m>0$, the behavior of $\omega_{\rm MW}(k)$ can be clearly seen if one expands the solution at $k\ll k_\nu$ and $k\gg k_\nu$:
\begin{eqnarray}
\label{eq:MW_nu}
  \omega_{\rm MW}\approx \frac{\omA^2}{2\Omega|\mu|}\times\left\{\begin{array}{lr}
  \displaystyle{ \left(2|\mu|-\sqrt{m_\star^2}\right)\left[-m + i\, \frac{k^2}{k_\nu^2}\frac{(m^2+m_\star^2)}{2\sqrt{m_\star^2}}\right], } & \quad k\ll k_\nu \\[4mm]
  \displaystyle{ \left(\sqrt{(m^2-m_\star^2)^2+16m^2\mu^2} - m^2 - m_\star^2\right)
   \left( -\frac{2m|\mu|}{\sqrt{(m^2-m_\star^2)^2+16m^2\mu^2}} \frac{k_\nu^4}{k^4}+\frac{ik_\nu^2}{2 k^2}\right) }, & \quad k\gg k_\nu.                                                                                                    \end{array}\right. \quad
\end{eqnarray}
\medskip
  
(2) $\nu=0$, $\eta\neq 0$, $N=0$: 
 \begin{align}
 \nonumber
    D_{\rm MW}(\omega) = & (4\Omega^2\mu^2+\eta^2 k^4)\omega^2+\left[8m\omA^2\Omega\mu^2 + i\eta k^2 \omA^2(m^2+m_\star^2) + 8 i \eta k^2\Omega^2\mu^2\right]\omega        \\
  &+ \, m^2\omA^4(4\mu^2-m_\star^2)-4\eta^2 k^4\Omega^2\mu^2+8i\eta k^2 m\omA^2 \Omega\mu^2 = 0.
 \end{align}
For further analysis given below it will be sufficient to use the solution at $k\ll k_\eta=(2\Omega|\mu|/\eta)^{1/2}$,
\beq
\label{eq:MW_eta}
   \omega_{\rm MW}\approx -\frac{m\omA^2}{2\Omega|\mu|}\left(2|\mu|-\sqrt{m_\star^2}\right) - i \eta k^2, 
   \qquad k\ll k_\eta.
\eeq
\medskip

(3) $\nu=0$, $\eta=0$, $N\neq 0$: 
\begin{eqnarray}
\label{eq:DMW_N}
    D_{\rm MW}(\omega)=  
 4\Omega^2\mu^2\omega^2+8m\omA^2\Omega\mu^2\omega+m^2\omA^4(4\mu^2-m_\star^2-\epsilon_N)=0,  
 \qquad  \epsilon_N\equiv \frac{k_\theta^2 N^2\omega}{k^2\omA^2\omega_\kappa} = \frac{k_N^2\omega}{k^2\omega_\kappa}.
 \end{eqnarray}
For $m=0$ the only solution is $\omega=0$, so below we consider $m>0$.
The buoyancy term $\epsilon_N$ weakly affects the dispersion relation as long as $|\epsilon_N|\ll 1$. In this case, the solution may be expanded in $\epsilon_N$:
 \beq
 \label{eq:MW_epsN}
   \omega_{\rm MW}=\frac{m\omA^2}{2\Omega|\mu|}\left[-2|\mu|\pm \sqrt{m_\star^2} \left(1+\frac{\epsilon_N}{2m_\star^2}\right) + {\cal O}(\epsilon_N^2)\right]
   \qquad {\rm if}
   \quad |\epsilon_N|\ll 1.
 \eeq
The small correction $\epsilon_N$ in the linear order can be found using iterations, i.e. using the zero-order solution for $\omega_{\rm MW}$ (with $\epsilon_N=0)$ to evaluate $\epsilon_N=k_N^2\omega/k^2\omega_\kappa$. The result also shows the range of $k$ where the regime of $|\epsilon_N|\ll 1$ holds:
 \beq
 \label{eq:eps_N}
    |\epsilon_N|\approx \left| \frac{k^2}{k_N^2}+\frac{i k^4}{m(-2|\mu| \pm \sqrt{m_\star^2})\, k_\kappa^4}\right|^{-1}\ll 1 \quad {\rm for} \quad k\gg k_1\equiv \min\{k_N,k_\kappa\}, \qquad  k_\kappa\equiv \left(\frac{k_N^2\omA^2}{2\Omega|\mu|\kappa}\right)^{1/4}.
\eeq
The opposite case of strong buoyancy effects, $|\epsilon_N|\gg 1$, can occur in two regimes: $k\ll k_N<k_\kappa$ and $k\ll k_\kappa<k_N$ (which correspond to small and large diffusivity $\kappa$, respectively). The solutions of \Eq~(\ref{eq:DMW_N}) in these two limits are
\beq
\label{eq:MW_N}
   \omega_{\rm MW}\approx \pm\frac{m\omA^2 k_N}{2\Omega |\mu| k} - \frac{i\kappa k^2}{2}, 
      \qquad k\ll k_N<k_\kappa,
\eeq
\beq
\label{eq:MW_kappa}
   \omega_{\rm MW}\approx \frac{\omA^2}{2\Omega|\mu|}(4\mu^2-m_\star^2)\left(-\frac{4|\mu|k^8}{k_\kappa^8}
    + \frac{ik^4}{k_\kappa^4}\right), 
      \qquad k\ll k_\kappa<k_N.
\eeq

In addition, the MW frequency $\omega_{\rm MW}(k)$ may be derived for the interval $\min\{k_N,k_\kappa\}\ll k \ll k_\nu$ including both viscosity and buoyancy effects, $\nu\neq 0$ and $N\neq 0$. Then, buoyancy terms may be considered as a small correction to $D_{\rm MW}(\omega)$ given in \Eq~(\ref{eq:DMW_nu}). In particular, in the main case of interest $m=1$, $m_\star=1$ and $|\mu|=1$, the dispersion relation becomes
\begin{eqnarray}
\label{eq:DMW_Nnu}
    D_{\rm MW}(\omega)=  
 4\Omega^2\omega^2+\left(2i\frac{k^2}{k_\nu^2}+8\right)\omA^2\Omega\omega+\omA^4(3-\epsilon_N) =0,  
 \qquad  \epsilon_N\equiv -\frac{\omega_N^2\omega}{\omA^4\omega_\kappa}\left(\omega_\nu\omega-\omA^2\right).
 \end{eqnarray}
Its solution can be expanded in $|\epsilon_N|\ll1$:
\beq
\label{eq:om_MW_nu_N}
\omega_{\rm MW}=\frac{\omA^2}{\om}\left[-1+\frac{\epsilon_N}{2}+i\frac{k^2}{k_\nu^2}(1-\epsilon_N)+{\cal O}(\epsilon_N^2)+{\cal O}\left(\frac{k^4}{k_\nu^4}\right)\right].
\eeq
Here one can use iteration: use the $\omega_{\rm MW}=(\omA^2/2\Omega)(-1+ik^2/k_\nu^2)$ found at $\epsilon_N=0$ to evaluate $\epsilon_N$ in the next order. This gives $\epsilon_N\approx -\omega_N^2/2\Omega\omega_\kappa$.
The resulting solution~(\ref{eq:om_MW_nu_N}) gives $\omega_{\rm MW}$ with the buoyancy correction; its imaginary part $\gamma_{\rm MW}$ is stated in \Eq~(\ref{eq:lowMWnu}).

\bigskip 
The obtained solutions for $\omega_{\rm MW}$ display the characteristic wavenumbers $k$ where perturbations become affected by viscosity, magnetic diffusivity, or stratification. The solutions become particularly simple in the case of main interest, $p=1$ and $|\mu|=1$, which is discussed in the main text. This Appendix extends the stability analysis to general $p$ and $\mu$. Note that the parameter $m_\star^2\equiv m^2-2(p\mu^2-1)$ may become negative for sufficiently large gradients of the background magnetic field $p=\partial \ln B_\phi/\partial \ln r$. The instability picture is qualitatively different for $m_\star^2>0$ and $m_\star^2<0$.

\subsection{Moderate gradients of the magnetic field ($m_\star^2>0$)}
\label{sec:moderate}

Magnetic configurations with $p=1$ provide the most important example of TI where all perturbations have $m_\star^2>0$. Then, as shown in Section~\ref{sec:RotatingTI}, the growth rate of MW instability can have two peaks with $\gamma\approx\omA^2/4\Omega$, at $k_\nu$ and $k_\kappa$. The peak at $k\sim k_\nu$ is shaped by viscosity while effects of magnetic diffusivity and buoyancy response are negligible. So, this instability peak is described by the solution $\omega_{\rm MW}(k)$ found with $\eta=0$ and $N=0$ for $m>0$ (\Eq~\ref{eq:MW_nu}). Instability requires $\gamma_{\rm MW}\equiv Im(\omega_{\rm MW})>0$, which occurs (at both $k<k_\nu$ and $k>k_\nu$) if $m>0$ and
\beq
\label{eq:cond_mu}
   4\mu^2>m_\star^2 \qquad \Leftrightarrow \qquad \mu^2>\frac{m^2+2}{4+2p}.
\eeq

The other peak of MW instability can only exist in the case of fast buoyancy diffusion, $k_\kappa<k_N$.  It occurs at $k\sim k_\kappa$ and is shaped by the diffusive buoyancy response while the effects of viscosity and magnetic diffusivity are negligible. This peak is described by the solution $\omega_{\rm MW}(k)$ found with $\nu=0$ and $\eta=0$ with $m>0$ (\Eqs~(\ref{eq:MW_epsN}) and (\ref{eq:MW_kappa})). The root with positive imaginary part $\gamma_{\rm MW}$ is
\begin{eqnarray}
\label{eq:MW_kappa1}
  \omega_{\rm MW}\approx \frac{\omA^2}{2\Omega|\mu|}\times\left\{\begin{array}{lr}
  \displaystyle{ (4\mu^2-m_\star^2)\left(-\frac{4|\mu| k^8}{k_\kappa^8}
    + \frac{ik^4}{k_\kappa^4}\right), } & \quad k\ll k_\kappa \\[4mm]
  \displaystyle{\left(2|\mu| - \sqrt{m_\star^2}\right)\left( -1 + \frac{i m\, k_\kappa^4}{ 2\sqrt{m_\star^2} \, k^4}\right) }, & \quad k\gg k_\kappa                                                                                                        \end{array}\right. \quad
\end{eqnarray}
One can see that $\gamma_{\rm MW}>0$ if $4\mu^2>m_\star^2$, same as in \Eq~(\ref{eq:cond_mu}). 

Thus, the conditions on $p$, $\mu$, and $m>0$ for the two peaks of MW instability are the same  (\Eq~\ref{eq:cond_mu}). In particular, when $p=1$, the instability condition can only be satisfied for $m=1$ (since $\mu^2\leq 1$) and requires $\mu^2>1/2$. The condition also requires $p>(m^2+2)/2\mu^2-2>-1/2$.

\subsection{Strong gradients of the magnetic field ($m_\star^2<0$)}

Magnetic configurations with a sufficiently large gradient $p$ allow perturbations with $m_\star^2<0$. Then, instability easily occurs without any effects of viscosity, magnetic diffusivity, or stratification. Indeed, the MW dispersion relation with $\nu=0$, $\eta=0$, and $N=0$ gives a solution with $\gamma_{\rm MW}>0$:
\beq
\label{eq:MW_large_p}
  \omega_{\rm MW}=\frac{m\omA^2}{2\Omega|\mu|}\left(-2|\mu|+i\,\sqrt{-m_\star^2}\right)  
  \qquad    (\nu=0,\, \eta=0,\, N=0),
\eeq
which is constant for all $k$. When $\nu\neq 0$, $\eta\neq 0$, or $N\neq 0$, the simple constant solution approximately holds in a limited range $k_1<k<k_2$. It changes where viscosity, magnetic diffusion, or stratification become important. 

The effects of viscosity $\nu$ and magnetic diffusivity $\eta$ become important at large $k$. 
Consider first the correction due to $\nu\neq 0$ (\Eq~\ref{eq:MW_nu}). At $k\ll k_\nu$, $\omega_{\rm MW}$ remains close to the ideal MHD result (\Eq~\ref{eq:MW_large_p}), with a small correction. The solution strongly changes at $k\sim k_\nu$.
 At $k\gg k_\nu$, the solution no longer contains $\sqrt{-m_\star^2}$, so the negative $m_\star^2$ no longer generates an imaginary part in $\omega_{\rm MW}$ and the instability becomes similar to that for $m_\star^2>0$: its growth rate at $k\gg k_\nu$ becomes independent of $\Omega$ and decreases as $\gamma_{\rm MW}\propto k^{-2}$.
Next, consider the correction due to magnetic diffusivity $\eta\neq 0$ (\Eq~\ref{eq:MW_eta}). Here, the correction $-i\eta k^2$ weakly affects the instability growth rate $\gamma_{\rm MW}\approx m\omA^2\sqrt{-m_\star^2}/2\Omega|\mu|$ at $k\ll \tilde{k}_\eta\equiv (\omA/2\Omega|\mu|)\, m^{1/2}(-m_\star^2)^{1/4} k_\eta$. At $k>\tilde{k}_\eta$, $Im(\omega_{\rm MW})$ changes sign, so the instability is completely suppressed.
If both $\eta\neq 0$ and $\nu\neq 0$, the deviation of $\omega_{\rm MW}$ from the ideal MHD solution (\Eq~\ref{eq:MW_large_p}) occurs at $k_2\equiv \min\{k_\nu,\tilde{k}_\eta\}$.

Buoyancy effects due to $N\neq 0$ become important at low $k$. In particular, in the regime of a large diffusivity $\kappa$ ($k_\kappa<k_N$), the solution for $\omega_{\rm MW}$ is summarized in \Eq~(\ref{eq:MW_kappa1}); it shows that buoyancy effects are negligible at $k\gg  k_\kappa$ and strongly reduce $\gamma_{\rm MW}$ at $k<k_\kappa$. In the regime of a small $\kappa$ ($k_\kappa>k_N$), buoyancy effects become important at $k\sim k_N$, as one can see from \Eqs~(\ref{eq:MW_epsN}) and (\ref{eq:MW_N}). At $k<k_N$, the imaginary part of $\omega_{\rm MW}$ becomes negative (\Eq~\ref{eq:MW_N}), i.e. instability is completely suppressed. The regimes of $k_\kappa>k_N$ and $k_\kappa<k_N$ can be summarized together as follows: buoyancy weakly affects $\omega_{\rm MW}$ if $k>k_1\equiv \min\{k_\kappa,k_N\}$.


\section{Instability peak in the fifth branch}
\label{ap:fifth_root}

The dispersion relation~(\Eq~\ref{eq:SphericalFullIntro}) can be written in a standard form $F(\omega)=0$ where $F(\omega)\equiv\omega_\kappa D(\omega)$ is a fifth-order polynomial. It has five roots. In the fast rotation regime $\Omega\gg\omA^2$ and with non-zero diffusivities, all five roots may be important. We have discussed above two pairs of roots: $\omega_{\rm IW}=\omega_{1,2}$ and $\omega_{\rm MW}=\omega_{3,4}$. In addition, there is a fifth branch, which has a small frequency $\omega_5$. Numerical exploration shows that the fifth branch can develop a significant instability, with $\gamma_5\equiv Im(\omega_5)$ becoming in some cases comparable to the peaks of IW and MW instabilities. This occurs when the effects of viscosity $\nu$ and diffusivity $\kappa$ are small while magnetic diffusivity and buoyancy effects are essential. Consider the full dispersion relation~(\Eq~\ref{eq:SphericalFullIntro}) with $\kappa=0$ and rearrange terms using $\omega_\eta k_\theta^2N^2/k^2=\omega\omega_N^2+i\eta k^2\omega_N^2$:
\beq
\label{eq:F_no_kappa}
 F(\omega)= \omega\left[\left(\omega_\nu\omega_\eta-m_\star^2\omA^2-\omega_N^2\right)\left(\omega_\nu\omega_\eta-m^2\omA^2\right)-4\mu^2\left(\Omega \omega_\eta+m\omA^2\right)^2\right]-i\eta k^2\omega_N^2\left(\omega_\nu\omega_\eta-m^2\omA^2\right)=0.
\eeq
One can see that the fifth root $\omega_5\neq 0$ appears only when both $\eta\neq 0$ and $\omega_N\neq 0$.

At low frequencies of interest for the fifth branch, $F(\omega)$ can be further simplified by neglecting the terms $\omega_\nu\omega_\eta\ll \omA,\omega_N$ (this approximation may fail when the term $i\nu k^2$ in $\omega_\nu$ is large; the viscosity correction is discussed in the end of this appendix). Then the dispersion relation becomes
\beq
\label{eq:F_3}
  F(\omega)=\frac{m^3\omA^6}{2\Omega|\mu|}\left\{ \tilde\omega\left[m_\star^2+\frac{k_N^2}{k^2}-\left(\tilde \omega+2|\mu|+iW\frac{ k^2}{k_N^2}\right)^2\right]+iW\right\}=0, 
  \qquad W\equiv \frac{2\Omega|\mu|\eta k_N^2}{m\omA^2},
\eeq
where $\tilde\omega\equiv 2\Omega|\mu|\omega/m\omA^2$ is a dimensionless frequency and $W$ is the dimensionless parameter containing the effects of buoyancy and magnetic diffusivity. 
The cubic equation~(\ref{eq:F_3})  describes the low-frequency roots $\omega_{3,4}$ and $\omega_5$ of the full dispersion relation~(\ref{eq:F_no_kappa}). In particular, at $W=0$ one finds
\begin{equation}
\label{eq:MW_ideal_apendix}
    \omega_{3,4}=\frac{m\omA^2}{\om|\mu|}\left( -2|\mu| \pm
    \sqrt{m_\star^2+\frac{k_N^2}{k^2}} \right),\quad \omega_5=0 \qquad (W=0).
\end{equation}
We wish to find $\omega_5$ at $W\neq 0$ and check if there is an instability, $\gamma_5>0$. Using marginal stability analysis similar to \cite{spruit1999differential}, \cite{zahn2007magnetic} and \cite{ma2019angular}, one can show that $\gamma_5$ can be positive when $W\lesssim {\cal O}(1)$ (see Appendix~\ref{ap:PreviousProofs} for details). Let us examine how $\omega_5$ is changed from zero by a small $W\ll 1$. The small $\omega_5\neq 0$ at $W\neq 0$ can be calculated using the expansion of $F(\omega)$ about $\omega=0$:

\beq
\label{eq:om5_exp}
   F(\omega)=F(0)+F'(0)\, \omega  +\frac{F''(0)}{2}\, \omega^2 + {\cal O}(\omega^3)=0,
\eeq
where the derivatives of $F(\omega)$ at $\omega=0$ are
\beq
\label{eq:F_derivatives}
  F(0)=i \eta k_N^2 m^2  \omA^4 + {\cal O}(W^2), \quad\;
  F'(0)=m^2\omA^4\left(m_\star^2-4\mu^2+\frac{k_N^2}{k^2}\right) + {\cal O}(W), \quad\;
  F''(0)=-16\Omega\mu^2 m\omA^2 + {\cal O}(W).
\eeq
\begin{figure}
    \label{fig:Fifthroot}
    \centering
	\includegraphics[width=\linewidth]{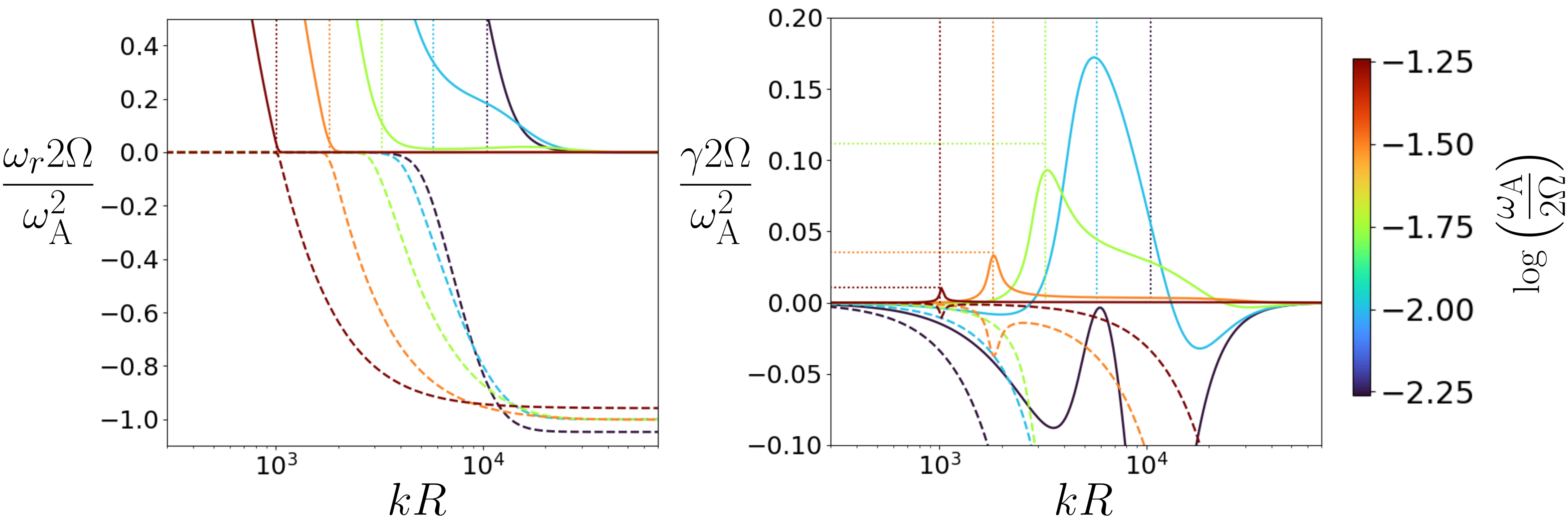}
    \caption{ 
    Numerical solutions for the fifth branch $\omega_5(k)$ (solid curves) and the stable MW branch $\omega_3(k)$ (dashed curves). The real part of $\omega$ is shown in the left panel and the imaginary part in the right panel. The solutions are plotted for perturbations with $m=1$ in the polar regions $|\mu|\equiv|\cos\theta|\approx 1$ of magnetic configurations with gradient $p=1$. Other parameters are $\nu=\kappa=0$, $R k_\theta N/\om=10^2$, and $\eta k_\theta^2/\om =10^{-12}$. Different colors correspond to different values of $\omA/2\Omega$, as indicated in the color bar on the right. In the ideal MHD limit ($W=0$) the branches $\omega_3$ and $\omega_5=0$ would cross at $k_0=k_N/\sqrt{3}$. A finite $W\neq 0$ eliminates the degeneracy at $k_0$, creating a finite gap between the two continuous branches and showing that the fifth branch continually transforms into a MW-like wave at $k<k_0$. The fifth branch is unstable, $\gamma_5>0$, when the parameter $\omA/2\Omega$ exceeds a critical value $\omA^c/\om\approx 10^{-2.25}$, which corresponds to $W\approx 11$. With increasing $\omA/2\Omega$ (which corresponds to a decreasing $W$), the peak of  $\gamma_5(k)$ first increases, reaching a maximum value at $W\approx 1$, and then decreases. At small $W\ll 1$, $\omega_5(k)$ follows the analytical solution~(\ref{eq:om5}): the peak of $\gamma_5$ is located at $k_0=k_N/\sqrt{3}$ (indicated by vertical dotted lines) and approximately follows  $\gamma_5^{\max}$ given in \Eq~(\ref{eq:gam5_max}) (indicated by horizontal dotted lines).
    }
\end{figure}
We have expanded $F(\omega)$ up to the second order in $\omega$ to include the case of $F'(0)=0$. From \Eq~(\ref{eq:om5_exp}) we find
\beq
\label{eq:om5_sol}
   \omega_5=\frac{F'(0)}{F''(0)}\left[-1+\sqrt{1-\frac{ 2 F(0)F''(0)}{[F'(0)]^2}}\right]
   =\left\{\begin{array}{lr}
   \displaystyle{ -\frac{F(0)}{F'(0)} }, & \qquad Q\ll 1 \\ [4mm]
   \displaystyle{ \sqrt{-\frac{2 F(0)}{F''(0)} } }, & \qquad Q\gg 1
             \end{array}\right.
\eeq
where the root is chosen so that $\omega_5=0$ when $\eta=0$, and we have defined
\beq
  iQ\equiv -\frac{F(0)F''(0)}{[F'(0)]^2}=\frac{8iW |\mu| }{(k_N^2/k^2-f)^2}, \qquad f\equiv 4\mu^2-m_\star^2.
\eeq
The solution~(\ref{eq:om5_sol}) holds when $W\ll 1$. It shows that $\gamma_5(k)$ has a narrow peak near $k=k_N/\sqrt{f}$ when $f>0$. The peak corresponds to $Q\gg 1$ and gives 
\beq
\label{eq:gamma5_peak}
   \gamma_5^{\max}\approx Im\left(\sqrt{-\frac{ 2 F(0)}{F''(0)} }\right)=\sqrt{ \frac{\eta k_\theta^2 N^2 m}{16\Omega\mu^2}}
   \qquad (W\ll 1, \; f>0).
\eeq
It should be compared with the peak growth rate of the IW  mode at $k_\eta$ or the MW mode at $k_\nu$ (in the branch $\omega_2$ or $\omega_4$) where $\gamma_{\max}\approx \omA^2/4\Omega$. Note that the condition $\kappa<\eta$, required for the fifth branch instability, implies no MW instability at $k_\kappa$. However, the $k_\nu$ peak of MW instability or the $k_\eta$ peak of IW instability can be present simultaneously with $\gamma_5>0$. Rewriting \Eq~(\ref{eq:gamma5_peak}) as $\gamma_5^{\max}\approx (m\omA^2/4\Omega|\mu|)\sqrt{W/2|\mu|}$, one can see that $\gamma_5^{\max}$ becomes competitive with $\gamma_{\max}$ only when $W\sim1$, i.e. near marginal stability.

Note that the result stated in \Eq~(\ref{eq:om5_sol}) is based on an expansion in $W$, which breaks at low and high $k$ far from the instability peak at $k_0$. In particular, the expansion assumes $Wk^2/k_N^2\ll 1$, which breaks at $k\sim k_N/\sqrt{W}$. Therefore, \Eq~(\ref{eq:om5_sol}) does not describe the boundaries of the instability interval $k_1<k<k_2$. The boundaries $k_1$ and $k_2$ (where $\gamma_5$ changes sign) are given by the marginal stability analysis (Appendix~\ref{ap:PreviousProofs}) similar to the previous analysis of the TI criteria by \cite{spruit1999differential}. The instability interval exists if $k_2>k_1$, which requires $W<W_\star={\cal O}(1)$. Numerically, we find $W_\star\approx 11$ for $p=1$, $m=1$, and $\mu=1$. The value of $\gamma_5^{\max}(W)$ peaks at $W\sim 1$ and decreases to zero at $W=W_\star$ (Figure~\ref{fig:Fifthroot}).

It is straightforward to include the effects of viscosity $\nu$ and diffusivity $\kappa$ in the above analysis of the fifth branch. In particular, for $\nu\neq 0$ and $\kappa=0$ we find from \Eq~(\ref{eq:F_no_kappa}) that $F(0)$ and $F'(0)$ remain the same as in \Eq~(\ref{eq:F_derivatives}), and $F''(0)$ changes to
\beq
  F''(0)=-16\Omega\mu^2 m\omA^2 -2i\nu k^2\omA^2\left(m^2+m_\star^2 + \frac{k_N^2}{k^2}\right) + {\cal O}(W).
\eeq
One can see that viscosity weakly affects the $\gamma_5$ peak at $k\approx  k_N/\sqrt{f}$ when $k_N\ll k_\nu=(2\Omega|\mu|/\nu)^{1/2}$. 


\section{Previous analysis of Tayler instability with fast rotation}
\label{ap:PreviousProofs}
TI in the fast rotation regime $\Omega\gg\omA$ was previously investigated by \cite{spruit1999differential}, \cite{zahn2007magnetic}, and \cite{ma2019angular}. Their analysis assumed $\nu=0$ and so did not show instability of MW enabled by viscosity, which exists only when $\nu\gg \eta$. They also disregarded the IW branch of the dispersion relation and so missed the IW instability enabled by magnetic diffusivity when $\eta\gg\nu$. Under these limitations, one can still find the MW instability enabled by diffusivity $\kappa$, with the canonical growth rate $\gamma_{\max}\sim\omA^2/4\Omega$ at $k\sim k_\kappa$ \citep{spruit2002dynamo}. However, the TI criteria presented by \cite{spruit1999differential}, \cite{zahn2007magnetic}, and \cite{ma2019angular} show instability at $\kappa=0$. Hence, their criteria do not describe the MW branch. We have checked their analysis and found that they implicitly switched from MW to the fifth branch $\omega_5(k)$ when considering stability conditions. The fifth branch can indeed be unstable at $\nu=\kappa=0$, although normally with a small growth rate $\gamma_5^{\max}$ (Appendix~\ref{ap:fifth_root}), which makes it less important than the IW instability.

A basic fact missed in the previous work is the existence of more than one unstable branches. Let us start with the full dispersion relation~(\ref{eq:SphericalFullIntro}) with $\nu=0$ and $\kappa=0$ at $|\cos\theta|\approx 1$:
\beq
\label{eq:disp_5}
    F(\omega)=\left(\omega^2\omega_\eta-m_\star^2\omA^2\omega-\omA^2\frac{k_N^2}{k^2}\omega_\eta\right)\left(\omega\omega_\eta-m^2\omA^2\right)-4\left(\Omega \omega_\eta+m\omA^2\right)^2\omega=0 \qquad (\nu=\kappa=0).
\eeq
$F(\omega)$ is a fifth-order polynomial with five roots. When $\omA\ll\om$, there are two IW roots ($\omega_{1,2}\approx \pm \om$), two MW roots $\omega_{3,4}\sim\omA^2/\om$, and a fifth root $\omega_5$. Previous works focused on the roots with small $|\omega|\ll\omA\ll\Omega$. Then, terms with high powers of $\omega$ become negligible and the dispersion relation~(\ref{eq:disp_5}) simplifies to a cubic equation, 
\beq
\label{eq:D_simpleMW_appendix}
    F(\omega)=m^2m_\star^2\omA^4\omega+m^2\omA^4\frac{k_N^2}{k^2}\omega_\eta-4\left(\Omega \omega_\eta+m \omA^2\right)^2\omega=0 \qquad (|\omega|\ll\omA\ll\Omega).
\eeq
It gives roots $\omega_{\rm MW}=\omega_{3,4}$ and $\omega_5$. There are no unstable solutions at $m=0$, so only perturbations with $m>0$ are considered below.

Instead of directly finding the roots and examining $Im(\omega)$, previous works employed a marginal stability analysis \citep{chandra1961}, which may be stated as follows. As instability develops in a limited range of wavenumbers $k$, both stable ($Im(\omega)\leq 0$) and unstable ($Im(\omega)>0$) perturbations are present in the unstable branch of the dispersion relation $\omega(k)$. Then, there must exist a mode with $Im(\omega)=0$ at some $k$ (the boundary of the unstable region), and the condition for instability becomes the existence of real solutions $\omega$ of $F(\omega)=0$.

The marginal stability method is used in Appendix~A in \cite{spruit1999differential}, Section~3 of \cite{zahn2007magnetic},
and Appendix~C2 in \cite{ma2019angular} to derive the condition for TI. We repeat it below and for easy comparison define quantities $\alpha=\omega\Omega/m\omA^2$ and $h=\eta\Omega/m\omA^2$; in the most relevant case of $m=1$ these quantities are the same as $\alpha$ and $h$ in \cite{spruit1999differential}. We will retain our notation for the poloidal wavevector components $k_\theta$ and $k_z\approx k_R\approx k\gg k_\theta$; \cite{spruit1999differential} denoted them $l$ and $n$, respectively. The dispersion relation $F=0$ requires $Re(F)=0$ and $Im(F)=0$. For a real $\alpha$, this gives two real equations:

\begin{align}
\label{eq:spruit_re}
   Re(F)&= 4m^2\omA^4\alpha\left[ -(\alpha+1)^2+ \frac{m_\star^2}{4} + \frac{k_N^2}{4k^2}+h^2k^4 \right] = 0,\\
   Im(F)&= 8m^2\omA^4 \eta k^2\left[-\alpha(\alpha+1) + \frac{k_N^2}{8k^2}\right] = 0.
\label{eq:spruit_im}
\end{align}
Instability exists if both equations are satisfied at some $\alpha$ and $k$. They are not satisfied at $\alpha=0$. The required $\alpha\neq  0$ can be expressed in terms of $k$ in a few ways:
\begin{align}
\label{eq:alpha2}
   \alpha^2 &= 1- \psi, \qquad \psi\equiv \frac{m_\star^2}{4} + h^2k^4=\frac{m^2+2(1-p)}{4}+h^2k^4, \\
\label{eq:alpha3}
   \alpha &= x+\psi-1,   \qquad x\equiv \frac{k_N^2}{8k^2},  \\
\label{eq:alpha1}
   \alpha &= \frac{1}{2} \left( -1 + \epsilon \sqrt{1+4x} \right), \qquad  
    \epsilon=\pm 1. 
\end{align}
\Eqs~(\ref{eq:alpha2}) and (\ref{eq:alpha3}) are linear combinations of \Eqs~(\ref{eq:spruit_re}) and (\ref{eq:spruit_im}) while
\Eq~(\ref{eq:alpha1}) is obtained by solving \Eq~(\ref{eq:spruit_im}) for $\alpha$. Note that the root $\epsilon=-1$ corresponds to the MW branch, as it gives $\alpha=-1$ when $x\rightarrow 0$ (i.e. at large wavenumbers $k\gg k_N$). The root $\epsilon=1$ corresponds to the fifth branch $\omega_5(k)$ (Appendix~\ref{ap:fifth_root}), as it vanishes at $x\rightarrow 0$. A valid solution has $\epsilon={\rm sign}(\psi)$, so that $|\alpha|<1$ if $\psi>0$, as required by \Eq~(\ref{eq:alpha2}).

Magnetic configurations of main interest have moderate gradients $p<m^2/2+1$, which give $m_\star^2>0$ and $\psi>0$. Then, real solutions of $F(\alpha)=0$ exist only on the fifth branch $\omega_5(k)$. This is consistent with the absence of instability on the MW branch at $m_\star^2>0$ and $\nu=\kappa=0$ (Appendix~\ref{ap:MW}), and the existence of instability in the fifth branch (Appendix~\ref{ap:fifth_root}).

Comparing \Eqs~(\ref{eq:alpha3}) and (\ref{eq:alpha1}) with Equation~(A20) in \cite{spruit1999differential} and \Eq~ (\ref{eq:alpha1}) with the chosen solution of Equation~(12) in \cite{zahn2007magnetic}, one can see that both \cite{spruit1999differential} and \cite{zahn2007magnetic} chose the root with $\epsilon=1$, and so their discussion of instability conditions at $\kappa=\nu=0$ applies to the fifth branch rather than MW. \cite{spruit1999differential} found a necessary condition $p>m^2/2-1$, equivalent to $m_\star^2/4<1$. This condition is seen from  \Eq~(\ref{eq:alpha2}), which always requires $\psi<1$ and hence $m_\star^2/4<1$. \cite{ma2019angular} focused on magnetic configurations $B_\phi\propto \sin\theta$, which have $p=1$ near the polar axis, and so have $m_\star^2>0$ and $\psi>0$. The stability analysis in \cite{ma2019angular} closely followed \cite{spruit1999differential}; they also chose $\epsilon=1$ (see their their Appendix~C2) and thus also ended up applying marginal stability analysis to the fifth branch.

The instability in the fifth branch has a peak at $k_0= k_N/\sqrt{4-m_\star^2}$ with $\gamma_5^{\max}\approx(\omA^2/4\Omega)\sqrt{W/2}$ (Appendix~\ref{ap:fifth_root}), as long as $W<1$. One can use the marginal stability analysis to find the interval of $k$ for unstable perturbations at $W<1$. At $k<k_0$, one finds $h^2k^4<[W/2\mu(4-m_\star^2)]^2\ll 1$, and so $\psi\approx m_\star^2/4$. Then, \Eqs~(\ref{eq:alpha2}) and (\ref{eq:alpha3}) yield $\alpha\approx \sqrt{1-m_\star^2/4}$ and $x\approx \sqrt{1-m_\star^2/4} +1 - m_\star^2/4$. This solution gives the $k$-threshold for instability, $k>k_1=k_N/\sqrt{8x}$. In particular, for $p=1$ and $m=1$ one finds $k_1^2=k_N^2/(4\sqrt{3}+6)$, same as obtained by \cite{zahn2007magnetic}{ and \cite{ma2019angular}.
There is also an upper bound $k<k_2$ for unstable perturbations. It satisfies $k_2\gg k_N$ when $W\ll 1$, and so the upper boundary is at $x\ll 1$.  The corresponding real solution of $F(\alpha)=0$ is $\alpha=x\ll 1$ (from \Eq~(\ref{eq:alpha1})); then one finds $\psi\approx 1$ from \Eq~(\ref{eq:alpha1}), which gives $k_2^2\approx \sqrt{4-m_\star^2}/2h=k_N^2\sqrt{4-m_\star^2}/W$. Note that $k_2\gg k_1$ when $W\ll 1$, so $\gamma_5(k)$ has a small positive value in a broad interval of $k$, in addition to the sharp peak of width $\Delta k/k_0\sim \sqrt{W}$ at $k_0=k_N/\sqrt{4-m_\star^2}$ where it reaches $\gamma_5^{\max}\approx (\omA^2/4\Omega)(\sqrt{W/2})$  (Appendix~\ref{ap:fifth_root}). The instability interval $k_1<k<k_2$ disappears in the case of $W\gg 1$ because it gives $k_2<k_1$.

\cite{spruit1999differential} also used the marginal stability analysis to examine the regime of fast buoyancy diffusion $\kappa\gg\eta$. In this case, he obtained the instability condition $\omA^2>{\cal O}(1)\times \Omega\eta^2k_N^2/\kappa$. It is consistent with our condition~(\ref{eq:MWNS2}) for the existence of the $k_\kappa$ peak of instability in the MW branch.


\bibliography{refs}{}
\bibliographystyle{aasjournal}



\end{document}